\documentclass[apj,twocolumn,twocolappendix,numberedappendix,appendixfloats]{openjournal}
\usepackage{amsmath}
\usepackage[dvipsnames]{xcolor} 
\usepackage[breaklinks,colorlinks,citecolor=blue,urlcolor=blue,linkcolor=blue]{hyperref}
\usepackage{orcidlink}
\usepackage{graphicx}
\usepackage{amssymb}
\usepackage[dvipsnames]{xcolor}
\usepackage{subfigure}

\newcommand{\UF}{\affiliation{$^{1}$Department of Astronomy, University of Florida, Bryant Space Science Center, Gainesville, FL 32611-2055, USA}}
\newcommand{\CfA}{\affiliation{$^{2}$Center for Astrophysics \textbar{} Harvard \& Smithsonian, 60 Garden Street, Cambridge, MA 02138-1516, USA}}

\def\Angstrom{\textup{\AA}}
\def\Msun{{\rm M}_\odot}
\def\Rsun{{\rm R}_\odot}

\begin{document}

\title{\bf Type Ia-CSM Supernovae. I. Uniform Light-Curve Characterization and Inference on the Underlying Type Ia Explosions}
\shorttitle{Type Ia-CSM Supernovae. I.}
\shortauthors{Hiramatsu et al.}

\author{\href{https://orcid.org/0000-0002-1125-9187}{Daichi~Hiramatsu}$^{1}$\orcidlink{0000-0002-1125-9187}}
\author{\href{https://orcid.org/0000-0002-6347-3089}{Daichi~Tsuna}$^{2}$\orcidlink{0000-0002-6347-3089}}
\author{\href{https://orcid.org/0000-0002-1417-8024}{Morgan~MacLeod}$^{2}$\orcidlink{0000-0002-1417-8024}}

\UF
\CfA

\email{Corresponding author: dhiramatsu@ufl.edu}

\begin{abstract}
We present the largest uniform study to date of 50 thermonuclear Type Ia supernovae interacting with circumstellar medium (SNe~Ia-CSM). As the first paper in a series delineating the population properties of this class, this work focuses specifically on a uniform analysis of their light curves. Our results show that SNe~Ia-CSM display a significantly more uniform light-curve morphology than interacting core-collapse SNe~IIn, closely resembling the luminous-slow subgroup. The Ia-CSM sample spans approximately an order of magnitude in peak luminosity, total timescale, and radiated energy, with a median optical peak luminosity of $\approx2\times10^{43}$\,erg\,s$^{-1}$ ($\approx-20$ mag). We find a distinct bimodality in the radiated energy distribution: a predominant population of energetic events with $\approx 2\times10^{50}$\,erg and characteristic timescales of $\approx100$ days, and a secondary population that is roughly an order of magnitude less energetic with shorter timescales of $\approx50$ days. Decomposition of the luminosity contributions from underlying SN~Ia and CSM interaction by the Ia-CSM spectral classification reveals a diverse range of underlying SN~Ia subtypes spanning from luminous 1991T-like to faint 1991bg-like. Both SN~Ia and CSM properties are broadly consistent with the core-degenerate channel resulting from common-envelope evolution followed by core merger, where the presence of dense CSM is dictated by the delay time between the common-envelope phase and the thermonuclear explosion. Subsequent papers in this series will utilize numerical simulations, semi-analytical modeling, spectroscopy, and host-galaxy environments to further constrain the progenitor and explosion pathways to SNe~Ia-CSM.
\end{abstract}

\keywords{
\href{https://vocabs.ands.org.au/repository/api/lda/aas/the-unified-astronomy-thesaurus/current/resource.html?uri=http://astrothesaurus.org/uat/1668}{Supernovae (1668)}; 
\href{https://vocabs.ands.org.au/repository/api/lda/aas/the-unified-astronomy-thesaurus/current/resource.html?uri=http://astrothesaurus.org/uat/1728}{Type Ia supernovae (1728)}; 
\href{https://vocabs.ands.org.au/repository/api/lda/aas/the-unified-astronomy-thesaurus/current/resource.html?uri=http://astrothesaurus.org/uat/1731}{Type II supernovae (1731)}; 
\href{https://vocabs.ands.org.au/repository/api/lda/aas/the-unified-astronomy-thesaurus/current/resource.html?uri=http://astrothesaurus.org/uat/1799}{White dwarfs (1799)}; 
\href{https://vocabs.ardc.edu.au/repository/api/lda/aas/the-unified-astronomy-thesaurus/current/resource.html?uri=http://astrothesaurus.org/uat/655}{Giant stars (655)};
\href{https://vocabs.ardc.edu.au/repository/api/lda/aas/the-unified-astronomy-thesaurus/current/resource.html?uri=http://astrothesaurus.org/uat/1661}{Supergiant stars (1661)};
\href{https://vocabs.ardc.edu.au/repository/api/lda/aas/the-unified-astronomy-thesaurus/current/resource.html?uri=http://astrothesaurus.org/uat/241}{Circumstellar matter (241)};
\href{https://vocabs.ardc.edu.au/repository/api/lda/aas/the-unified-astronomy-thesaurus/current/resource.html?uri=http://astrothesaurus.org/uat/2154}{Common envelope evolution (2154)}
\href{https://vocabs.ardc.edu.au/repository/api/lda/aas/the-unified-astronomy-thesaurus/current/resource.html?uri=http://astrothesaurus.org/uat/2157}{Stellar mergers (2157)}
}

\maketitle

\section{Introduction}\label{sec:intro}

Type~Ia supernovae (SNe~Ia) are the thermonuclear explosions of carbon-oxygen (CO) white dwarfs (WDs; \citealt{Whelan1973ApJ...186.1007W,Nomoto1982ApJ...253..798N,Nomoto1982ApJ...257..780N,Nomoto1984,Iben1984ApJS...54..335I,Webbink1984ApJ...277..355W,Woosley1986ApJ...301..601W,Pakmor2010Natur.463...61P,Nugent2011Natur.480..344N,Bloom2012ApJ...744L..17B}, see also \citealt{Howell2011Ia,Maoz2014Ia,Maguire2016Ia,Taubenberger2017hsn..book..317T,Maeda2022hxga.book...75M,Ruiter2025A&ARv..33....1R} for reviews). Despite their historical cosmological utility in discovering the accelerating expansion of the Universe and constraining its energy contents \citep{Riess1998,Perlmutter1999}, the exact progenitor systems and explosion mechanisms of SNe~Ia remain fundamental open questions. Three leading theoretical frameworks exist: the single-degenerate scenario \citep{Whelan1973ApJ...186.1007W, Nomoto1982ApJ...253..798N,Nomoto1982ApJ...257..780N,Nomoto1984}, where a CO WD accretes material from a non-degenerate companion; the double-degenerate scenario \citep{Iben1984ApJS...54..335I,Webbink1984ApJ...277..355W,Woosley1986ApJ...301..601W}, involving the accretion from or merger with a companion WD; and the core-degenerate (CD) scenario \citep{Sparks1974ApJ...188..149S,Livio2003ApJ...594L..93L,Kashi2011MNRAS.417.1466K}, in which a CO WD merges with the stellar core of a non-degenerate companion through common-envelope (CE) evolution. This CE can potentially act as a dense, confined circumstellar medium (CSM) if the merger-driven explosion promptly follows the CE phase.

CSM interaction serves as the primary power source for core-collapse Type~IIn supernovae (SNe~IIn), which are characterized by narrow Balmer-series emission lines in their optical spectra (\citealt{Schlegel1990MNRAS.244..269S,Filippenko1997ARA&A..35..309F}; see also \citealt{Smith2017} for a review). However, severe CSM contamination and pronounced observational heterogeneity (e.g., \citealt{Nyholm2020,Hiramatsu2026ApJ..1005...82H}) obscure the precise nature of the underlying explosions and their progenitor systems.
In this context, SNe~Ia-CSM occupy a compelling intersection between classic SNe~Ia and SNe~IIn, displaying prominent Balmer emission lines superimposed on a diluted SN~Ia-like continuum (see \citealt{Silverman2013_sample,Sharma2023ApJ...948...52S} for previous sample studies). 
A subset of SNe~Ia-CSM show clear, Ia-dominated light-curve and spectral evolution at early phases -- exemplified by SN~2002ic \citep{Hamuy2003ic,Deng2004,Wood-Vasey2004} and PTF11kx \citep{Dilday2012,Silverman2013}, as well as the more recent cases of SN~2020eyj \citep{Kool2023Natur.617..477K} and SN~2020aeuh \citep{Tsalapatas2025A&A...704A.135T}, which feature hydrogen-poor/helium-rich and hydrogen/helium-poor CSM, respectively -- unambiguously confirming a thermonuclear origin and variation in the CSM composition. Conversely, the majority of SNe~Ia-CSM are dominated by strong CSM interaction throughout their evolution without exhibiting pristine Ia signatures (e.g., \citealt{Leloudas2015,Hiramatsu2026arXiv260712040H}). This spectral masking leaves open the possibility of a hydrogen/helium-poor core-collapse origin for some events, as debated in the case of SN~2012ca (e.g., \citealt{Fox2015,Inserra2016}, but see also \citealt{Leloudas2015} suggesting that these misclassifications could only occur with low signal-to-noise spectra).
Establishing a definitive thermonuclear origin for SNe~Ia-CSM is of paramount importance: if they indeed arise from SNe~Ia, the presence of dense CSM makes them the premier observational candidates for the CD channel (e.g., \citealt{Sparks1974ApJ...188..149S,Taam2000,Livio2003ApJ...594L..93L,Chugai2004AstL...30...65C,Hachisu2008ApJ...679.1390H,Kashi2011MNRAS.417.1466K,Soker2015,Soker2019,Ablimit2021}).

In an effort to explore and map the diverse landscape of progenitor, CSM, and explosion properties within this class, we present the largest sample study to date of SNe~Ia-CSM, comprising 50 events with well-sampled light curves and confirmed spectroscopic classifications. In this first paper of the series, we focus on a uniform analysis of their optical light-curve properties to extract empirical features -- such as peak luminosities, characteristic timescales, and radiated energies -- and evaluate their statistical distributions and correlations. Subsequent papers in this series will present a population-level comparative analysis against numerical light-curve simulations of SNe~Ia-CSM (e.g., \texttt{STELLA} models; \citealt{Hiramatsu2026arXiv260712040H}); detailed semi-analytical modeling of individual events to constrain explosion and CSM parameters (e.g., \texttt{TransFit-CSM}; \citealt{Zhang2026ApJ...999..186Z}); a comprehensive spectroscopical analysis of the SN~Ia continuum and SN-CSM shock interaction lines; and an investigation into their host galaxies and local environments.

This paper (Paper I) is structured as follows. In \S\ref{sec:sample}, we outline our methodology for constructing the largest sample of SNe~Ia-CSM. In \S\ref{sec:phot}, we describe the observational data sources compiled for light-curve construction. In \S\ref{sec:mea}, we introduce a uniform Gaussian Process (GP) regression framework used to generate standardized, rest-frame and time-sampled light curves, and detail the measurements extracted from these GP light curves. In \S\ref{sec:ana}, we analyze the resulting light-curve properties and investigate potential correlations and multimodalities among the measured and calculated features. Finally, we summarize our key findings and discuss their broader physical implications in \S\ref{sec:sum}, and offer concluding remarks in \S\ref{sec:conc}.

\section{Sample Construction} 
\label{sec:sample}

\begin{deluxetable}{cc}
\tablecaption{Sample Selection Criteria\label{tab:cut}}
\tablehead{
\colhead{Selection Criterion} & \colhead{Sample Size}
}
\startdata
Ia-CSM on TNS/WISeREP + literature & 45+14=59 \\
Verified Ia-CSM classification & 56 \\
Light-curve peak coverage in $g$/$r$ bands & 50 \\
\enddata
\end{deluxetable}

We construct our sample of SNe~Ia-CSM from transients discovered between 2016-01-01 and 2025-12-31 that had been classified as ``Ia-CSM'' by our last query date on 2026-10-22 on the Transient Name Server\footnote{\url{https://www.wis-tns.org/}} (TNS) and/or the Weizmann Interactive Supernova Data Repository\footnote{\url{https://wiserep.weizmann.ac.il}} (WISeREP; \citealt{Yaron2012PASP..124..668Y}). The discovery window baseline is anchored by the official roll-out of TNS, while the end date is chosen to guarantee sufficient temporal baseline post-discovery ($\gtrsim300$ days) prior to our last query.
To verify each Ia-CSM classification with narrow (${\rm FWHM}\lesssim2$,$000$\,km\,s$^{-1}$) Balmer-series emission lines superimposed on a diluted, SN~Ia-like continuum with Fe~{\sc ii/iii}, Si~{\sc ii}, and/or S~{\sc ii} absorption alongside Ca~{\sc ii} emission \citep{Silverman2013_sample}, we require that each event has at least one publicly available spectrum at any phase. This requirement enables visual inspection and cross-correlation against standard supernova spectral template libraries, specifically \texttt{Superfit} \citep{Howell2005ApJ...634.1190H} (along with its Python-based successor, \texttt{NGSF}; \citealt{Goldwasser2022TNSAN.191....1G}), \texttt{SNID} \citep{Blondin2007ApJ...666.1024B}, and \texttt{GELATO} \citep{Harutyunyan2008A&A...488..383H}.

As of our last query date, the initial search on TNS and WISeREP retrieved 45 events classified as Ia-CSM. We exclude three contaminants whose spectra are inconsistent with the Ia-CSM classification. We then incorporate additional Ia-CSM re-classifications from the recent literature: four events from \cite{Sharma2023ApJ...948...52S}\footnote{This is in addition to the eight events from that study already classified as Ia-CSM on TNS and/or WISeREP.}, nine from \cite{Hiramatsu2026ApJ..1005...82H}, and one from \cite{Aleo2023ApJS..266....9A}.
Finally, we enforce a light-curve quality cut requiring sufficient temporal sampling around the optical peak to facilitate robust parameter extraction (see \S\ref{sec:mea}), yielding a finalized sample of 50 SNe~Ia-CSM.
Our sequential sample selection criteria are summarized in Table~\ref{tab:cut}. Additional descriptions of the final sample are tabulated in Table~\ref{tab:sample}, while the omitted transients and their exclusion rationales are detailed in Table~\ref{tab:exc}.

\section{Photometry}
\label{sec:phot}

We compiled publicly available photometry up to our last query date of 2026-10-22 (${\rm MJD}=61336$) from the Asteroid Terrestrial-impact Last Alert System (ATLAS; \citealt{Tonry2018PASP..130f4505T,Smith2020PASP..132h5002S}), Zwicky Transient Facility (ZTF; \citealt{Bellm2019PASP..131a8002B,Graham2019PASP..131g8001G}), and \textit{Gaia} Spacecraft \citep{Gaia2016A&A...595A...1G}.

\subsection{ATLAS}
\label{sec:atlas}

We retrieved ATLAS photometry in the cyan ($c$; 4,200--6,500$\,\Angstrom$) and orange ($o$; 5,600--8,200$\,\Angstrom$) bands from the ATLAS forced photometry server\footnote{\url{https://fallingstar-data.com/forcedphot/}} \citep{Shingles2021TNSAN...7....1S}. Following \citet{Young2024zndo..10978968Y}, we applied standard quality cuts with a $5\sigma$ clipping to filter out outlier points. For each SN and filter, we measured a background baseline by calculating the median flux of a minimum of 30 data points within temporal windows situated well before ($>500$ days) and/or after ($>1000$ days) the SN discovery date. All subsequent flux measurements were then corrected relative to this measured baseline.

\subsection{ZTF}
\label{sec:ztf}

We retrieved ZTF photometry in the $g$, $r$, and $i$ bands from the ZTF forced photometry service\footnote{\url{https://ztfweb.ipac.caltech.edu/batchfp.html}} \citep{Masci2019PASP..131a8003M,Masci2023arXiv230516279M}. Following \citet{Masci2023arXiv230516279M}, we enforced standard quality cuts to remove outlier data points. We applied additional $5\sigma$ clipping to further minimize photometric scatter. For each SN and unique combination of filter, ZTF field, and CCD quadrant, we corrected the flux measurements against a background baseline using a method analogous to our ATLAS reduction. Finally, we corrected the flux uncertainties using the reduced $\chi^{2}$ distribution of the point-spread function fits.

\subsection{Gaia}
\label{sec:gaia}

We retrieved \textit{Gaia} photometry in the broad $G$ band (3,300--10,500$\,\Angstrom$) from the \textit{Gaia} Science Alerts\footnote{\url{http://gsaweb.ast.cam.ac.uk/alerts/home}} \citep{Hodgkin2021A&A...652A..76H}. We discarded epochs flagged with ``null'' or ``untrusted'' magnitude measurements. With the lack of individual error in the alert light curves, we estimate the uncertainty for each magnitude measurement by adopting the median standard deviation at the specific magnitude, following the empirical calibration described in \citet[see their Figure~15]{Hodgkin2021A&A...652A..76H}.

\subsection{Additional Data Source}
\label{sec:add}

In addition, we supplement our data sets with publicly available Pan-STARRS1 \citep{PS12016arXiv161205560C} photometry in the $g$, $r$, $i$, and $z$ bands for three events in our sample (see Table~\ref{tab:sample}). We retrieved these data from the Young Supernova Experiment Data Release 1\footnote{\url{https://zenodo.org/records/7317476}} (YSE DR1; \citealt{Jones2021ApJ...908..143J,Aleo2023ApJS..266....9A}).

\subsection{Postprocessing}
\label{sec:post}

Throughout this work, we report all flux measurements in units $\mu$Jy. All photometry is uncertainty-weighted and stacked every 0.5 days in flux space to account for negative flux (if avaiable). The detection significance ($\sigma$) is determined from the ratio of measured flux ($F_\nu$) to its error ($F_{\nu,{\rm err}}$). For the measurements above and below $3\sigma$, we report their magnitudes ($-2.5\,{\rm log}_{10}(F_\nu)+{\rm ZP}$) and $3\sigma$ upper limits ($-2.5\,{\rm log}_{10}(3\times F_{\nu,{\rm err}})+{\rm ZP}$), respectively, where ``ZP'' is the zero-point in the AB magnitude system, i.e., $-2.5\,{\rm log}_{10}(3631\,{\rm Jy}/\mu{\rm Jy})=23.9$ mag.

We assume a standard $\Lambda$CDM cosmology with $H_0=71.0$\, km\,s$^{-1}$\,Mpc$^{-1}$, $\Omega_{\Lambda}=0.7$, and $\Omega_m=0.3$, and convert the redshift to a luminosity distance. Due to the lack of spectra with the resolved host-galaxy Na~{\sc i}~D doublet  absorption for the bulk of the sample, we only account and correct for the Milky Way extinction from \cite{Schlafly2011ApJ...737..103S}, assuming the \cite{Fitzpatrick1999PASP..111...63F} reddening law with $R_V=3.1$.

\section{GP Measurements}
\label{sec:mea}

To achieve uniform measurements across a large sample with diverse survey systematics (e.g., cadences, filters, and depths) in a model-independent way, we perform a two-dimensional (2D) GP interpolation on the light curves across both epochs and frequencies. We utilize \texttt{GaussianProcessRegressor} from the \texttt{scikit-learn} package \citep{Pedregosa2011JMLR...12.2825P}, closely following the methodology developed by \citet[hereafter H26]{Hiramatsu2026ApJ..1005...82H} for the SN~IIn sample study (see their Figure 1).

We perform measurements directly on the 2D GP light curves in the well-constrained phase above $3\sigma$ threshold ($F_\nu/F_{\nu, {\rm err}}\geq3$). For each SN and filter, we find a maximum flux and measure the timescale at each available 10\% flux level on the rise and decline with respect to the maximum value (see Figure~2 of H26). For example, $t_{{\rm rise}50}$ (and $t_{{\rm decl}50}$) corresponds to the duration between $50\%$ flux to (from) the maximum on the rise (decline). To ensure that the maximum flux is indeed the peak flux, we require that both $t_{{\rm rise}90}$ and $t_{{\rm decl}90}$ are properly measured with uncertainty. With this requirement for the rest-frame peak coverage in the $g$ and/or $r$ bands, we exclude 6 objects from the 56 with verified Ia-CSM classification in our sample (Table~\ref{tab:exc}).

Using the 2D GP light curve for each sample, we also construct bolometric and pseudo-bolometric (i.e., optical: 3,250--8,900$\,\Angstrom$) light curves by fitting and integrating a blackbody spectral energy distribution (SED)\footnote{This assumption is reasonable given a typical SN Ia-CSM SED dominated by a blue continuum with a narrow H$\alpha$ emission line, except during late phases ($\gtrsim100$ days) when the H$\alpha$ begins to dominate the SED. Because of its narrow FWHM ($\lesssim 100\,\Angstrom$), the H$\alpha$ contribution is largely smeared out within broad-band photometry.} at every epoch containing GP interpolations above $3\sigma$ in at least three filters.\footnote{Excluding \textit{Gaia} $G$ band due to its broad wavelength coverage.} We then perform the same set of measurements on the bolometric and optical light curves (see Figure~3 of H26). 
We note that SEDs generally peak blueward of the optical coverage in the early phase (i.e., $\gtrsim10$,$000$\,K), and the fitted blackbody temperatures may be underestimated by up to $\sim10$,$000$\,K alongside the radius overestimations, which translates to a factor of up to $\sim5$ in the bolometric luminosity  (e.g., \citealt{Valenti2016MNRAS.459.3939V,Arcavi2022ApJ...937...75}). Given this potential uncertainty in the bolometric luminosity at early times, we center our analysis on the optical luminosity.

\section{Analysis} 
\label{sec:ana}

In the following analysis, we use the rest-frame measurements in the $r$-band and pseudo-bolometric (optical) light curves, unless otherwise noted. Where appropriate, we also compare our Ia-CSM sample with the H26 sample of 490 SNe~IIn. We note that the overall trends and correlations are similar in the other band and bolometric light curves, and we provide all the measurements as the data behind Figure~\ref{fig:Mr_z}. 

To quantify potential multimodality in the Ia-CSM sample distributions, we apply a Gaussian mixture model (GMM) composed of $k$-independent Gaussian distributions (each with its own mean and variance), and find the best combination of $k$ Gaussians with the lowest Bayes information criterion (BIC). We do so by utilizing \texttt{GaussianMixture} from \texttt{scikit-learn}. Following \cite{Nyholm2020}, we consider $k=1,2,3$ and the model with $\Delta {\rm BIC}\geq6$ from the surrounding models to be significant, and plot  it on the 1D distributions.

\subsection{Peak Brightness versus Redshift}
\label{sec:z}

\begin{figure}
    \centering
    \includegraphics[width=0.49\textwidth]{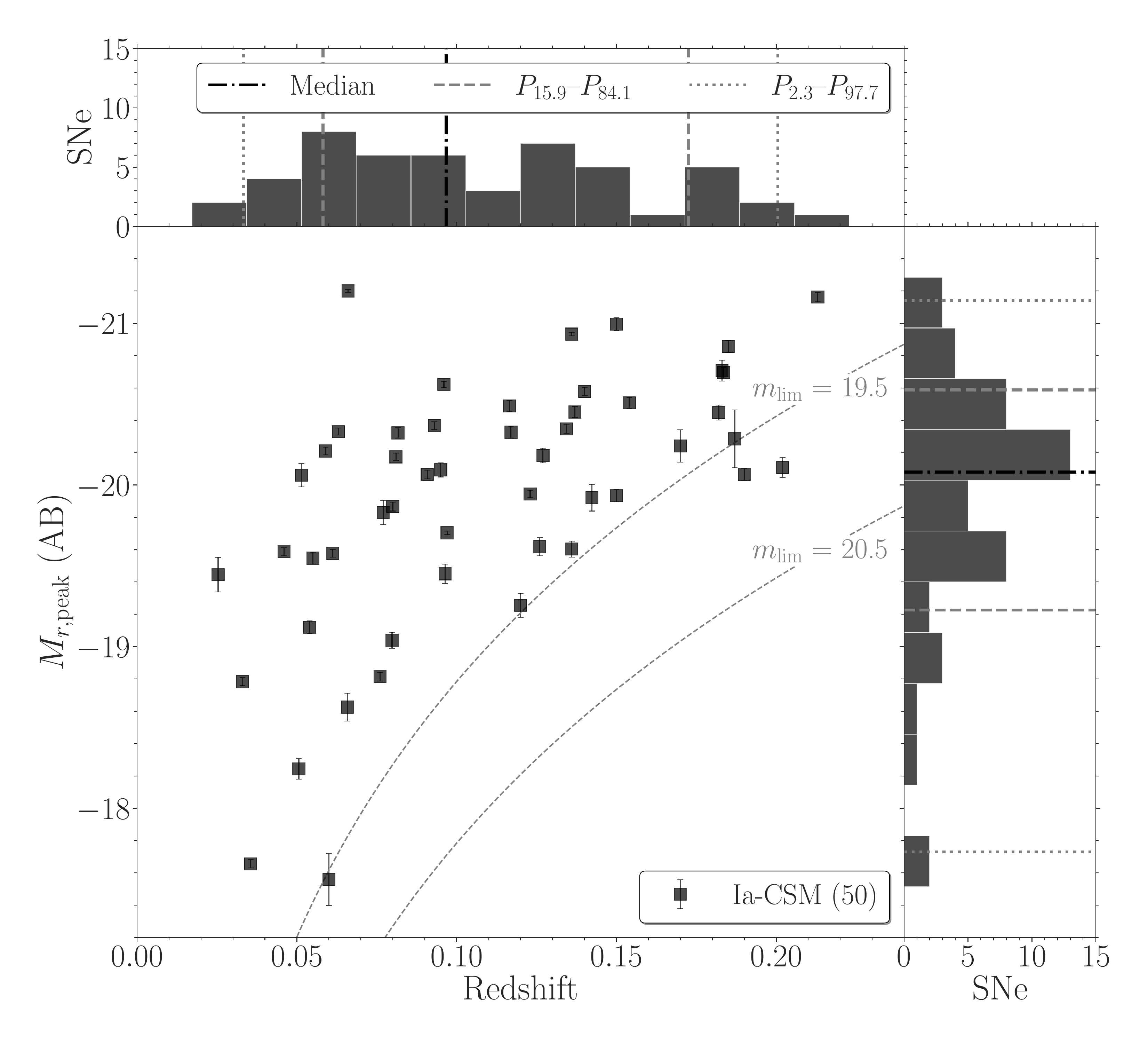}
\caption{Peak $r$-band absolute magnitude vs redshift with projected histograms of the Ia-CSM sample.
The error bars denote $1\sigma$ uncertainties. The dashed curves correspond to a range of survey limiting magnitudes: $m_{\rm lim}\approx19.5$ and $20.5$ for ATLAS and ZTF, respectively.
(The data used to create this figure and all other measurements are available in machine-readable format.)
 }
    \label{fig:Mr_z}
\end{figure}


\begin{figure*}
    \centering
    \includegraphics[width=0.49\textwidth]{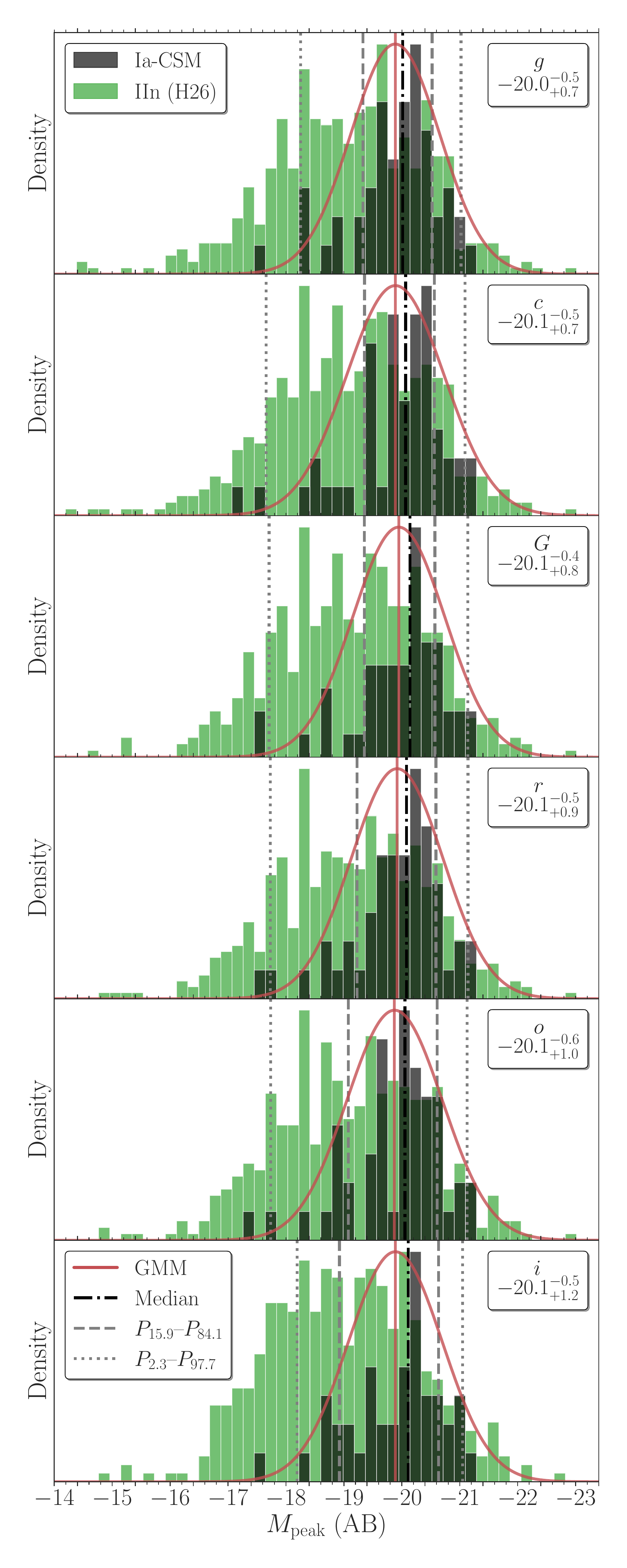}
    \includegraphics[width=0.49\textwidth]{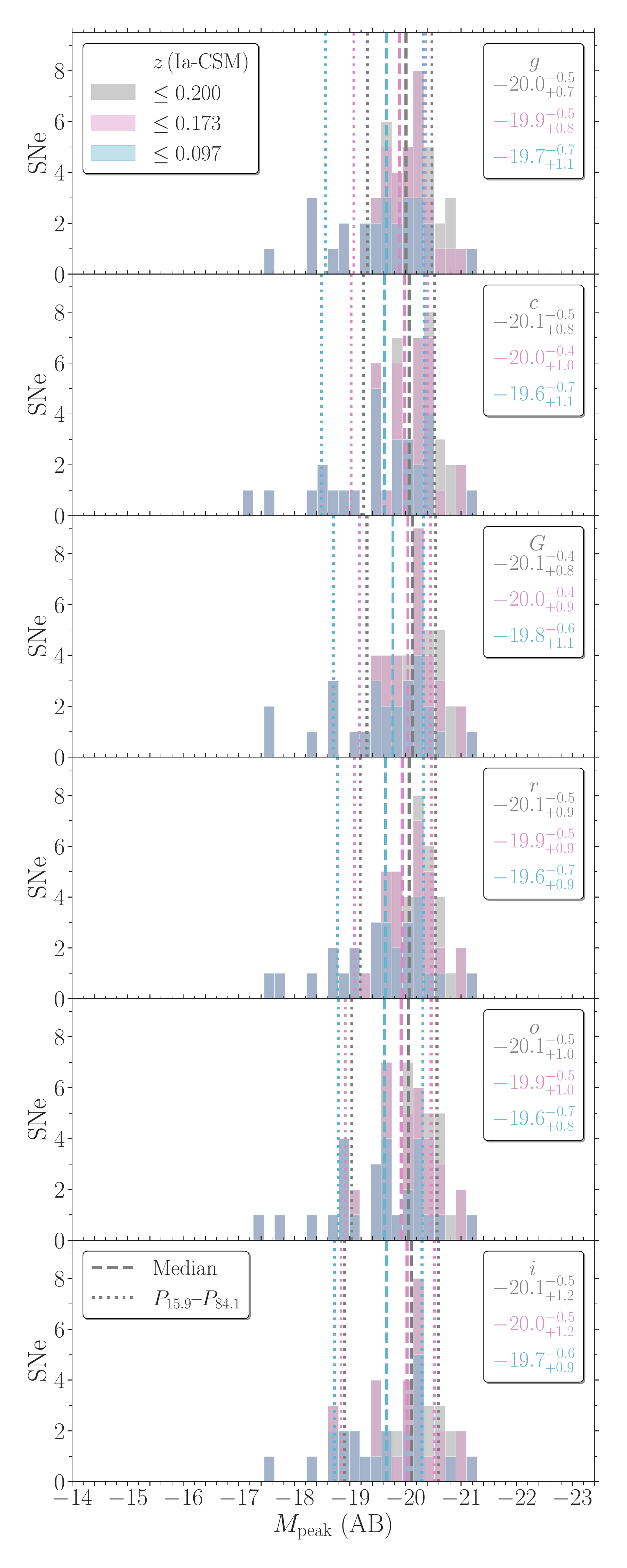}
\caption{\textit{Left}: Histograms of the peak absolute magnitude of the Ia-CSM sample in each filter ($g$, $c$, $G$, $r$, $o$, and $i$ from \textit{top} to \textit{bottom}), along with the IIn sample from \citet[H26]{Hiramatsu2026ApJ..1005...82H}. The black and gray vertical lines mark the median ($\widetilde{M}_{{\rm peak}}$ with its value on the top right) and $P_{15.9}$--$P_{84.1}$ and $P_{2.3}$--$P_{97.7}$ ranges of the Ia-CSM sample in each filter, while the red single Gaussian shows the GMM with the lowest BIC. The peak distributions span $\approx1.4$ and $\approx3.2$ mag in all filters at the $P_{15.9}$--$P_{84.1}$ and $P_{2.3}$--$P_{97.7}$ ranges, respectively.
\textit{Right}: Similar to \textit{left}, but for the Ia-CSM sample with upper redshift cuts at the median, $P_{84.1}$, $P_{97.9}$ redshift values (Figure~\ref{fig:Mr_z}). The vertical lines mark each subsample's $\widetilde{M}_{{\rm peak}}$ and $P_{15.9}$--$P_{84.1}$ range in each filter, which shift $<0.5$ mag with the redshift cuts.
}
    \label{fig:bandP}
\end{figure*}

\begin{figure*}
    \centering
    \includegraphics[width=0.49\textwidth]{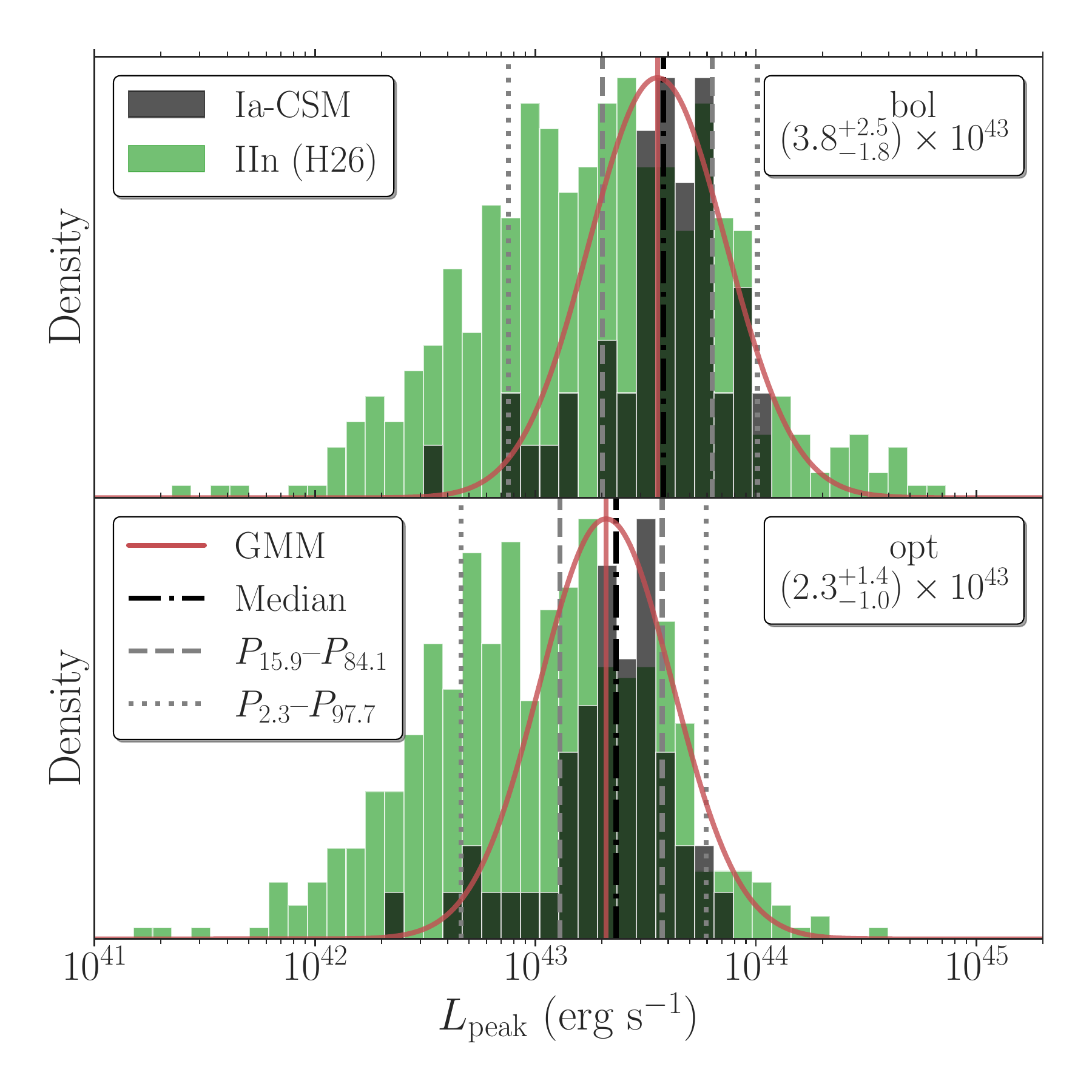}
    \includegraphics[width=0.49\textwidth]{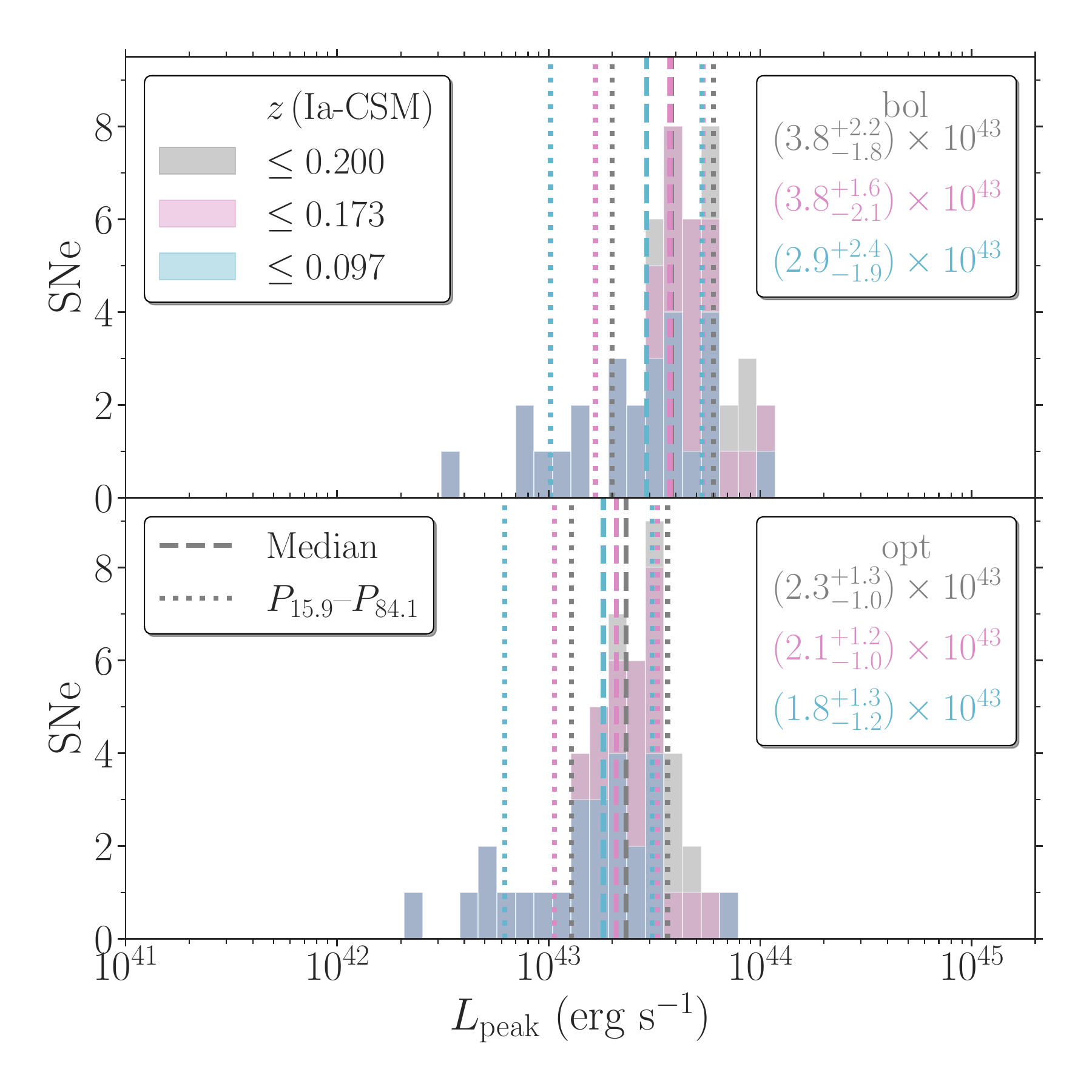}
\caption{Similar to Figure~\ref{fig:bandP}, but for the bolometric (\textit{top}) and optical (\textit{bottom}) luminosity. The $P_{15.9}$--$P_{84.1}$ and $P_{2.3}$--$P_{97.7}$ ranges of the Ia-CSM sample are about factors of 3 and 13, respectively. The median offset between the optical and bolometric peaks of the Ia-CSM sample is about a factor of $1.6$, which can be used as a rough proxy for the overall optical-bolometric correction of the class. 
}
    \label{fig:bolP}
\end{figure*}

In Figure~\ref{fig:Mr_z}, we plot the peak $r$-band absolute magnitude ($M_{r,{\rm peak}}$) versus redshift for the SN~Ia-CSM sample. The median redshift is $0.097$ with the 15.9 and 84.1th percentiles (which we denote as $P_{15.9}$ and $P_{84.1}$ in this work) at $0.058$ and $0.173$, respectively, and the 2.3 and 97.7th percentiles ($P_{2.3}$ and $P_{97.7}$) at $0.033$ and $0.200$, respectively. The median $\widetilde{M}_{r,{\rm peak}}$ is $-20.08$ mag with the $P_{15.9}$--$P_{84.1}$ and $P_{2.3}$--$P_{97.7}$ spreads of $-19.23$ to $-20.59$ mag and $-17.73$ to $-21.14$ mag, respectively. 

In addition, we show histograms of $M_{\rm peak}$ in each filter in the left panels of Figure~\ref{fig:bandP}. $\widetilde{M}_{\rm peak}$ values are $-20.0^{-0.5}_{+0.7}$, $-20.1^{-0.5}_{+0.7}$, $-20.1^{-0.4}_{+0.8}$, $-20.1^{-0.5}_{+0.9}$, $-20.1^{-0.6}_{+1.0}$, and $-20.1^{-0.5}_{+1.2}$ mag in the $g$, $c$, $G$, $r$, $o$, and $i$ bands, respectively, with the upper/lower bounds corresponding to the $P_{15.9}$--$P_{84.1}$ range. 
The sample spans $\approx 3.2$ mag in all filters at the $P_{2.3}$--$P_{97.7}$ range, which is narrower than the IIn sample range of $\approx5.0$ mag (H26), occupying the bright end of the IIn distribution but without exceeding $M_{\rm peak} < -21.2$ mag. We find no significant multimodality in $M_{\rm peak}$ in any filters when applying GMM.

To examine the sample Malmquist bias, we apply upper redshift cuts at the median, $P_{84.1}$, and $P_{97.7}$ redshift values in the right panels of Figure~\ref{fig:bandP}. These cuts correspond to $M_{r, {\rm lim}}\leq-17.7$, $-19.0$, and $-19.4$ mag at the ZTF limiting magnitude (or $\approx P_{98}$, $P_{88}$, $P_{82}$ in the $M_{\rm peak}$ distribution), respectively (Figure~\ref{fig:Mr_z}). This results in $<0.5$ mag shifts in $\widetilde{M}_{{\rm peak}}$ across all filters, given the rather flat sample redshift distribution. Thus, we do not correct the sample for Malmquist bias, but such a correction should marginally lower the median, well within the $P_{15.9}$--$P_{84.1}$ spread (e.g., \citealt{Nyholm2020}).

In the left panels of Figure~\ref{fig:bolP}, we show histograms of the peak bolometric and optical luminosities ($L_{\rm peak}$). The median $\widetilde{L}_{\rm peak}$ values are $(3.8^{+2.5}_{-1.8})\times10^{43}$ and $(2.3^{+1.4}_{-1.0})\times10^{43}$\,erg\,s$^{-1}$ in bolometric and optical, respectively, with an offset of $\approx 1.6$ between them. This offset can be used as a rough proxy for the overall optical-bolometric correction of the Ia-CSM sample (although we reiterate the potential underestimation of the bolometric peak luminosity as noted in \S\ref{sec:mea}). As similarly seen in the filter peaks, the sample spans, without significant multimodality, about a factor of 13, i.e., $2.5{\rm log}_{10}(13)=2.8$ mag, in both bolometric and optical peaks at the $P_{2.3}$--$P_{97.7}$ range. The spread is smaller than the IIn sample range of a factor of 100 (H26) and resides in the luminous end of the IIn distribution but with an upper limit at $L_{\rm peak} < 10^{44}$\,erg\,s$^{-1}$.
In the right panels, we show the subsamples with the redshift cuts as in Figure~\ref{fig:bandP}. The $L_{\rm peak}$ and $P_{15.9}$--$P_{84.1}$ range shift by $<25\%$, again due to the rather flat redshift distribution (Figure~\ref{fig:Mr_z}).

\subsection{Rise and Decline Timescales} 
\label{sec:time}

\begin{figure*}
    \centering
    \includegraphics[width=0.49\textwidth]{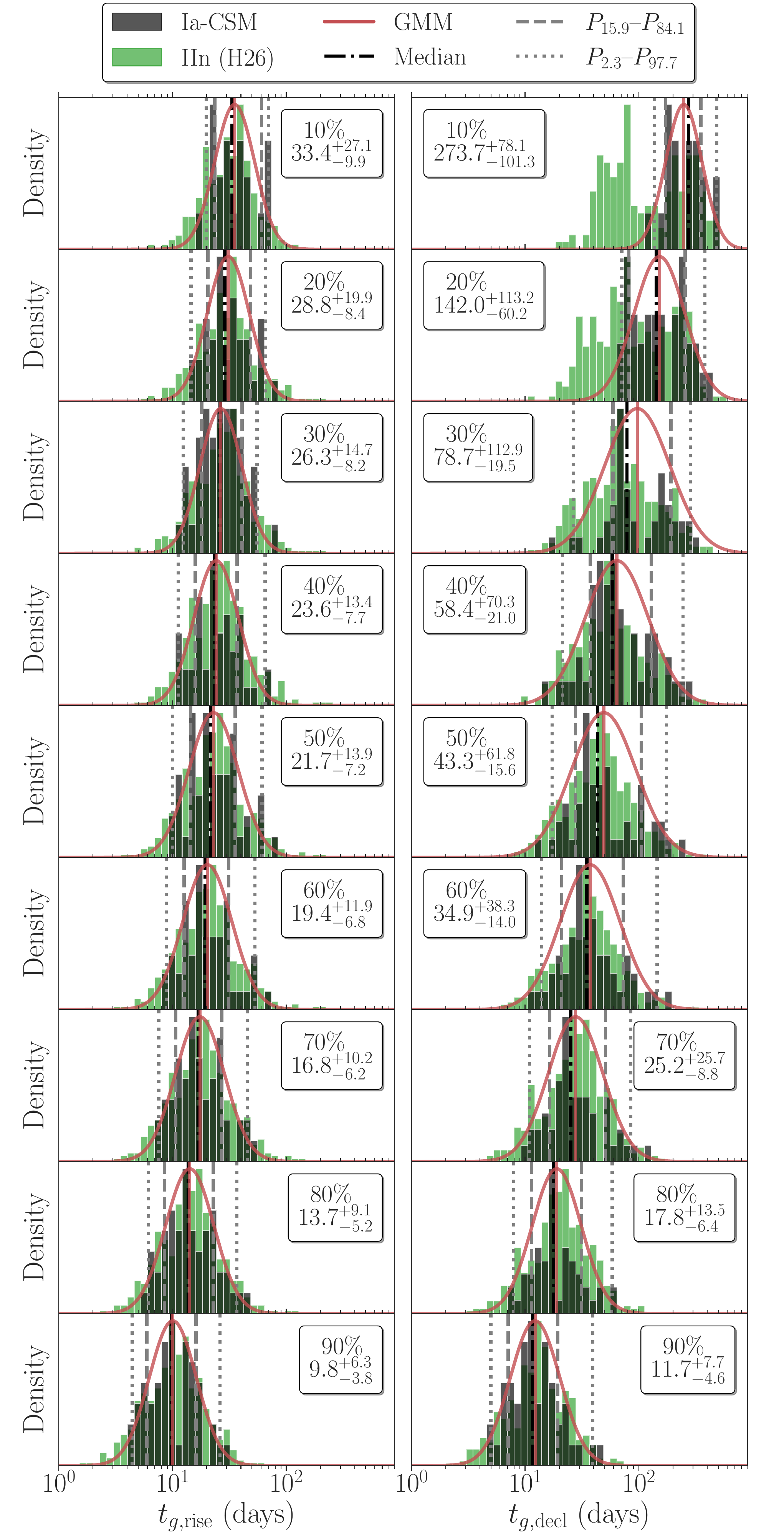}
    \includegraphics[width=0.49\textwidth]{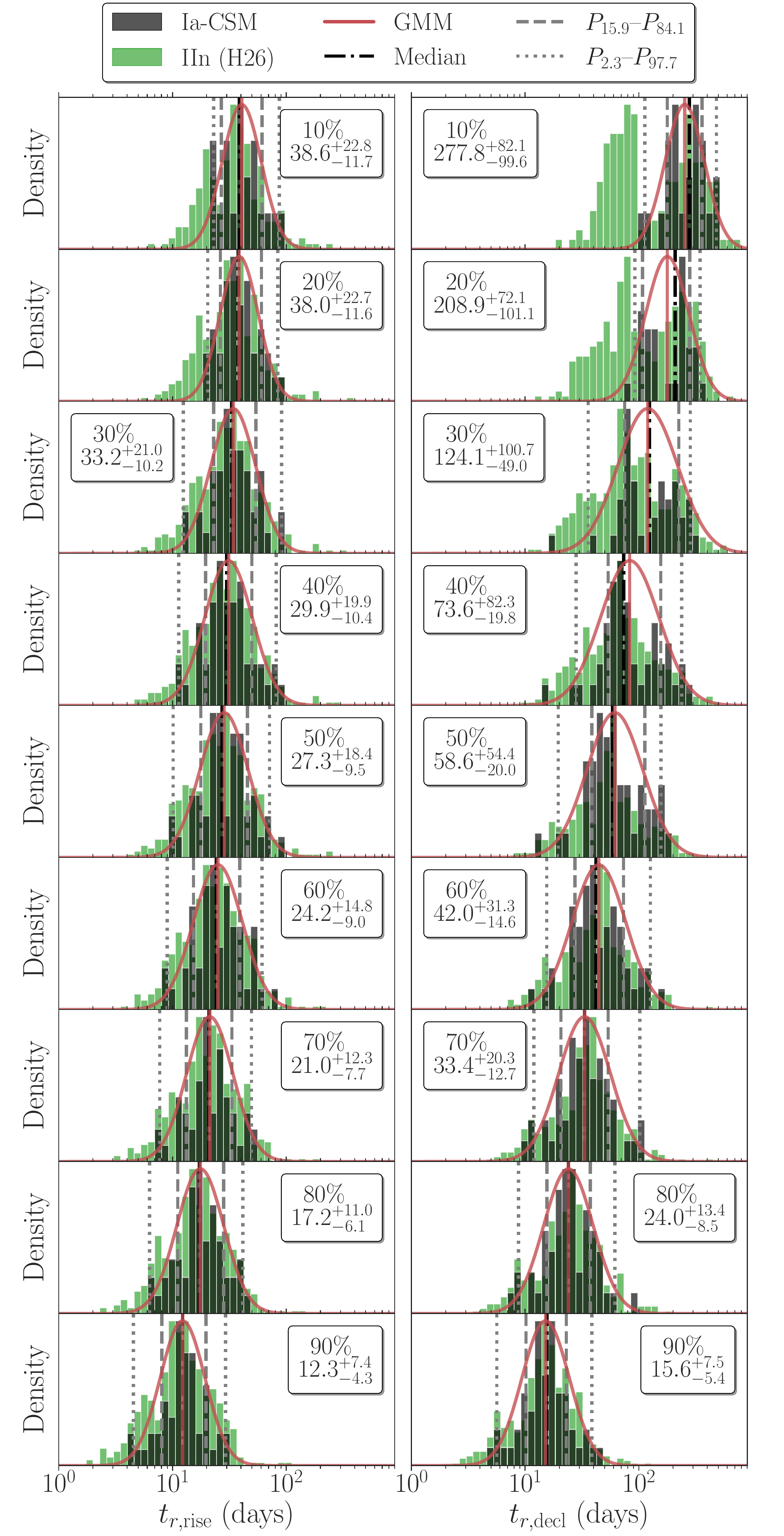}
\caption{Histograms of the $g$-band (\textit{left}) and $r$-band  (\textit{right}) timescales above each 10\% flux level relative to peak on the rise and decline. By the measurement definition, the temporal progression along the light curve is counterclockwise from the left-top to right-top panels, with the decreasing rise and increasing decline timescales. The timescales are faster in the $g$ than $r$ band, especially on the decline.
}
    \label{fig:rT}
\end{figure*}

\begin{figure*}
    \centering
    \includegraphics[width=0.46\textwidth]{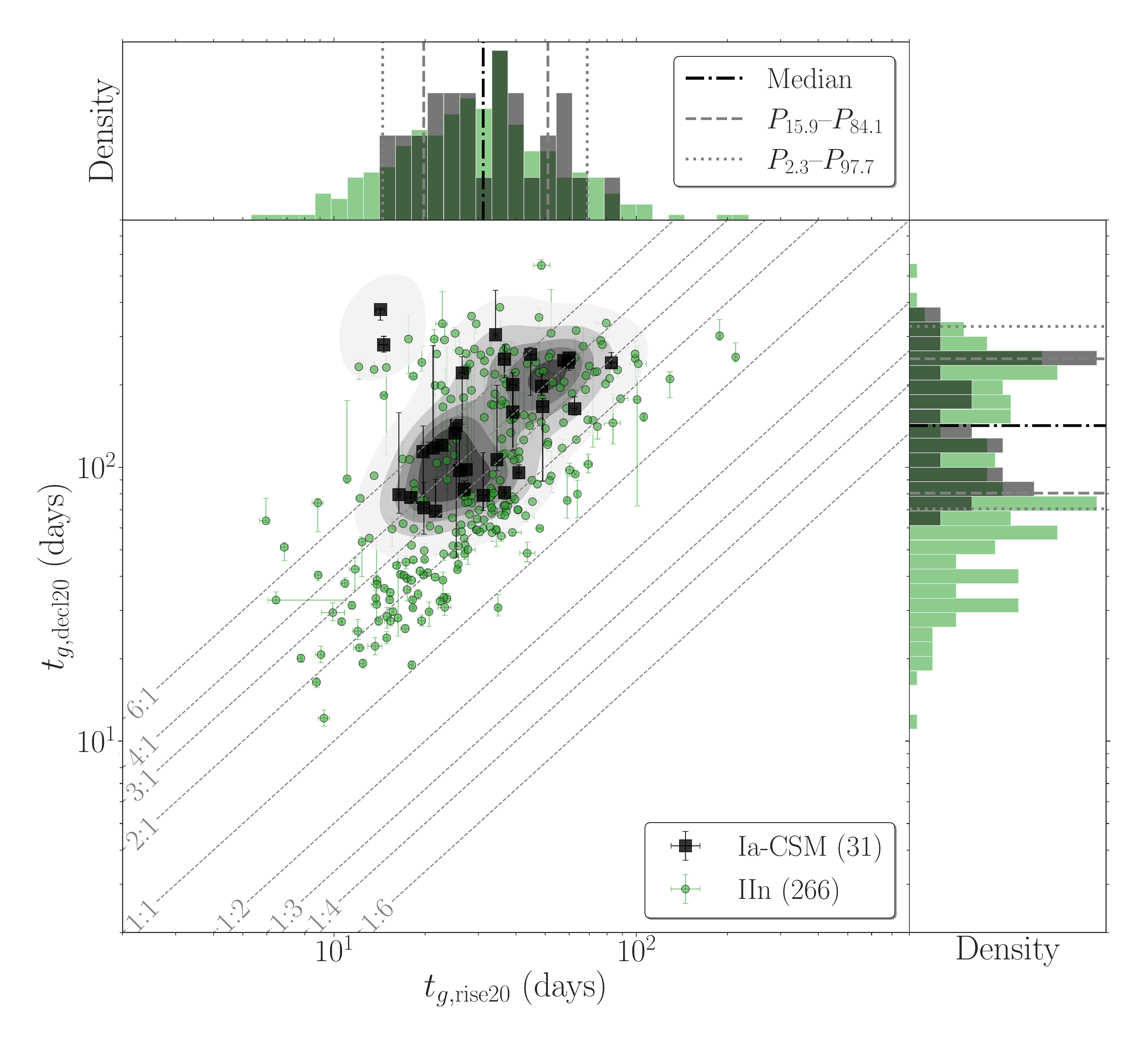}
    \includegraphics[width=0.46\textwidth]{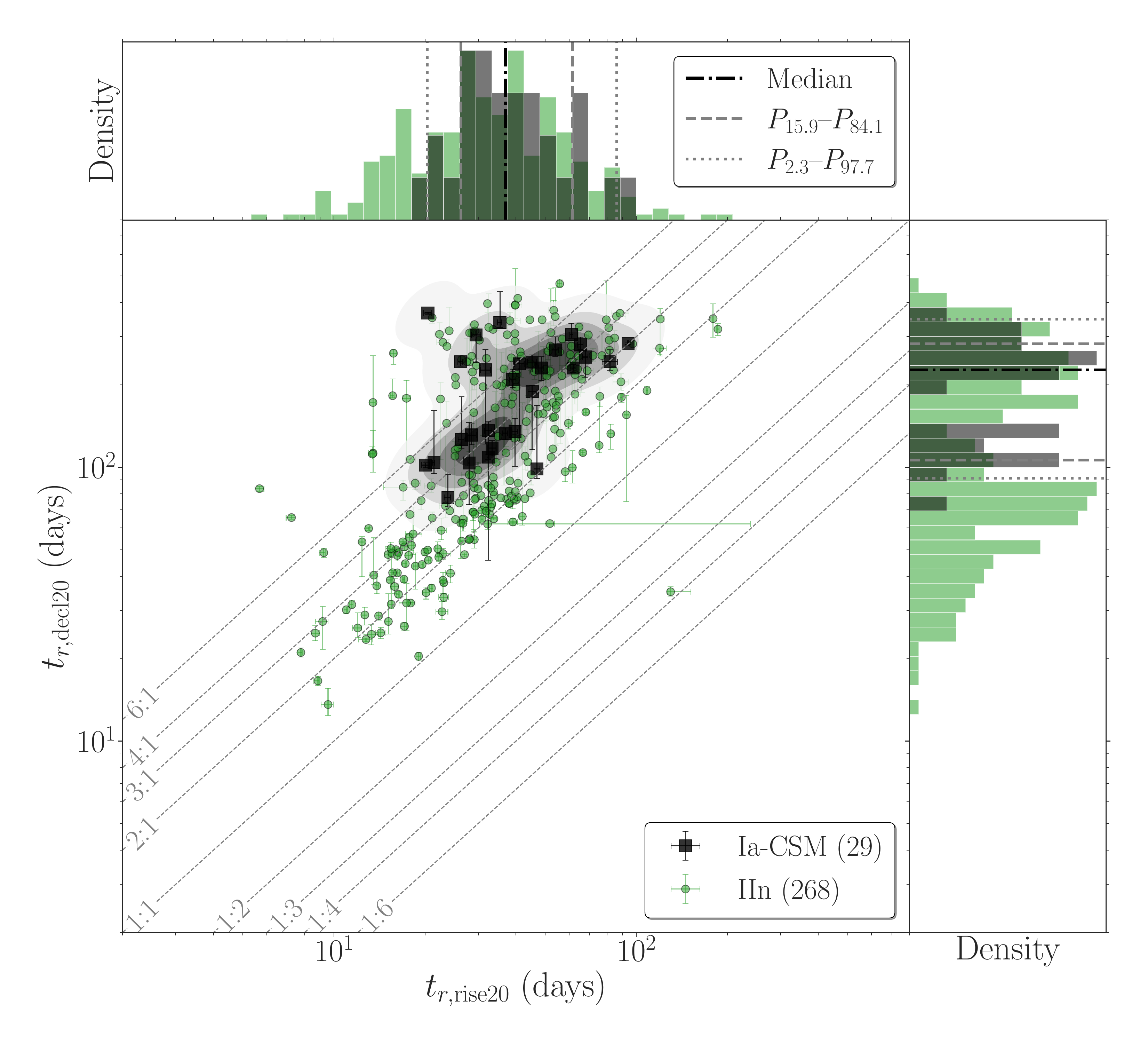}
    \includegraphics[width=0.46\textwidth]{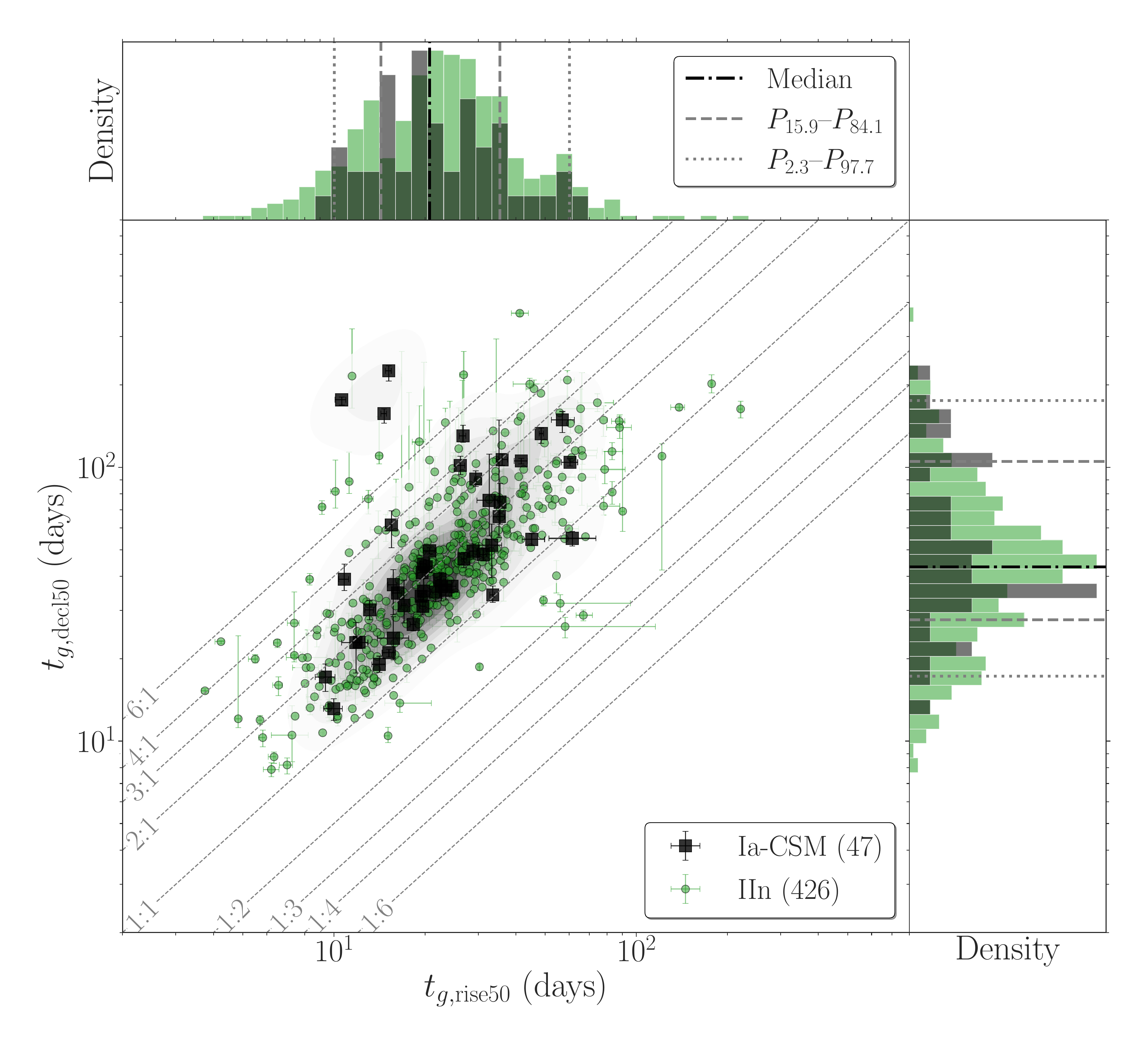}
    \includegraphics[width=0.46\textwidth]{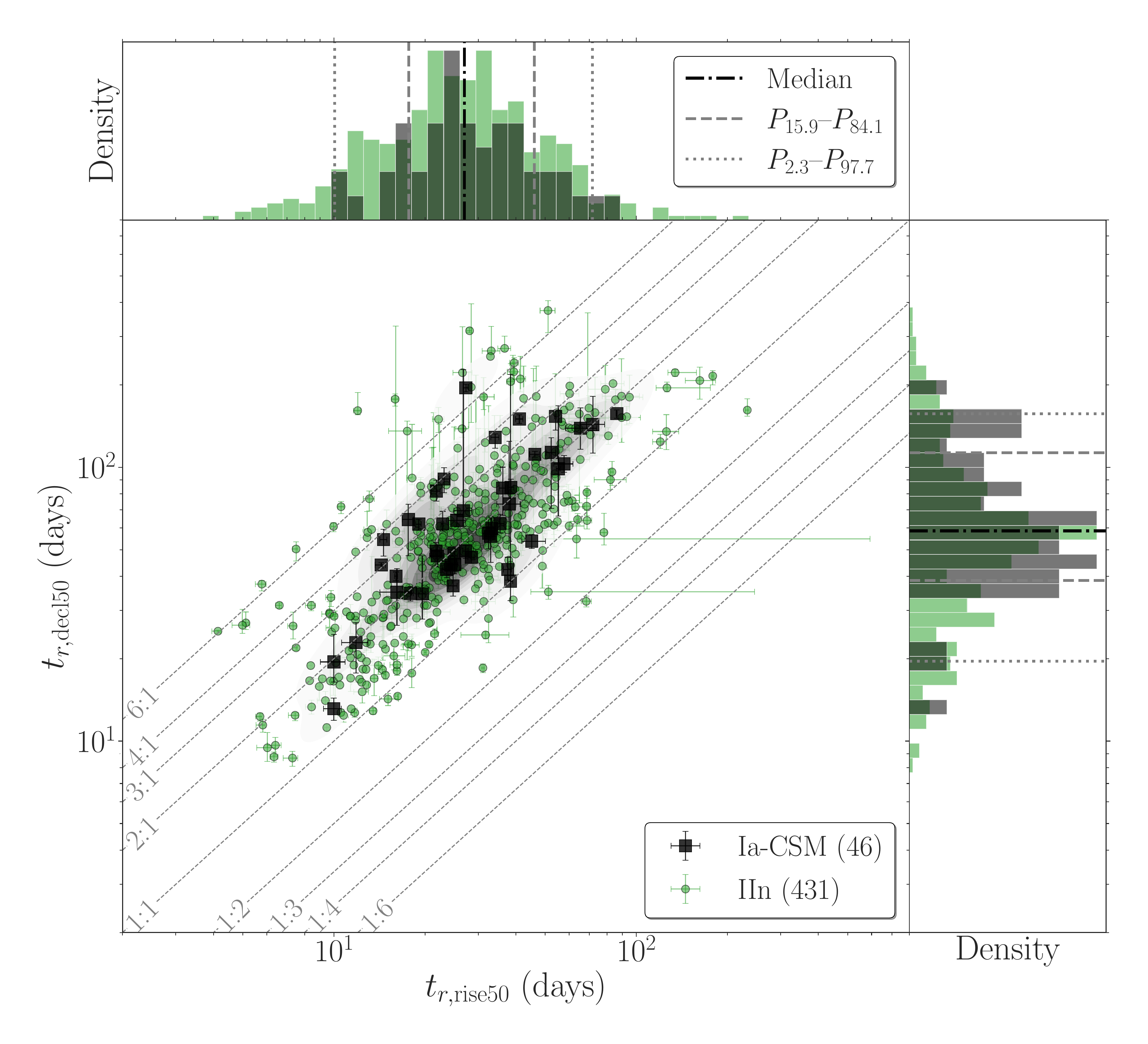}
    \includegraphics[width=0.46\textwidth]{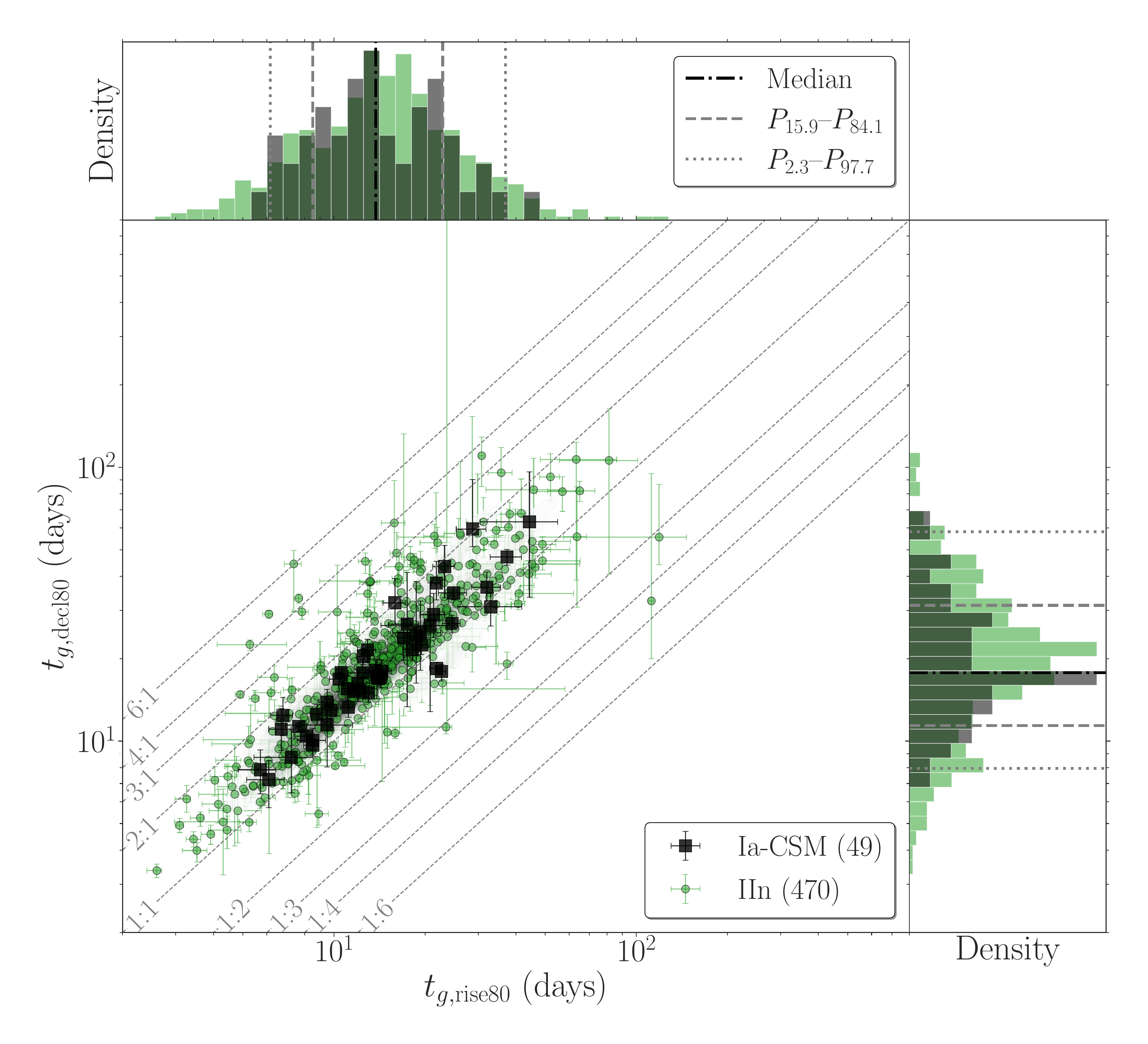} 
    \includegraphics[width=0.46\textwidth]{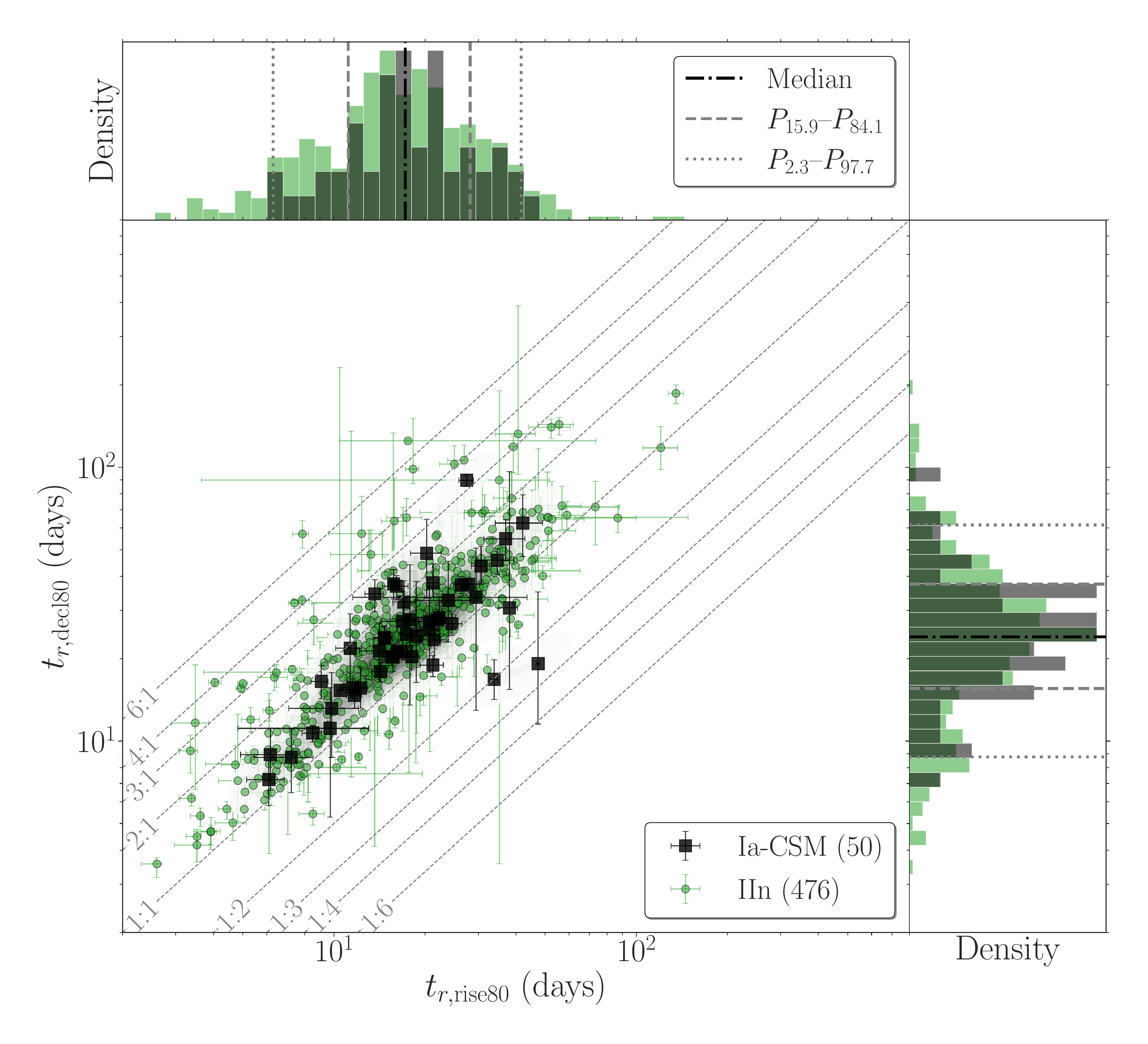}
\caption{Comparisons of the $g$-band (\textit{left}) and $r$-band (\textit{right}) timescales on the rise and decline of the Ia-CSM sample above three flux levels: 20\% (\textit{top}), 50\% (\textit{middle}), and 80\% (\textit{bottom}), along with the IIn sample from \cite{Hiramatsu2026ApJ..1005...82H}. The underlying gray shaded regions represent the density maps of the Ia-CSM sample. Both the $g$ and $r$-band light curves progressively become more symmetric around the peak; $t_{{\rm decl}20} \approx 4 \times t_{{\rm rise}20}$, $t_{{\rm decl}50} \approx 2 \times t_{{\rm rise}50}$, and $t_{{\rm decl}80} \approx 1.2 \times t_{{\rm rise}80}$.
 }
    \label{fig:rT205080}
\end{figure*}

In Figure~\ref{fig:rT}, we plot histograms of the $g$ and $r$-band rise and decline timescales above each 10\% flux level relative to peak. By the measurement definition, the timescales become shorter at higher fractional flux levels. At the 50\% flux level, the median timescales on the rise and decline in the $r$ (and $g$) bands are $\widetilde{t}_{{\rm rise50}}=27.3^{+18.4}_{-9.5}$ ($21.7^{+13.9}_{-7.2}$) and $\widetilde{t}_{{\rm decl}50}=58.6^{+54.4}_{-20.0}$ ($43.3^{+61.8}_{-15.6}$) days, respectively. The $P_{2.3}$--$P_{97.7}$ and $P_{0.2}$--$P_{98.8}$ ranges of $t_{{\rm rise}50}$ (and $t_{{\rm decl}50}$) are $10-71$ and $9-85$ days ($20-157$ and $14-192$ days), respectively, in the $r$ band, and $10-61$ and $9-73$ days ($17-175$ and $14-220$ days), respectively, in the $g$ band. The overall distribution is similar to those of SNe~IIn (H26), except at the low ($\lesssim30\%$) flux levels where the timescales shift towards the long end of the IIn distribution. We find no significant multimodality in timescales at any flux level when applying GMM.

At the 50\% flux level, the median total timescale in the $r$ (and $g$) bands are $t_{{\rm rise}50}+t_{{\rm decl}50}=87.0^{+75.5}_{-28.1}$ ($64.1^{+81.6}_{-20.3}$) days, with the $P_{2.3}$--$P_{97.7}$ and $P_{0.2}$--$P_{98.8}$ ranges of $30-222$ and $24-241$ ($27-205$ and $23-237$) days, respectively. The evolution is faster in the $g$ than $r$ band, especially on the decline. This could be due to the combination of sustained H$\alpha$ emission from CSM interaction in the $r$ band and Fe~{\sc ii/iii} absorption from the diluted SN~Ia continuum (e.g., \citealt{Uno2023ApJ...944..203U}) with declining photospheric temperature in the $g$ band.

In Figure~\ref{fig:rT205080}, we plot the decline versus rise timescales in the $g$ and $r$ bands for the 20\%, 50\%, and 80\% flux levels. 
For both filters, the rise and decline timescales are linearly correlated with $t_{{\rm decl}50} \approx 4\times t_{{\rm rise}20}$, $t_{{\rm decl}50} \approx 2\times t_{{\rm rise}50}$, and $t_{{\rm decl}80} \approx 1.5\times t_{{\rm rise}80}$. This demonstrates that the overall light-curve shape becomes progressively more symmetric around the peak regardless of wavelengths, as similarly found for SNe~IIn (H26). This could be driven by photon diffusion through dense CSM (e.g., \citealt{Chevalier1982ApJ...258..790C,Moriya2013MNRAS.435.1520M,Khatami2024ApJ...972..140K}). More quantitatively, the correlation at the 50\% flux level, $t_{r50,{\rm decl}} \approx 2 \times t_{r50,{\rm rise}}$, matches well with the findings of 2D numerical simulations of CSM interaction \citep{Suzuki2019ApJ...887..249S}. Interestingly, the simulations show a weak dependence of timescale correlation on the exact CSM density and geometry (i.e., spherical versus disk).

For completeness, we also show the timescale evolution and correlation in the bolometric and optical luminosity in Figures~\ref{apfig:optbolT} and \ref{apfig:optbolT205080}, where similar trends and correlations can be seen. The timescales are faster in the bolometric than optical luminosity on the rise due to rapid temperature evolution, but they become comparable on the decline when the SED begins to peak in the optical.

\subsection{Light-Curve Morphology}
\label{sec:lc}

\begin{figure*}
    \centering
    \includegraphics[width=0.49\textwidth]{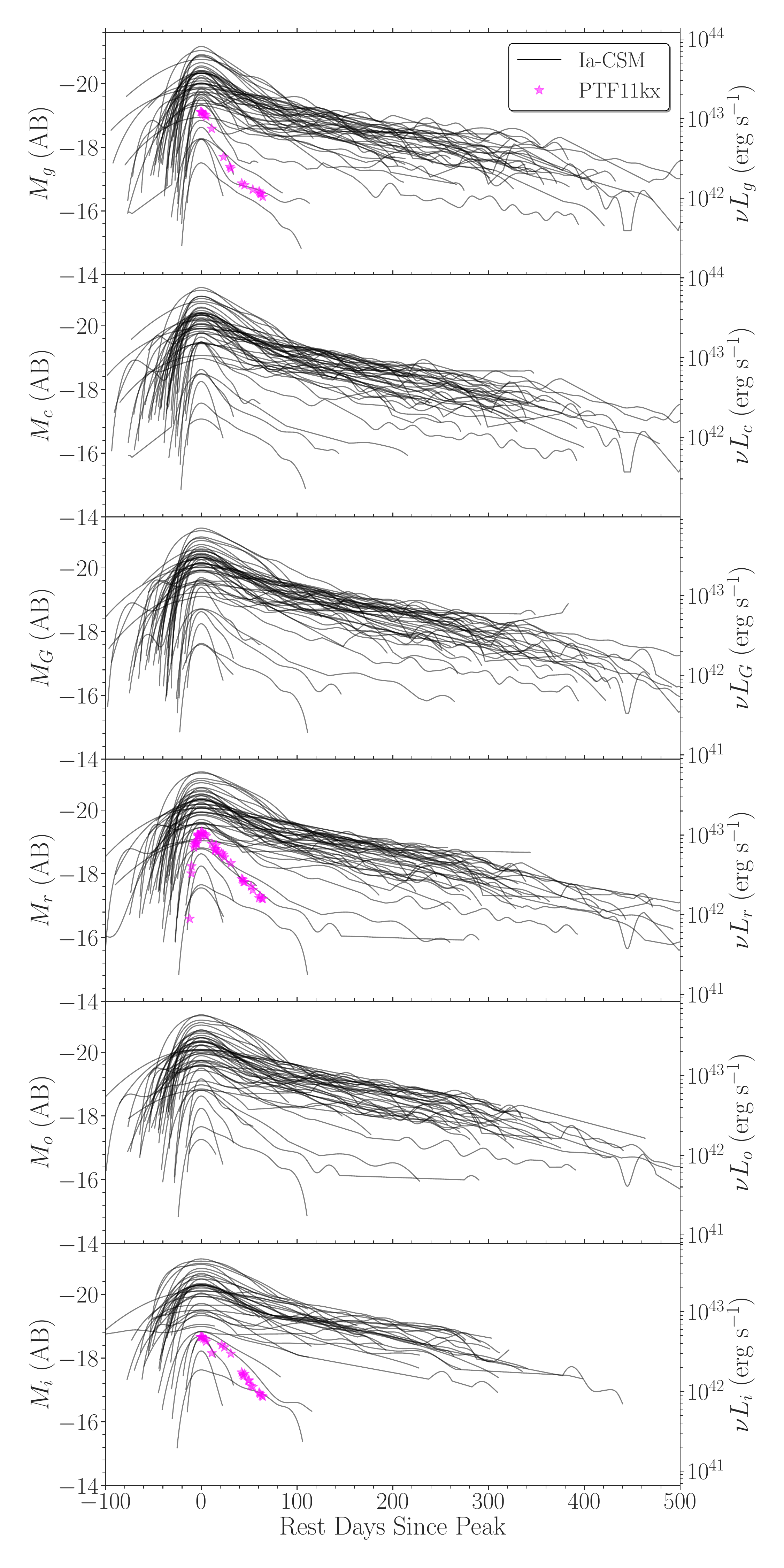}
    \includegraphics[width=0.49\textwidth]{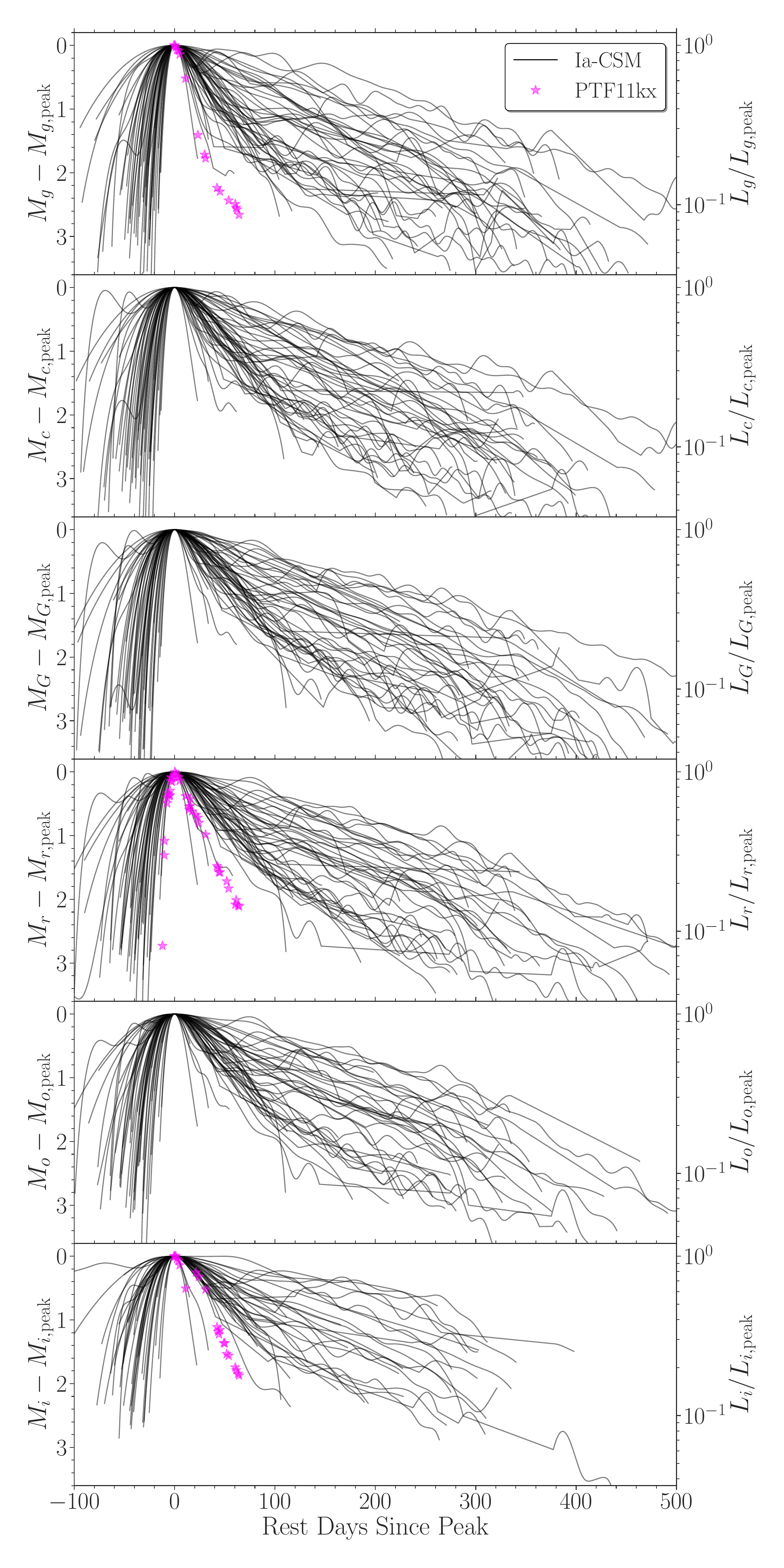}
\caption{{\it Left:} GP light curves of the Ia-CSM sample in absolute magnitude in each filter ($g$, $c$, $G$, $r$, $o$, and $i$ from \textit{top} to \textit{bottom}), alongside PTF11kx with early Ia-dominated evolution (e.g., narrow peak at $-19.3$ mag and post-peak $i$-band bump) followed by delayed CSM interaction \citep[][see also Figure~\ref{fig:bolLC}]{Dilday2012}.  
{\it Right:} Same light curves, but normalized at peak.
(The data used to create this figure are available in machine-readable format.)
}
    \label{fig:bandLC}
\end{figure*}

\begin{figure*}
    \centering
    \includegraphics[width=0.49\textwidth]{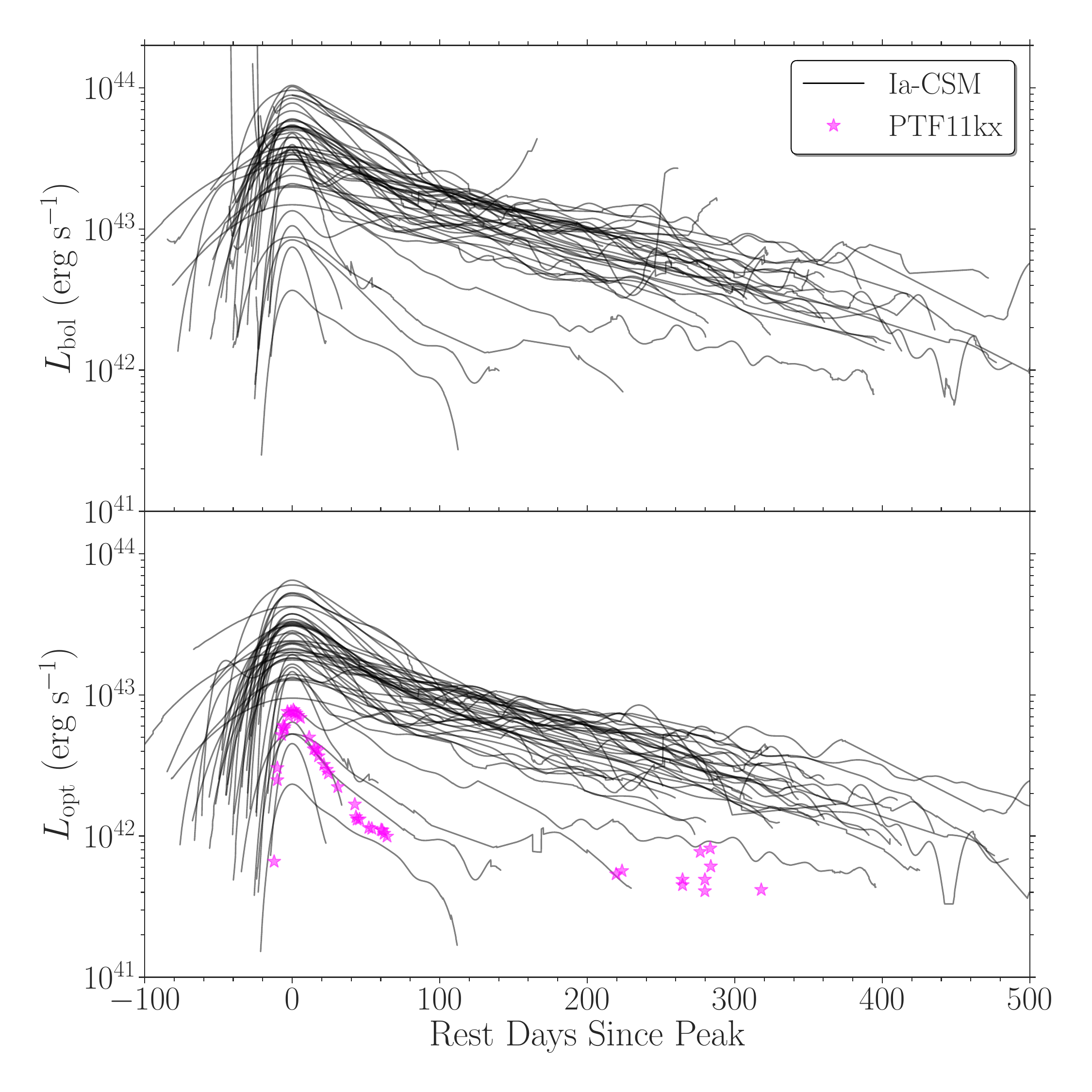}
    \includegraphics[width=0.49\textwidth]{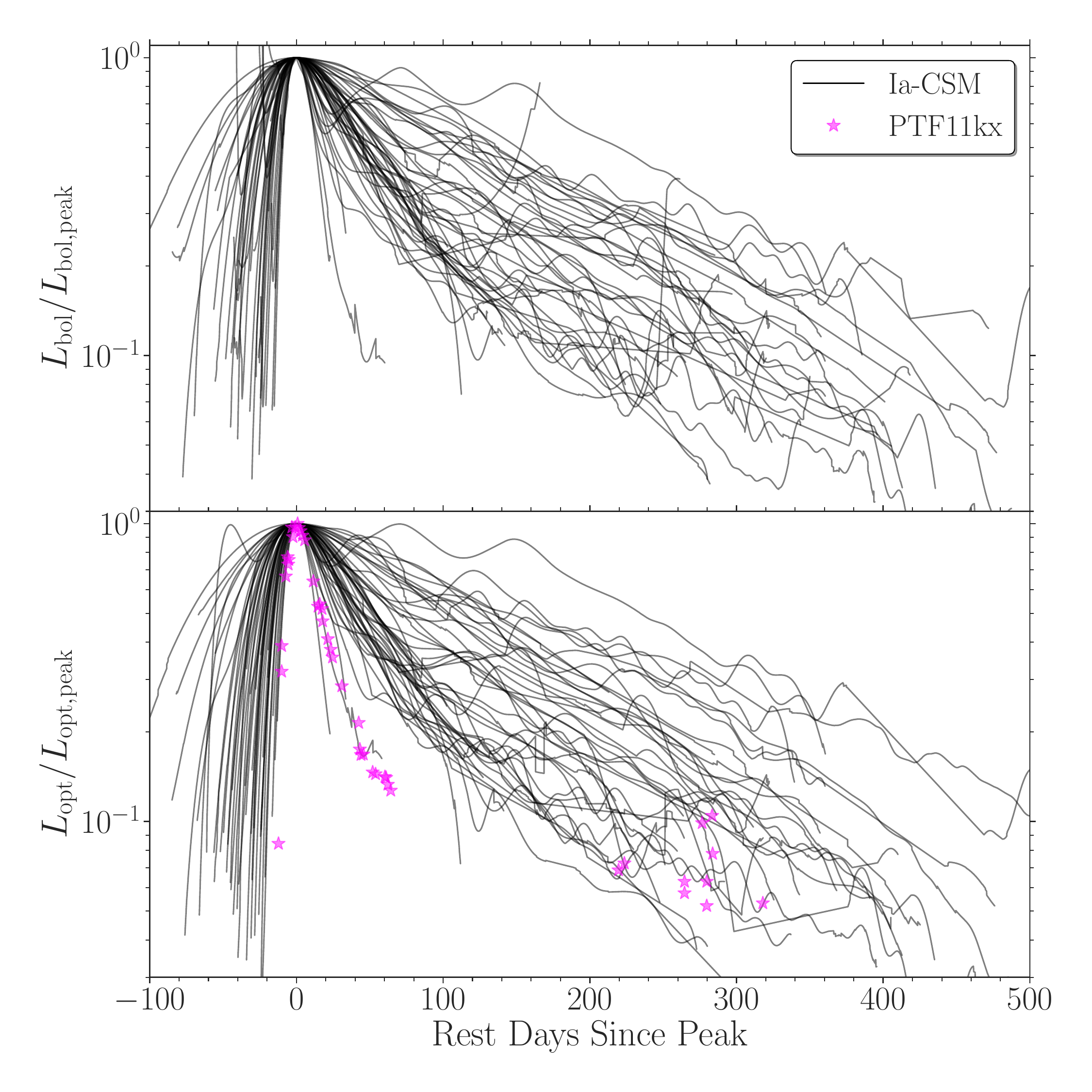}
\caption{Similar to Figure~\ref{fig:bandLC}, but in luminosity space.
(The data used to create this figure are available in machine-readable format.)
}
    \label{fig:bolLC}
\end{figure*}

\begin{figure*}
    \centering
    \includegraphics[width=0.49\textwidth]{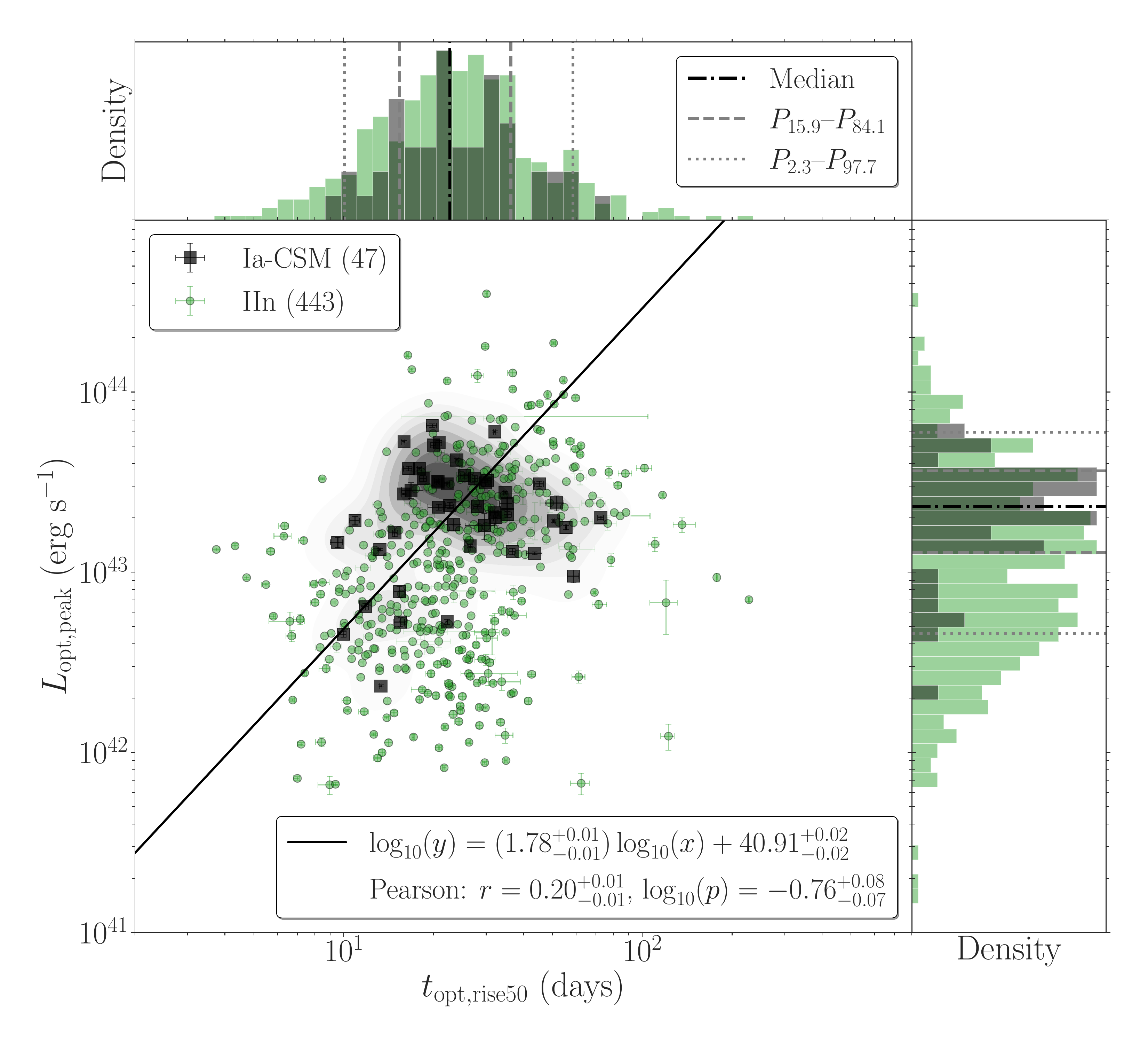}
    \includegraphics[width=0.49\textwidth]{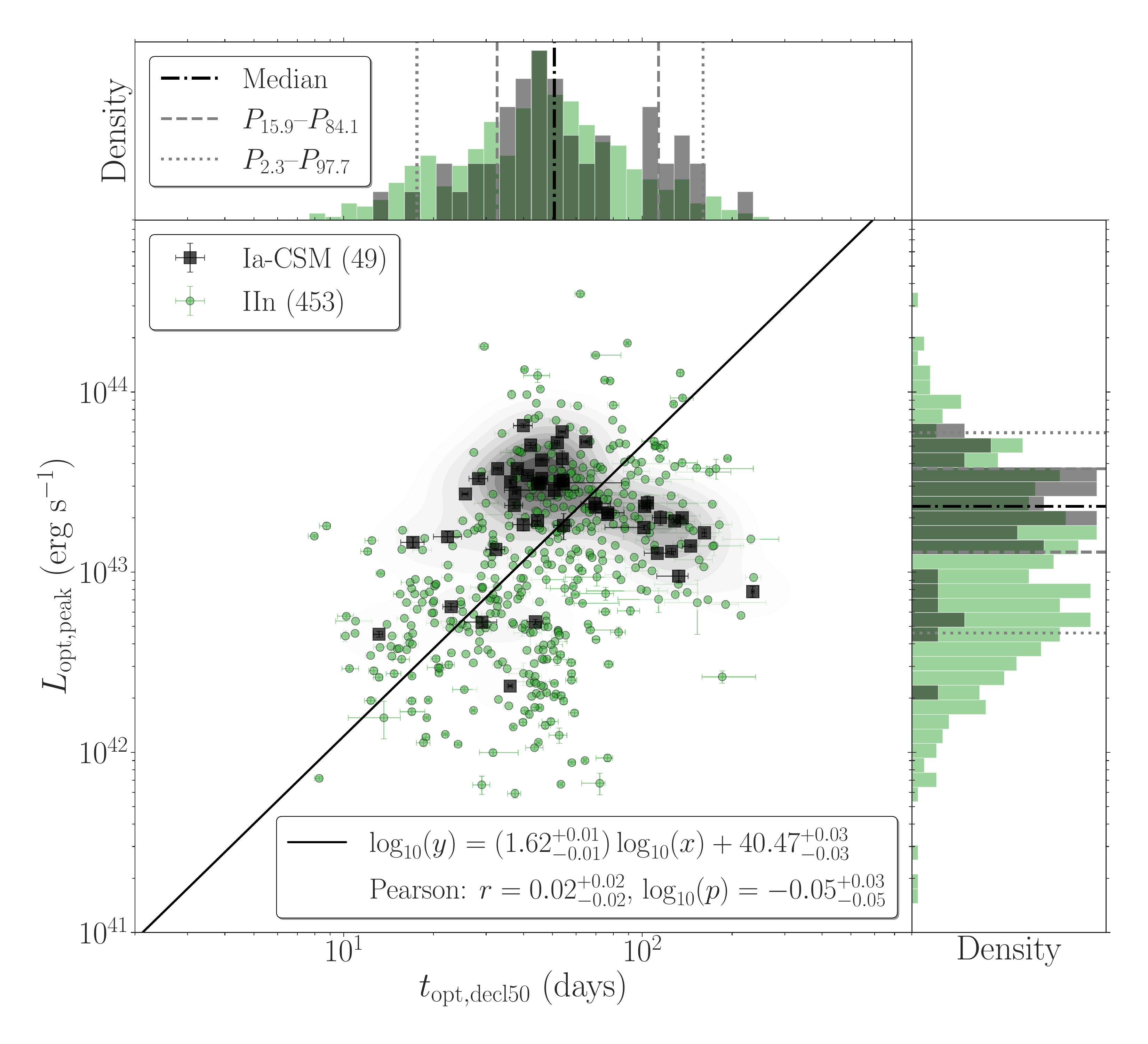}
\caption{Distributions of the peak optical luminosity and 50\% rise (\textit{left}) and decline (\textit{right}) timescales. In each panel, we show the correlation between the logarithmic of peak luminosity and timescale (black line) and provide the best fit and Pearson correlation test values (coefficient $r$ and log p-value) for the Ia-CSM sample, which indicate moderate and weak correlations of the peak luminosity with rise and decline timescales, respectively.
}
    \label{fig:Lopt_T50}
\end{figure*}

In Figure~\ref{fig:bandLC} and \ref{fig:bolLC}, we plot all the ensemble of GP light curves in each filter, and in bolometric and optical luminosity, respectively. We also show PTF11kx as a representative example that exhibits early Ia-dominated light-curve and spectral evolution around peak followed by delayed CSM interaction \citep{Dilday2012}. Although the sample spans wide ranges of peak brightness (\S\ref{sec:z}) and timescale (\S\ref{sec:time}), the majority of them share the morphology that resembles the luminous-slow subgroup of SNe~IIn, with some exceptions ($\sim10\%$ of the sample with fainter light-curve peaks than PTF11kx) residing within the faint-fast subgroup regime (see Figure~7 of H26). The majority shows rather linear post-peak decline, which is referred as ``superlinear" decline and suggested as a potential populational property of SNe~Ia-CSM by \cite{Hiramatsu2026arXiv260712040H}.

When normalized at peak, the timescale is better visualized, revealing a mostly continuous distribution on the rise of the light curve, and a wider spread on the decline (see also Figures~\ref{fig:rT} and \ref{apfig:optbolT}). In addition, there is an unoccupied region in duration below the shortest timescale of $\lesssim24$ days above the 50\% flux level (see \S\ref{sec:time}).


Using the peak brightness and timescale measurements, we also explore their correlations in the optical luminosity in Figure~\ref{fig:Lopt_T50}. We find moderate and weak correlations of the peak brightness with rise and decline timescales, respectively, with Pearson $r$ coefficients of $\approx -0.20$ (rise time) and $\approx -0.02$ (decline time), and log-probability (i.e., chance correlation) values of $\approx -0.76$ (rise time) and $\approx -0.05 $ (decline time). The correlation is linear between the logarithmic of peak luminosity and timescales.
The moderate correlation observed between the peak luminosity and rise time is responsible for the mostly continuous distribution on the rise of light curves (Figures~\ref{fig:bolLC}). On the other other hand, the weak correlation on the decline demonstrates the wider spread seen in the light-curve morphology.

\subsection{Radiated Energy} 
\label{sec:energy}

\begin{figure*}
    \centering
    \includegraphics[width=0.49\textwidth]{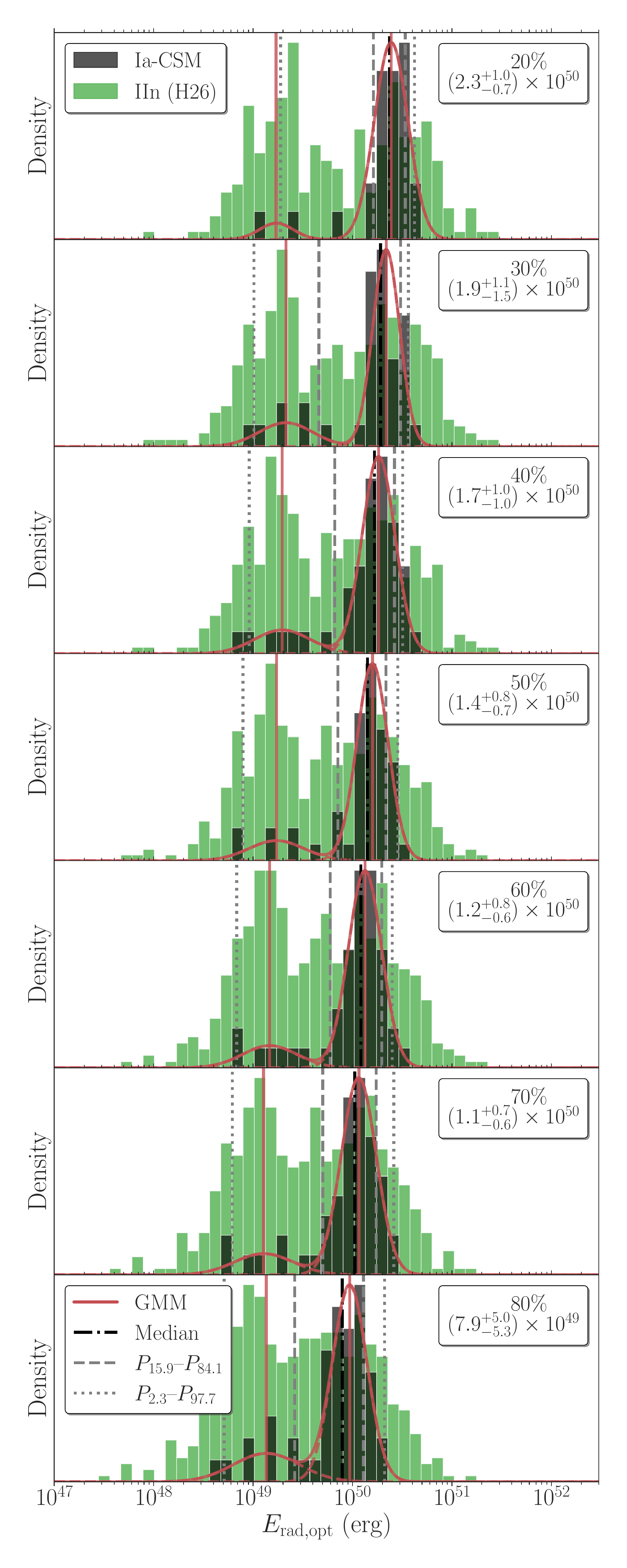 }
    \includegraphics[width=0.49\textwidth]{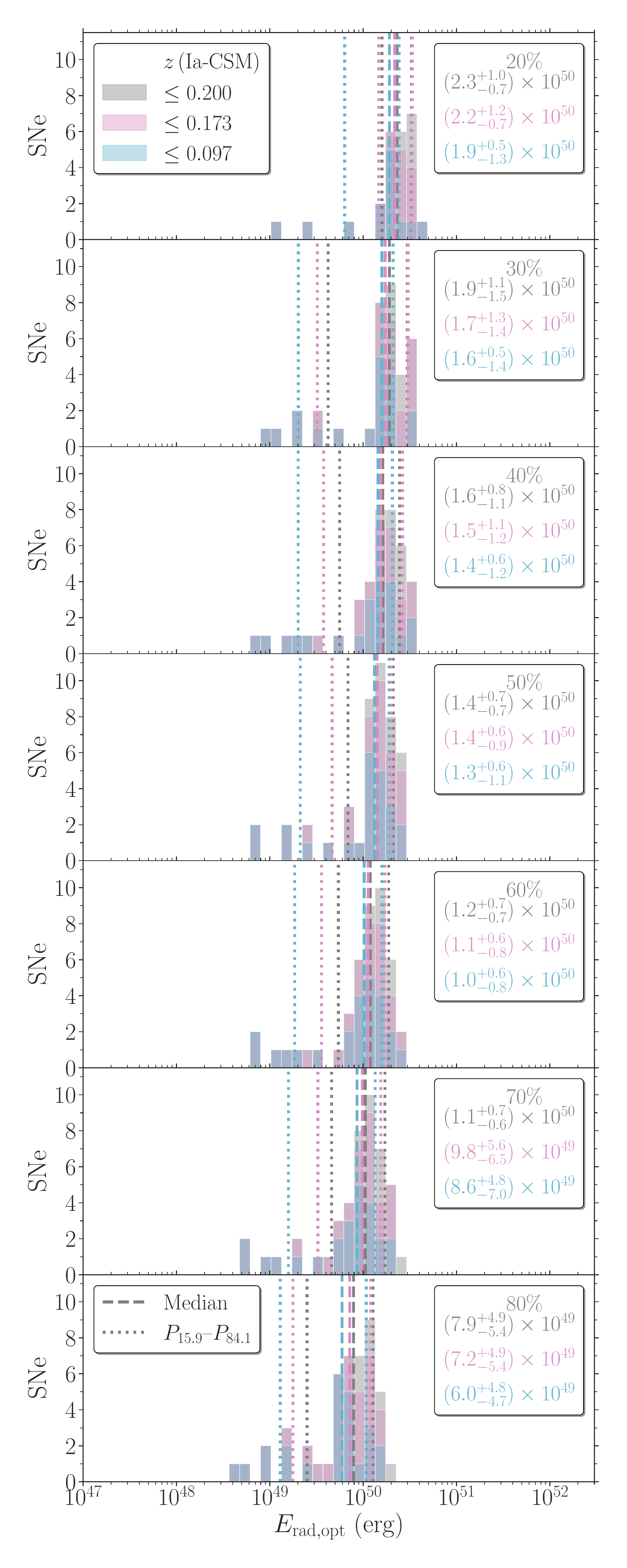 }
\caption{\textit{Left}: Histograms of the optical radiated energy ($E_{\rm rad,opt}$) above each 10\% flux level. The bulk of the Ia-CSM sample resides on the energetic end of the IIn distribution (i.e., the luminous-slow subgroup); however, the more extended distributions below median result in the bimodality at all flux levels with significance $\Delta{\rm BIC}>14$ according to GMM.
\textit{Right}: Similar to \textit{left}, but with the upper redshift cuts as in Figures~\ref{fig:bandP} and \ref{fig:bolP}. The median $E_{\rm rad,opt}$ and $P_{15.9}$--$P_{84.1}$ ranges shift only by $\lesssim20\%$ with the redshift cuts.
}
    \label{fig:Eradopt}
\end{figure*}

\begin{figure*}
    \centering
    \includegraphics[width=\textwidth]{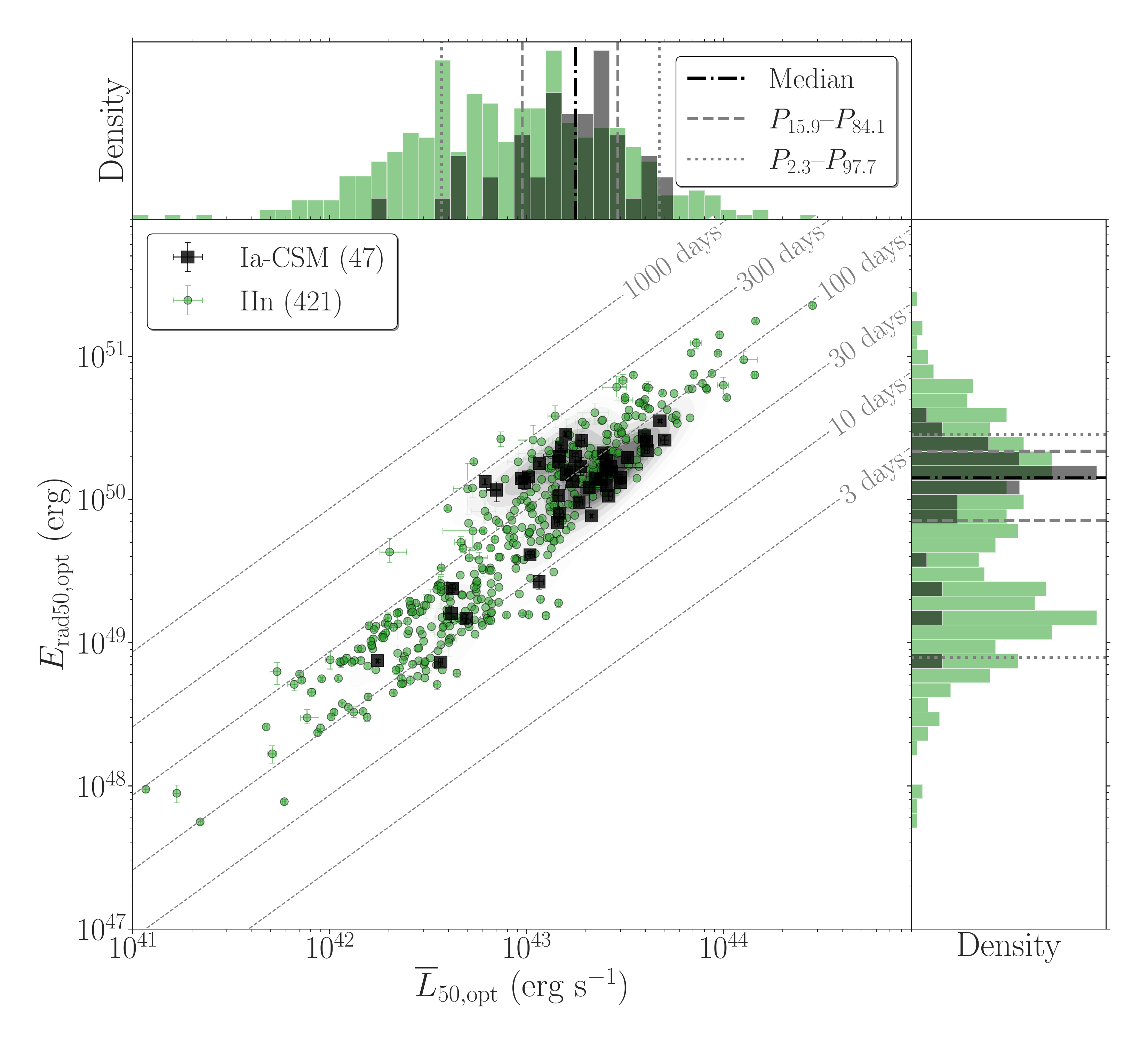}
\caption{Optical radiated energy ($E_{{\rm opt, rad}50}$) vs mean luminosity ($\overline{L}_{50, {\rm opt}}$) above the 50\% flux level. The diagonal dashed lines mark characteristic timescales, i.e., $E_{{\rm rad}50, {\rm opt}}/\overline{L}_{50, {\rm opt}}$. The Ia-CSM sample has a characteristic timescale of $\approx 100$ days, while the IIn sample show the luminous-slow subgroup with a similar characteristic timescale of $\approx 100$ days and the faint-fast subgroup that bifurcates to $\approx 40$ and $\approx 70$ days.}
    \label{fig:Erad_mL50}
\end{figure*}

We calculate the radiated energy by integrating the bolometric and optical light curves over the total duration ($t_{\rm rise}+t_{\rm decl}$) at the various fractional flux levels. In the left panels of Figure~\ref{fig:Eradopt}, we show the histograms of the optical radiated energy ($E_{\rm rad, opt}$) above each 10\% flux level. At the 50\% flux level, the median $\widetilde{E}_{{\rm rad}50, {\rm opt}}$ is $(1.4^{+0.8}_{-0.7})\times10^{50}$\,erg, with the $P_{2.3}$--$P_{97.7}$ range of $(0.08-2.85)\times10^{50}$\,erg, a factor of $\approx36$ that is smaller than the IIn sample range of a factor of $\approx250$. The Ia-CSM sample mostly occupies the energetic end of the IIn distribution, i.e., the luminous-slow subgroup (H26), but without exceeding $E_{\rm rad, opt}\gtrsim5\times10^{50}$\,erg even at the lowest flux level. 

However, the sample spread is larger below the median than above (i.e., $P_{50}-P_{2.3} > P_{97.7}-P_{50}$), extending to the faint-fast IIn subgroup regime, due to the exceptional events as noted in \S\ref{sec:lc} ($\sim10\%$ of the sample; Figure~\ref{fig:bolLC}). Their presence is significant enough to result in the bimodal distributions with $\Delta {\rm BIC} > 14$ at all flux levels according to GMM. In the right panels of Figure~\ref{fig:Eradopt}, we show the subsample distributions with the upper redshift cuts as in Figures~\ref{fig:bandP} and \ref{fig:bolP}. The $\widetilde{E}_{{\rm rad}50, {\rm opt}}$ and $P_{15.9}$--$P_{84.1}$ ranges shift by $<20\%$, with a negligible effect below the median given their faint light-curve peaks (Figure~\ref{fig:Mr_z}). This demonstrates that the cause of the bimodality is not Malmquist bias (see \S\ref{sec:z}), suggesting intrinsic subgroupings within SNe~Ia-CSM whose relative population sizes are distinct from those of SNe~IIn with more faint-fast than luminous-slow events (see Figure~15 of H26).

In addition, we calculate the mean luminosities ($\overline{L}=E_{\rm rad}/(t_{\rm rise}+t_{\rm decl})$) above each 10\% flux level. In Figure~\ref{fig:Erad_mL50}, we plot the optical $E_{{\rm rad}50,{\rm opt}}$ versus $\overline{L}_{50,{\rm opt}}$ at the 50\% flux level, along with characteristic timescales (i.e., $E_{{\rm rad}50,{\rm opt}}/\overline{L}_{50,{\rm opt}}$). The Ia-CSM sample is characterized by a timescale of $\approx100$ days with some events extending down to $\approx50$ days, while the luminous-slow and faint-fast IIn subgroups exhibit characteristic timescales of $\approx100$ and $\approx40-70$ days, respectively. This summarizes all prior analysis on the peak brightness (\S\ref{sec:z}), timescale (\S\ref{sec:time}), and their correlations (\S\ref{sec:lc}), stemming from the rather uniform light-curve morphology across the Ia-CSM sample compared to the IIn sample. 

Finally, we note that $E_{\rm rad, opt}$ represents a lower bound on the bolometric quantity $E_{\rm rad, bol}$, where a rough transformation can be obtained by the overall optical-bolometric correction of $\sim1.6$ (Figure~\ref{fig:bolP}). For completeness, we also show the histograms of $E_{\rm rad, bol}$ in Figure~\ref{apfig:Eradbol}, where the similar distribution to $E_{\rm rad, opt}$ can be seen, with an upper limit at $E_{\rm rad, bol}\lesssim7\times10^{50}$\,erg .

\subsection{Light-Curve Templates} 
\label{sec:temp}

\begin{figure*}
    \centering
    \includegraphics[width=0.49\textwidth]{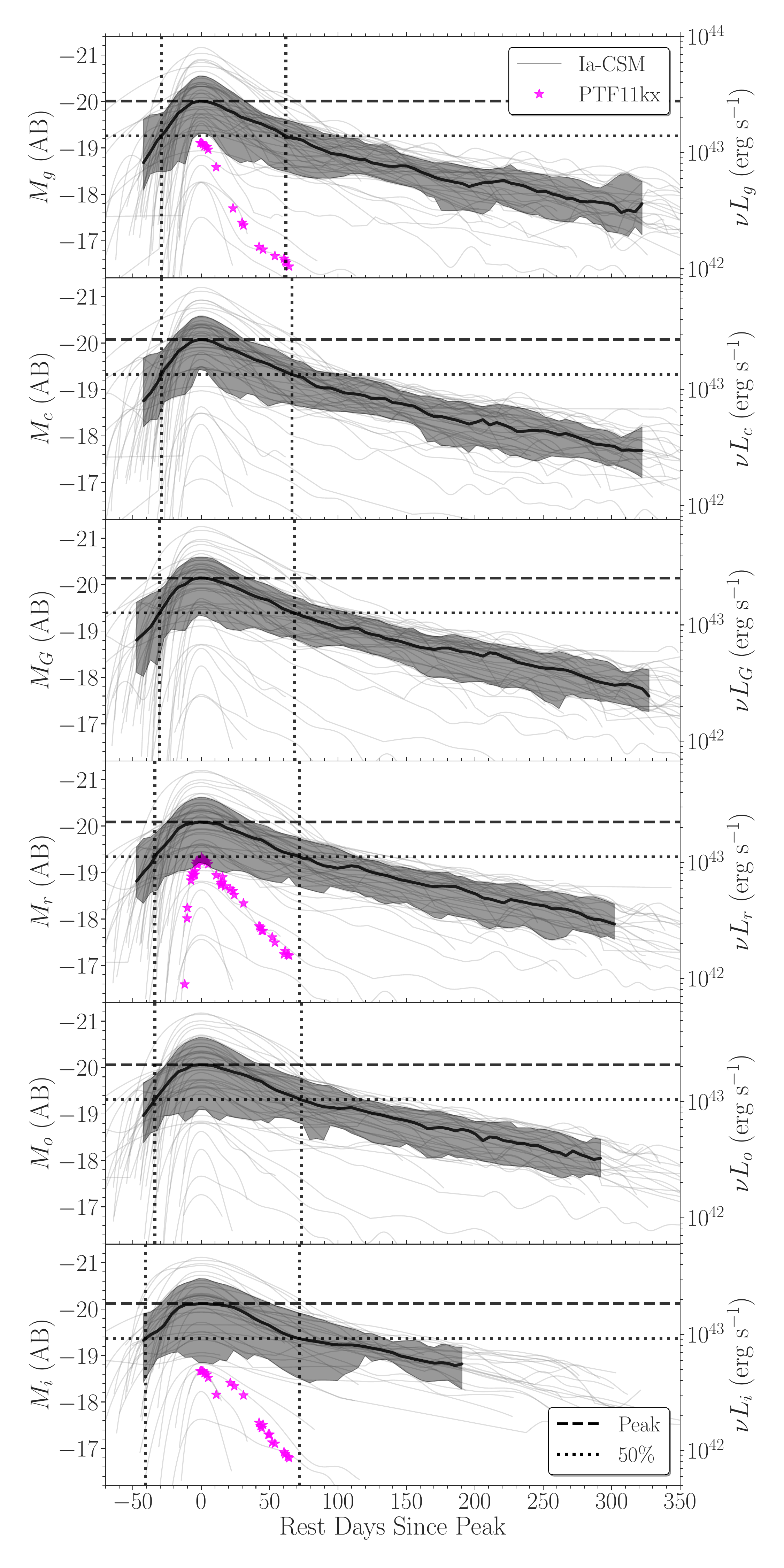}
    \includegraphics[width=0.49\textwidth]{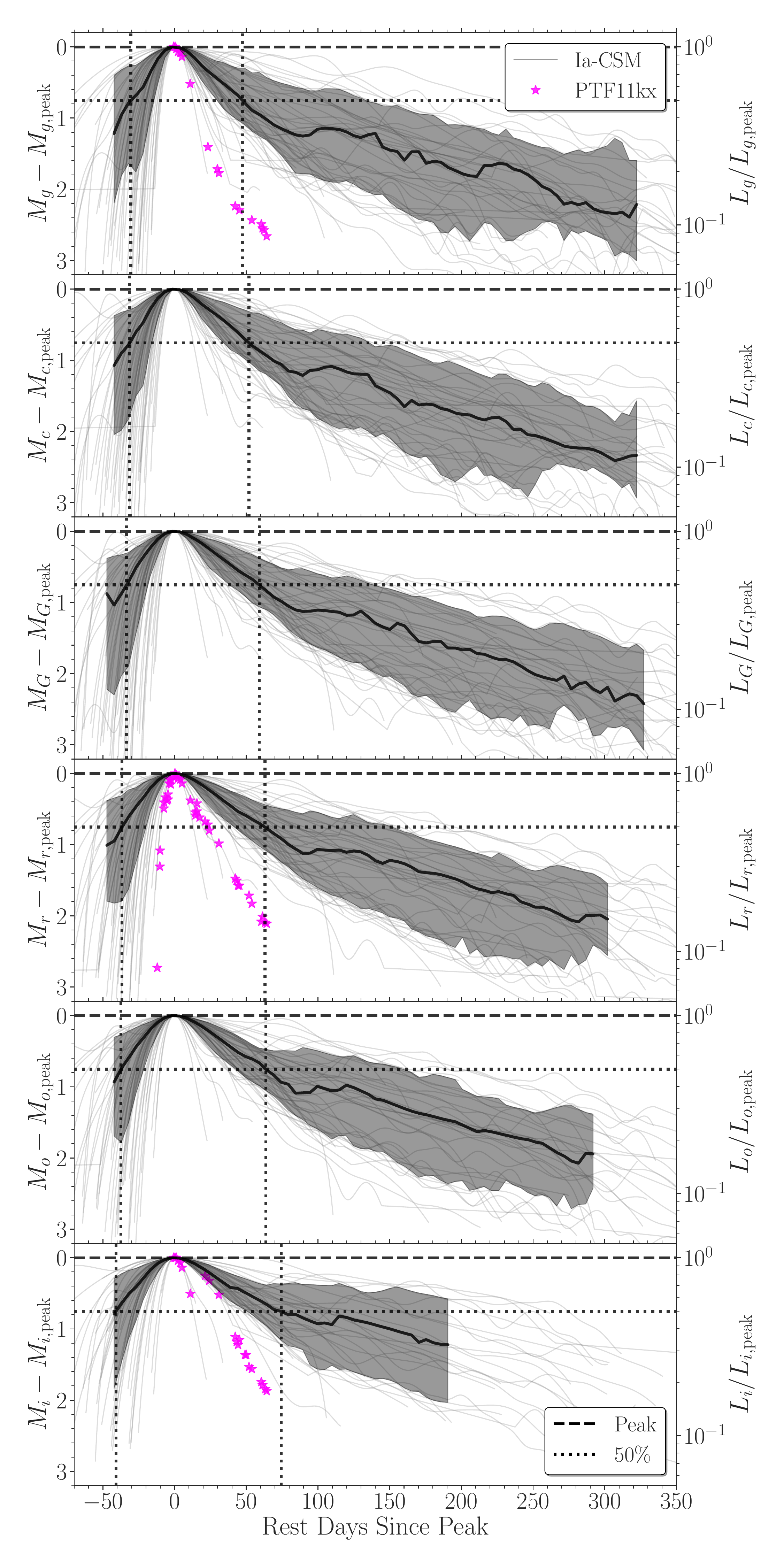}
\caption{\textit{Left}: Median light curves of the Ia-CSM sample in each filter ($g$, $c$, $G$, $r$, $o$, and $i$ from \textit{top} to \textit{bottom}) alongside PTF11kx. The ``superlinear" decline is evident in each filter. 
\textit{Right}: Same light curves, but normalized at peak. The peak and 50\% measurements are marked by the horizontal dashed and vertical dotted lines, respectively. The shaded regions represent $P_{15.9}$--$P_{84.1}$ ranges.
(The data used to create this figure are available in machine-readable format.)
 }
    \label{fig:LCtemp}
\end{figure*}

\begin{figure*}
    \centering
    \includegraphics[width=0.49\textwidth]{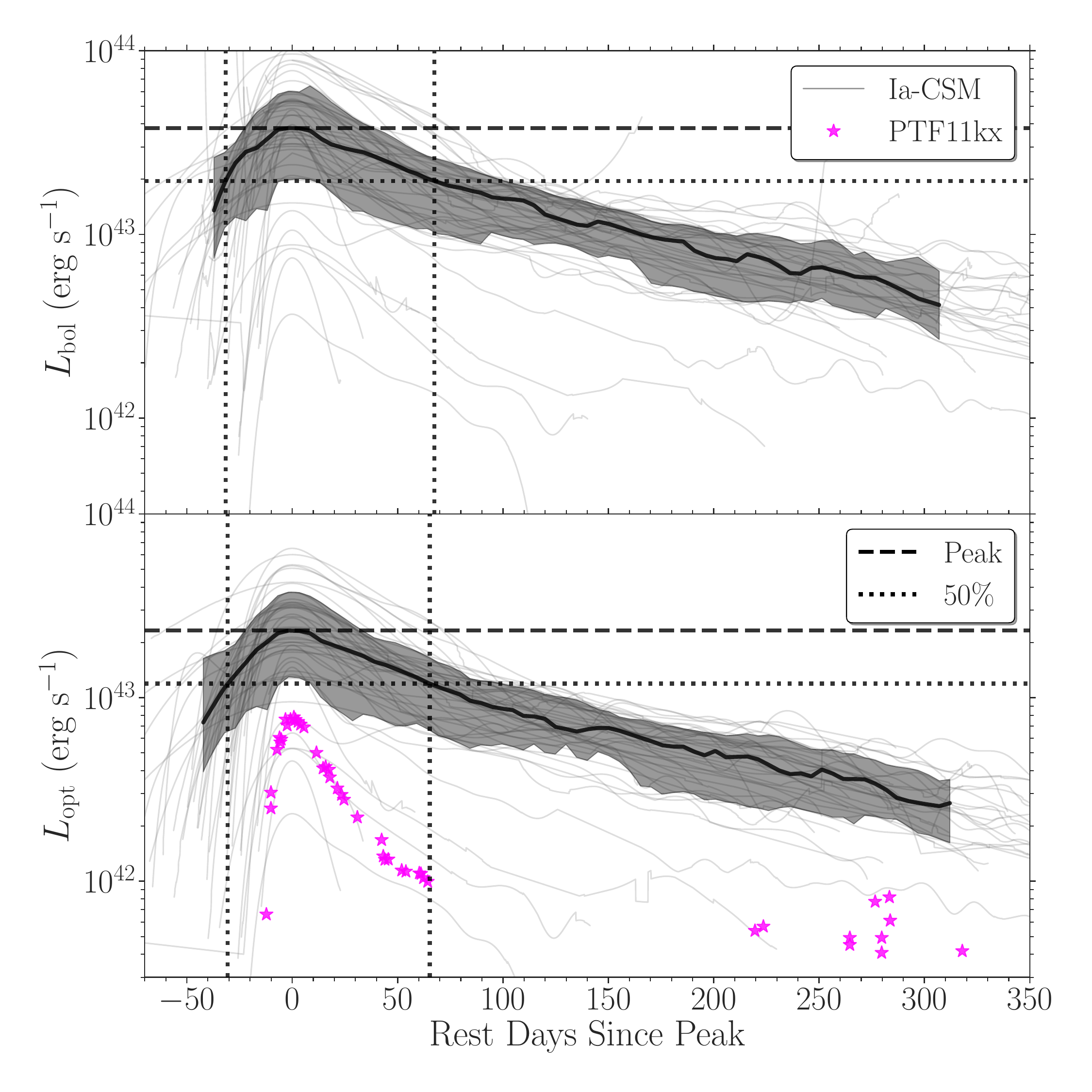}
    \includegraphics[width=0.49\textwidth]{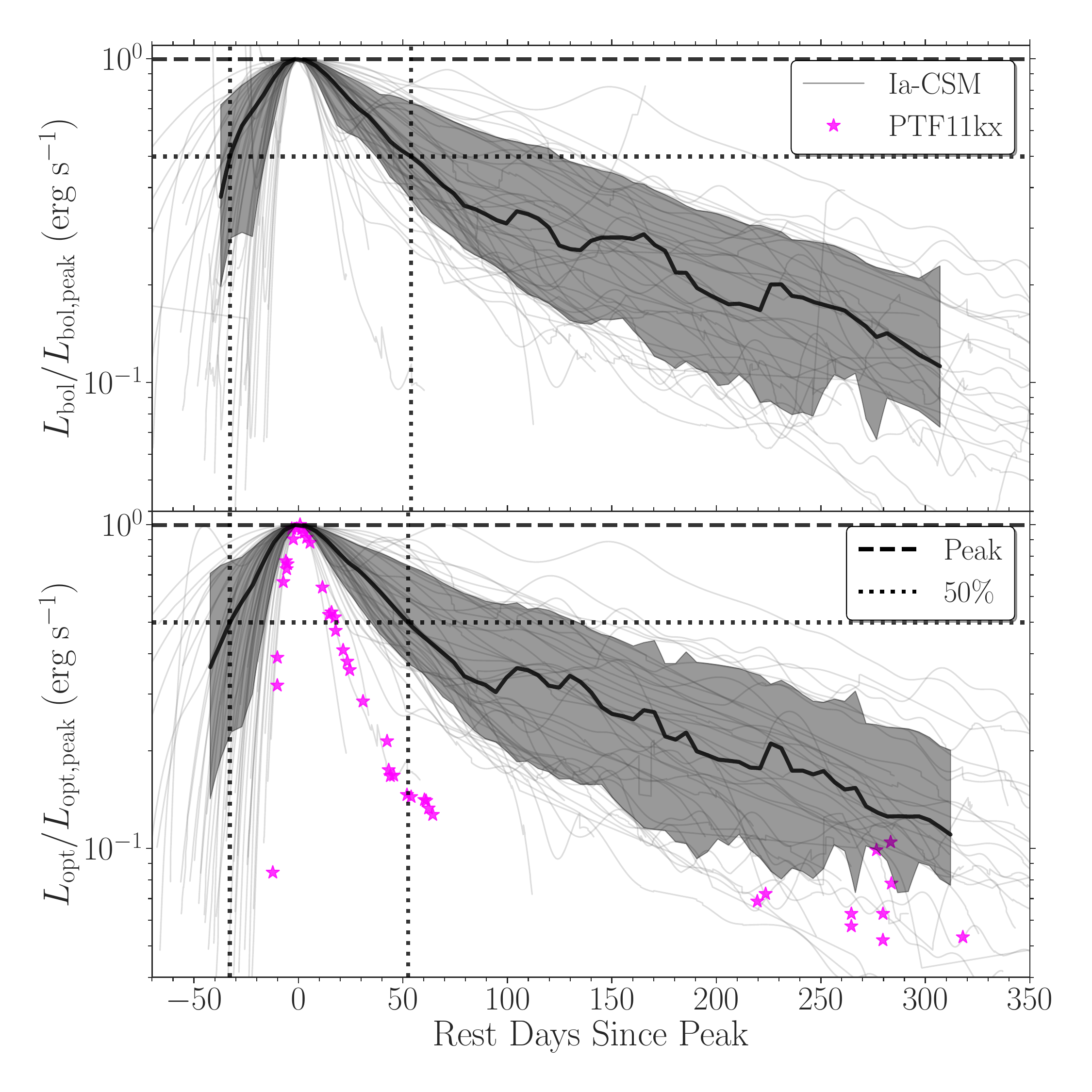}
\caption{Similar to Figure~\ref{fig:LCtemp}, but for the bolometric (\textit{top}) and optical (\textit{bottom}) luminosity. (The data used to create this figure are available in machine-readable format.)
}
    \label{fig:Lboltemp}
\end{figure*}

\begin{figure*}
    \centering
    \includegraphics[width=0.49\textwidth]{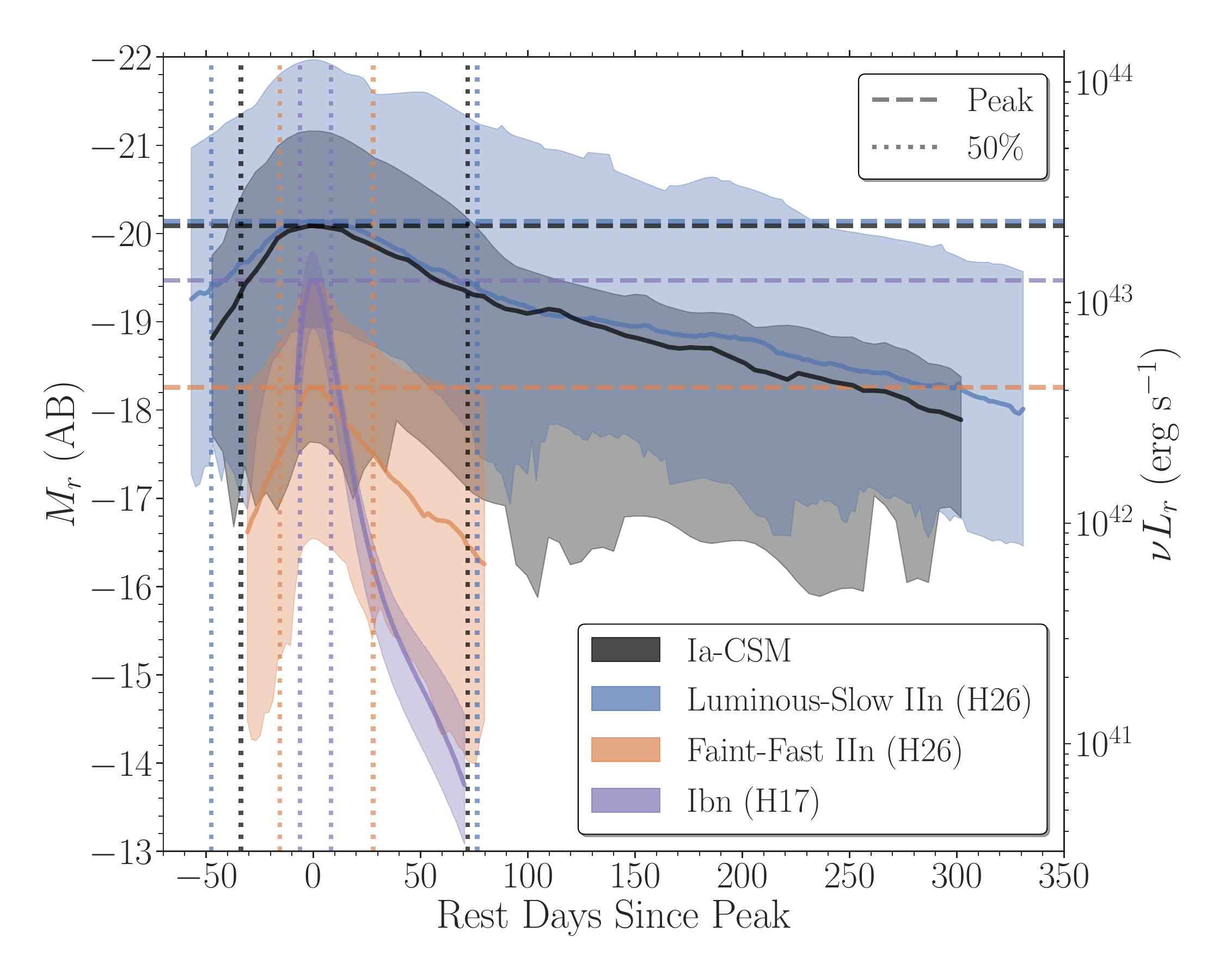}
    \includegraphics[width=0.49\textwidth]{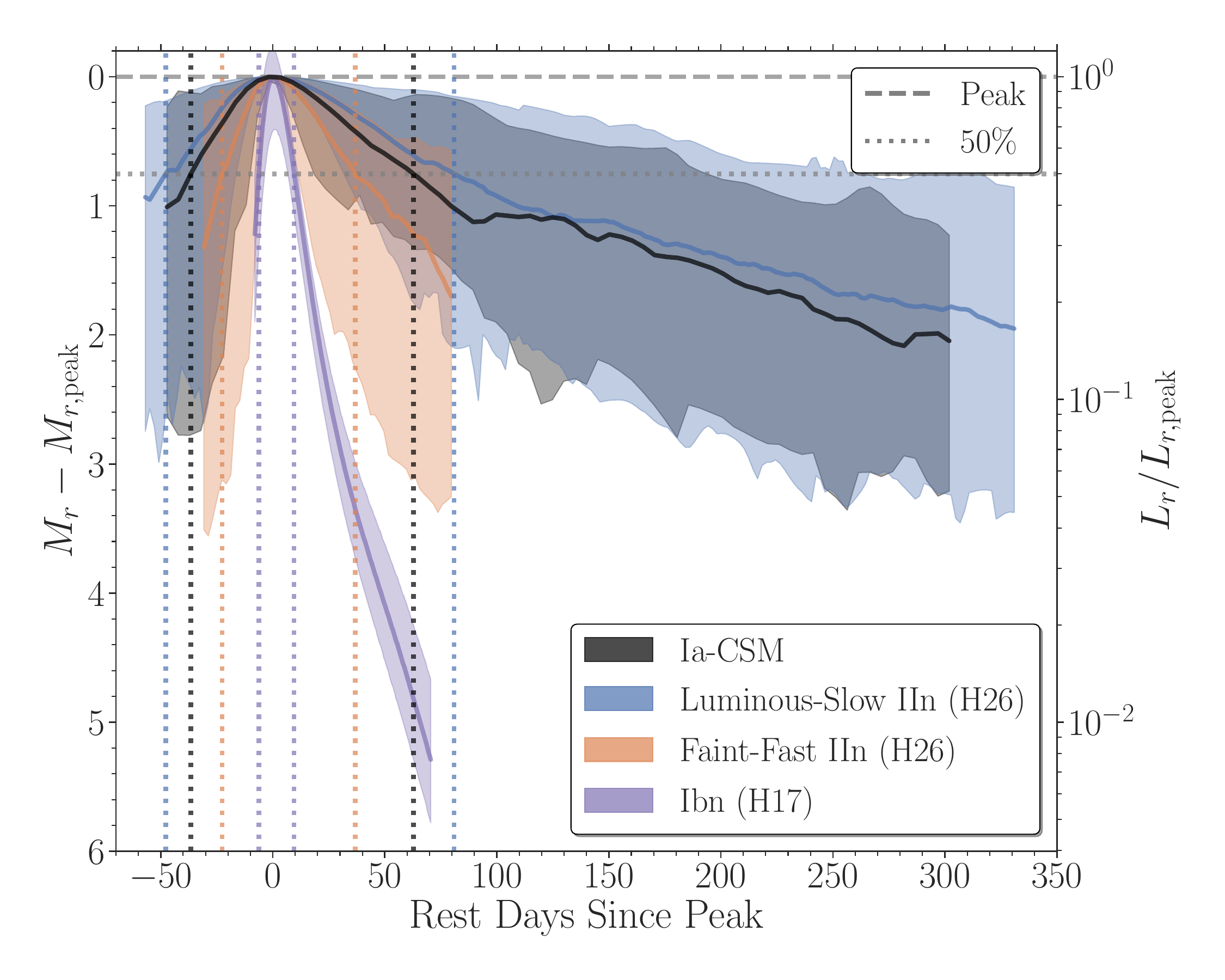}
\caption{\textit{Left}: Similar to Figure~\ref{fig:LCtemp}, but with the IIn templates from H26 and the Ibn template from \citet[H17]{Hosseinzadeh2017ApJ...836..158H} in the $r/R$-band. The shaded regions represent the $P_{2.5}-P_{97.5}$ (i.e., 95\%) range.
\textit{Right}: Same templates, but normalized at peak. 
The Ia-CSM template is almost identical to the luminous-slow IIn template and substantially different from the faint-fast IIn template, indicating a more uniform population for SNe~Ia-CSM than IIn.
}
    \label{fig:IInIbn}
\end{figure*}

With the largest SN~Ia-CSM sample to date, we construct median light curves, i.e., light-curve templates. Unlike the SN~IIn sample in H26, we do not divide the Ia-CSM sample based on the $E_{\rm rad}$ bimodality given the over-presence of more energetic events ($\sim90\%$) whose median is close to that of the whole sample (Figures~\ref{fig:Eradopt} and \ref{apfig:Eradbol}).
In each filter, we combine all the GP light curves and calculate the median flux and its variance in 5-day bins if more than 20 events with significant flux measurements ($F/F_{\rm err}\geq3$) are available in each bin. 
We also perform the same procedure in the bolometric and optical luminosities. We then construct normalized templates at peak by performing the same procedure on the normalized band, bolometric, and optical light curves.

In Figures~\ref{fig:LCtemp} and \ref{fig:Lboltemp}, we plot the resulting light-curve templates in each filter, and in bolometric and optical luminosity, respectively. The templates have peak absolute magnitudes of $\approx-20.1$ mag across the filters, while they have peak luminosities of $\approx2.3\times10^{43}$\,erg\,s$^{-1}$ in optical and $\approx3.8\times10^{43}$\,erg\,s$^{-1}$ in bolometric (see also Figures~\ref{fig:bandP} and \ref{fig:bolP}). Their $r$-band 50\% rise and decline timescales are $\approx34$ and $\approx72$ days, respectively, while those in both optical and bolometric luminosity are $\approx31$ and $\approx66$ days, respectively (see also Figures~\ref{fig:rT} and \ref{apfig:optbolT}). The superlinear decline (as noted in \S\ref{sec:lc}) manifests in the light-curve templates across filters, strengthening its utility as a populational property of SNe~Ia-CSM. 

In the left panel of Figure~\ref{fig:IInIbn}, we compare our $r$-band light-curve template of SNe~Ia-CSM with the templates of SNe~IIn from H26 and interacting, stripped-evelope SNe~Ibn from \citet{Hosseinzadeh2017ApJ...836..158H}. SNe~Ia-CSM show almost identical median light-curve evolution to the luminous-slow SNe IIn, but with a more skewed distribution towards the faint end, as also seen in the $E_{\rm rad}$ distributions (see \S\ref{sec:energy}).
In the right panel of Figure~\ref{fig:IInIbn}, we plot our normalized light-curve template of SNe~Ia-CSM, in comparison to the SN~IIn and Ibn templates. As noted in H26, SNe~Ibn fill in the unoccupied fast timescale regions of SNe~IIn, and from this work of SNe~Ia-CSM ($\lesssim24$ days; see \S\ref{sec:time}). This strengths the proposed continuum in timescales across different classes of interacting transients, from hydrogen-poor to hydrogen-rich SNe, i.e., from SNe Ibn, to faint-fast SNe~IIn, and to luminous-slow SNe~IIn and SNe~Ia-CSM. This could be attributed to the physical extent of the progenitor and CSM systems (e.g., \citealt{Moriya2016ApJ...824..100M, Khatami2024ApJ...972..140K}).

\section{Summary and Discussion} 
\label{sec:sum}

Before discussing the broader implications of our findings, we summarize the key observed properties of the SN~Ia-CSM sample identified in \S\ref{sec:ana}: 

\begin{itemize}
    \item Median peak absolute magnitudes of $\approx-20.1$ mag, with $P_{15.9}$--$P_{84.1}$ and $P_{2.3}$--$P_{97.7}$ spreads of $\approx1.4$ and $\approx3.2$ mag, respectively, in the $g$, $c$, $G$, $r$, $o$, and $i$ bands (Figure~\ref{fig:bandP}). 
    \item Median peak bolometric and optical luminosities of $\approx3.8\times10^{43}$ and $\approx2.4\times10^{43}$\,erg\,s$^{-1}$, respectively, with $P_{15.9}$--$P_{84.1}$ and $P_{2.3}$--$P_{97.7}$ spreads of about factors of 3 and 13 (Figure~\ref{fig:bolP}). 
    \item Median $r$-band rise (and decline) timescales at the 50\% flux level of $\approx 27$ ($\approx 59$) days with a 1:2 correlation between rise and decline, and $P_{15.9}$--$P_{84.1}$ and $P_{2.3}$--$P_{97.7}$ ranges of $\approx18-46$ and $\approx10-71$ days ($\approx39-113$ and $\approx20-157$ days), respectively (Figures~\ref{fig:rT} and ~\ref{fig:rT205080}). 
    \item Moderate and weak correlations of the peak luminosity with the rise and decline timescales, respectively (Figure~\ref{fig:Lopt_T50}).
    \item Median optical (and bolometric) radiated energies of $1.4\times10^{50}$ ($2.5\times10^{50}$) erg at the 50\% flux level and $P_{15.9}$--$P_{84.1}$ and $P_{2.3}$--$P_{97.7}$ ranges of $\approx(0.7-2.2)\times10^{50}$ and $\approx(0.07-2.8)\times10^{50}$ ($\approx(0.9-3.7)\times10^{50}$ and $\approx(0.1-4.9)\times10^{50}$) erg, respectively (Figures~\ref{fig:Eradopt} and \ref{apfig:Eradbol}).
    \item Distinct bimodality in the radiated energy distributions, characterized by a predominant population of energetic events with $\approx(1-5)\times10^{50}$\,erg and characteristic timescales of $\approx100$ days, alongside a secondary population of events that are an order of magnitude less energetic with characteristic timescales of $\approx50$ days (Figures~\ref{fig:Eradopt} and \ref{fig:Erad_mL50}).
    \item Significantly more uniform light-curve properties than SNe~IIn, bearing a close resemblance to the luminous-slow IIn subgroup and sharing the almost identical median light-curve evolution (Figure~\ref{fig:IInIbn}).
\end{itemize}

\subsection{Previous Sample Studies}
\label{sec:comp}

\cite{Silverman2013_sample} presented the first systematic sample study of 16 SNe~Ia-CSM, comprising eight events discovered by the Palomar Transient Factory (PTF)\footnote{PTF10htz, PTF10iuf, PTF10yni, PTF11dsb, PTF11hzx, PTF12efc, and PTF12hnr, alongside PTF11kx.} and eight compiled from the literature\footnote{SNe~1997cy, 1999E, 2005gj, 2008J, 2008cg, 2011jb, and CSS120327:110520-015205, alongside SN~2002ic.}. The PTF $R$--band peak magnitudes ($-19 \leq M_{R, {\rm peak}} \leq -21.3$) and rise timescales ($\approx20-40$ days) are well within the parameter space spanned by our larger sample (Figures~\ref{fig:bandP} and \ref{fig:rT}). While \cite{Silverman2013_sample} focused primarily on spectral line identification (e.g., Fe~{\sc ii/iii}, Si~{\sc ii}, and/or S~{\sc ii} absorption alongside Ca~{\sc ii} emission) and measurements -- noting systematically higher Balmer decrements and lower H$\beta$ equivalent widths for SNe~Ia-CSM compared to SNe~IIn -- we defer a comprehensive spectroscopic comparison to a follow-up paper in this series. In this work, we utilized these line diagnostics with visual inspection and spectral cross-correlation to verify the Ia-CSM classifications within our own sample (see \S\ref{sec:sample}). Furthermore, the host galaxies of the PTF sample are exclusively late-type spirals or dwarf irregulars, implying relatively young progenitor systems; we will provide a more detailed investigation of these environments in another follow-up paper focused on the host galaxies and local environments of SNe~Ia-CSM.

More recently, \cite{Sharma2023ApJ...948...52S} investigated 12 SNe~Ia-CSM from the ZTF Bright Transient Survey (BTS). Nine of these events are integrated into our sample, while the remaining three were excluded due to the lack of light-curve peak coverage (Tables~\ref{tab:sample} and \ref{tab:exc}). The BTS sample exhibits light-curve, spectroscopic, and host-galaxy properties closely matching those of the PTF sample. The superior photometric cadence of ZTF enabled the measurements of the total rise and decline timescales above the 50\% flux level ($t_{{\rm rise}50,r}+t_{{\rm decl}50,r}$) ranging from $\approx 30-180$ days. Benefiting from the controlled survey systematics of the BTS, they derived a volumetric rate of $29^{+27}_{-21}\,{\rm Gpc}^{-3}\,{\rm yr}^{-1}$ for SNe~Ia-CSM, which corresponds to roughly $\sim 0.02-0.2\%$ of the normal SN~Ia rate. Because our sample is inherently heterogeneous with different survey systematics, we cannot reliably derive an independent volumetric rate; consequently, we adopt the BTS rate as our baseline in the subsequent discussion regarding the underlying Ia explosions and possible progenitor channels for SNe~Ia-CSM.

\subsection{Underlying SN~Ia Explosions}
\label{sec:exp}

\begin{figure*}
    \centering
    \includegraphics[width=\textwidth]{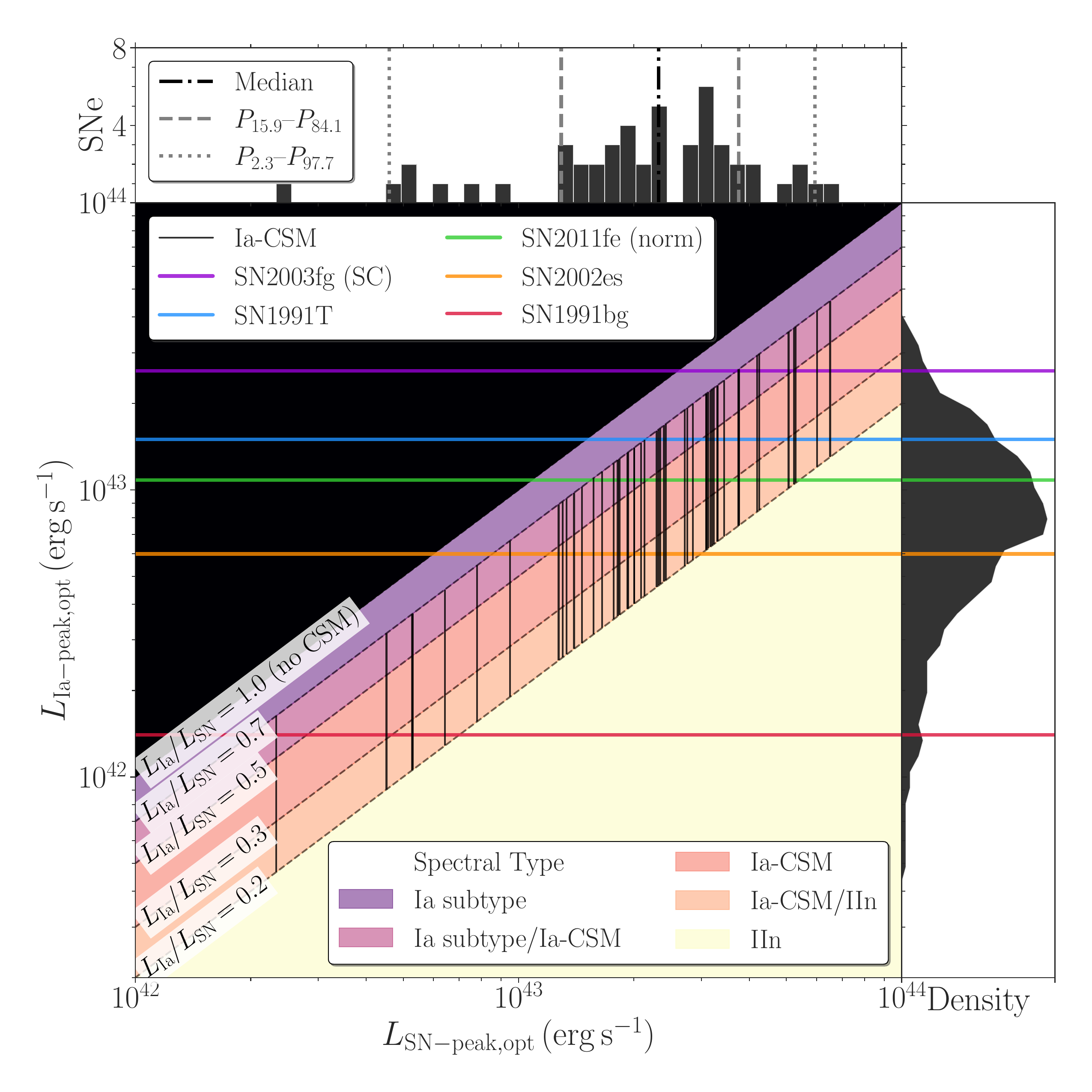}
\caption{
SN spectral types as a function of CSM interaction, parameterized by the ratio of the peak optical luminosity of the underlying Ia ($L_{\rm Ia-peak,opt}$) to the total SN ($L_{\rm SN-peak,opt}$; Figure~\ref{fig:bolP}), as adapted from \cite{Leloudas2015,Hiramatsu2026arXiv260712040H}. As $L_{\rm Ia}/L_{\rm SN}$ decreases (i.e., increasing CSM interaction), the spectral appearance evolves from an unattenuated Ia subtype to transitional Ia/Ia-CSM, Ia-CSM, transitional Ia-CSM/IIn, and finally interaction-dominated IIn, as color-coded between diagonal dashed lines. The $L_{\rm Ia}$ of prototypical Ia subtypes (e.g., normal and Super-Chandrasekhar; SC) are indicated by horizontal lines. The possible fractional luminosity ranges of the Ia-CSM sample ($0.2<L_{\rm Ia}/L_{\rm SN}<0.7$) are shown as vertical lines, with the corresponding projected density distribution in the right panel. The bulk of the Ia-CSM sample maps to underlying Ia subtypes ranging from 1991T-like to 2002es-like, extending to 2003fg-like (SC) and 1991bg-like at the extrema.
}
    \label{fig:fIa}
\end{figure*}

Together with the light-curve properties of the SN~Ia-CSM sample, we utilize spectral classification (see \S\ref{sec:comp}) to characterize their possible underlying Ia subtypes. In Figure~\ref{fig:fIa}, we compare the peak optical luminosities of the Ia-CSM sample ($L_{\rm SN}$), shown in the top histogram, to those of underlying SNe~Ia ($L_{\rm Ia}$); $L_{\rm Ia}$ of representative SN~Ia subtypes are plotted as horizontal lines (see e.g., \citealt{Taubenberger2017hsn..book..317T} and references therein). The fractional luminosity ($L_{\rm Ia}/L_{\rm SN}$) defines the parameter space of SN spectral types, deduced from the Monte Carlo simulations of \cite{Leloudas2015}. With increasing CSM interaction (decreasing $L_{\rm Ia}/L_{\rm SN}$), the resultant SN spectral types change continuously, as color-coded between diagonal lines in the main panel: from Type Ia subtypes ($L_{\rm Ia}/L_{\rm SN}<0.7$) to transitional Ia/Ia-CSM ($0.5<L_{\rm Ia}/L_{\rm SN}<0.7$), Ia-CSM ($0.3<L_{\rm Ia}/L_{\rm SN}<0.5$), transitional Ia-CSM/IIn ($0.2<L_{\rm Ia}/L_{\rm SN}<0.3$), and finally interaction-dominated IIn ($L_{\rm Ia}/L_{\rm SN}<0.2$). 

Given the verified Ia-CSM classification of our sample, we assume their fractional luminosities fall within $0.2<L_{\rm Ia}/L_{\rm SN}<0.7$. We plot these as vertical black lines in the main panel, with their projected sample $L_{\rm Ia}$ density displayed in the right panel.
The bulk of the sample is consistent with an $L_{\rm Ia}$ range within the luminous 1991T-like to subluminous 2002es-like, with the density peaking between normal Ia and 2002es-like. This population constitutes the main peak in the $E_{\rm rad}$ distribution around $(1-6)\times10^{50}$\,erg (Figures~\ref{fig:Eradopt} and \ref{apfig:Eradbol}). Assuming a canonical Ia kinetic energy of $E_{\rm KE}=1.3\times10^{51}$ erg \citep{Nomoto1984}, this $E_{\rm rad}$ range translates to a kinetic-to-radiative conversion efficiency of $\epsilon_{\rm rad}\approx0.1-0.6$, where $E_{\rm rad} = \epsilon_{\rm rad}E_{\rm KE}$. This high efficiency is consistent with the range of numerical light-curve simulations ($\epsilon_{\rm rad}\approx0.05-0.5$), which shows that a large mass ratio above unity of CSM ($\approx2-11\,\Msun$) to SN ejecta ($\approx1.4\,\Msun$, i.e., the Chandrasekhar mass of a CO WD) is required to produce the luminous peaks of SNe~Ia-CSM (see Figure~11 of \citealt{Hiramatsu2026arXiv260712040H}), with larger mass ratios leading to higher $\epsilon_{\rm rad}$ (e.g., \citealt{Tsuna2019ApJ...884...87T,Khatami2024ApJ...972..140K}). 

Moreover, the inferred $L_{\rm Ia}$ distribution extends down to the faint SN 1991bg-like regime. Objects in this fainter regime account for the secondary peak in the $E_{\rm rad}$ distribution around $(0.1-0.8)\times10^{50}$\,erg. Adopting a similar $\epsilon_{\rm rad}$ implies an intrinsically low $E_{\rm KE}\lesssim10^{51}$\,erg, which aligns with energy estimates for 1991bg-like events (e.g., \citealt{Turatto1996MNRAS.283....1T,Mazzali1997MNRAS.284..151M,Mazzali2011MNRAS.416..881M,Mazzali2012MNRAS.424.2926M}). Coupled with a potential extension to the Super-Chandrasekhar 2003fg-like regime at the luminous end, the overall SN~Ia-CSM sample spans the full diversity of $L_{\rm Ia}$. This demonstrates that a wide variety of underlying SN Ia subtypes can produce this interacting Ia-CSM class. Furthermore, it points to the possibility that events with a dominant CSM contribution ($L_{\rm Ia}/L_{\rm SN} \lesssim 0.2$) may observationally manifest as SNe~IIn (see also \citealt{Leloudas2015}), which could explain the apparent deficit of fainter SN~Ia subtypes within the Ia-CSM sample.

Although diverse, all the parameter distributions of the SN~Ia-CSM sample are notably more uniform than those of SNe~IIn. This relative homogeneity likely reflects the stricter energetic constraints governing thermonuclear Ia explosions compared to the highly diverse explosion mechanisms (e.g., neutrino driven, pulsational pair instability, jittering jet) and progenitor configurations of core-collapse SNe (as discussed in H26).

\subsection{Light-Curve Analytics}
\label{sec:LCana}

To gain baseline analytics, we lay out light-curve scaling equations of SN~Ia ejecta interacting with CSM of a mass $M_{\rm CSM}$ at a radius $R_{\rm CSM}$. We note that this analysis is independent of the exact physical origins of SN and CSM. When $M_{\rm CSM}$ exceeds the Ia ejecta mass, as typically anticipated if the CSM originates from the CE of a donor star, the kinetic energy of the Ia ejecta is almost entirely dissipated by the interaction at $R_{\rm CSM}$ (see also \S\ref{sec:exp}). After the interaction, the shocked CSM expands with velocity
\begin{eqnarray}
    v &\approx& \sqrt{\frac{2E_{\rm KE}}{M_{\rm CSM}}} \nonumber \\
    &\approx& 4,\!500\ {\rm km\ s^{-1}}\left(\frac{E_{\rm KE}}{10^{51}~{\rm erg}}\right)^{\!\!1/2}\left(\frac{M_{\rm CSM}}{5~\Msun}\right)^{\!-1/2},
\end{eqnarray}
and radiates over a diffusion timescale
\begin{eqnarray}
    t_{\rm diff}&\approx& \sqrt{\frac{\kappa M_{\rm CSM}}{4\pi c v}} \nonumber \\
    &\sim& 50~{\rm day}\left(\frac{M_{\rm CSM}}{5~\Msun}\right)^{\!\!3/4}\left(\frac{E_{\rm KE}}{10^{51}~{\rm erg}}\right)^{\!-1/4}
\end{eqnarray}
where $\kappa\approx 0.3\ {\rm cm^2\ g^{-1}}$ is the opacity which we adopt a typical value for the hydrogen-rich CSM. This diffusion roughly reproduces the characteristic timescales of SN Ia-CSM. Over this diffusion time, the CSM expands out to
\begin{eqnarray}
    vt_{\rm diff}\sim 2\times 10^{15}\ {\rm cm}\left(\frac{M_{\rm CSM}}{5~\Msun}\right)^{\!\!1/4}\left(\frac{E_{\rm KE}}{10^{51}~{\rm erg}}\right)^{\!1/4}.
\end{eqnarray}
If $R_{\rm CSM}<vt_{\rm diff}$, there is adiabatic expansion that leads to loss of internal energy generated by CSM interaction. The luminosity around peak is then roughly calculated as
\begin{eqnarray}
  L &\approx& \frac{E_{\rm KE}}{t_{\rm diff}}{\rm min}\left(1,\frac{R_{\rm CSM}}{t_{\rm diff}v}\right) \nonumber \\
  &\sim& 2\times 10^{43}\ {\rm erg\ s^{-1}}\left(\frac{M_{\rm CSM}}{5~\Msun}\right)^{\!\!-3/4}\left(\frac{E_{\rm KE}}{10^{51}~{\rm erg}}\right)^{\!5/4}\nonumber \\
  && {\rm min}\!\!\!\left[\!10, \!\left(\frac{R_{\rm CSM}}{3,\!000~\Rsun}\right)\!\!\!\left(\frac{M_{\rm CSM}}{5~\Msun}\right)^{\!\!\!-1/4}\!\!\!\!\left(\frac{E_{\rm KE}}{10^{51}~{\rm erg}}\right)^{\!\!\!-1/4}\!\right]
\end{eqnarray}
where the normalization factor of 10 inside the minimum function is chosen such that the function becomes unity for typical parameter values.
Thus, the typical Ia-CSM luminosity range of $10^{43}-10^{44}$\,erg\,s$^{-1}$ (Figure~\ref{fig:bolP}) indicates a fairly extended CSM spanning
$10^3-10^4~\Rsun$. While massive red supergiant (RSG) stars can reach such physical scales, these are an order of magnitude larger than the radii of typical intermediate-mass asymptotic giant branch (AGB) stars. Nonetheless, such a large interaction radius remains plausible for merger-driven explosions (see \S\ref{sec:prog}); the CE can expand continuously as the WD inspirals toward the donor core prior to the core merger and subsequent explosion \citep[e.g.,][]{MacLeod2018ApJ...868..136M,MacLeod2020ApJ...895...29M,Schroder20}.

\subsection{Possible Progenitor Channels}
\label{sec:prog}

An outstanding question is whether SN~Ia progenitor systems can produce such massive CSM ($\approx2-11\,\Msun$; \S\ref{sec:exp} and \ref{sec:LCana}). One viable channel is an interacting binary system consisting of a CO WD and either an AGB ($\lesssim8\,\Msun$) or RSG ($\gtrsim8\,\Msun$) companion star. This system undergoes CE evolution and explodes via a core merger inside the CE, i.e., the CD scenario (e.g., \citealt{Kashi2011MNRAS.417.1466K,Soker2015,Soker2019,Ablimit2021}). In this scenario, the initially more massive primary star evolves and fills its Roche lobe; after stable mass transfer, it is stripped down to a CO WD, while the secondary star accretes this material and grows into a massive O/B star. When this O/B companion subsequently evolves to the Hertzsprung gap or core helium-burning phase, it expands and outflows its own Roche lobe. Because the resulting mass ratio (WD to companion) is much less than unity, the mass transfer becomes highly unstable, forming a massive CE that engulfs the CO WD (see e.g., Figures~5 and 6 of \citealt{Ablimit2021}).

Binary population synthesis calculations by \cite{Ablimit2021} show that CE masses of $\sim2-7$ and $6-12\,\Msun$ are expected for AGB and RSG companion stars, respectively (see also \citealt{Liu2017A&A...606A.136L,Twum2026arXiv260210211T} for various other binary configurations and core-merger outcomes). Their respective delay-time distributions span $\sim0.1-10$ and $0.03-0.12$\,\,Gyr, with estimated event rates hovering around $0.1\%$  of the total SN~Ia rate. These properties align well with the host galaxy demographics and volumetric rates of SNe~Ia-CSM reported by \cite{Silverman2013_sample,Sharma2023ApJ...948...52S} (see \S\ref{sec:comp}). Therefore, the CD scenario offers a plausible mechanism for generating dense CSM, provided the merger-driven thermonuclear Ia explosion (e.g., \citealt{Dan2012MNRAS.422.2417D}) occurs promptly after or during the CE phase. 

Because the overall Ia-CSM sample spans the full diversity of  $L_{\rm Ia}$ (see \S\ref{sec:exp} and \ref{sec:LCana}), the CD scenario could potentially serve as a common evolutionary channel for not only SNe~Ia-CSM but also standard SNe~Ia. In this framework, the presence or absence of dense CSM -- and the resulting interaction signatures and onsets (e.g., the delayed CSM interaction observed in SN~2002ic and PTF11kx) -- would depend primarily on the exact delay time between the CE phase and subsequent Ia explosion, with longer delay times leading to cleaner SNe~Ia (see also \citealt{Kashi2011MNRAS.417.1466K,Soker2019}). 

However, detailed hydrodynamical simulations remain essential to confirm whether the final product of the CE phase is indeed a core merger capable of triggering a successful thermonuclear explosion. Furthermore, targeted population synthesis calculations are highly warranted to accurately constrain both the delay time distribution between the CE phase and the subsequent Ia explosion, and the relative rate of SNe Ia-CSM to standard SNe~Ia within this evolutionary channel.

\section{Conclusion} 
\label{sec:conc}

We have presented the largest uniform study to date of SNe~Ia-CSM, analyzing a sample of 50 objects. As the first paper in a series designed to delineate the broader population properties of this class, this work focuses specifically on a uniform analysis of their light curves. Our key findings and conclusions are that SNe~Ia-CSM:

\begin{itemize}
\item Span approximately an order of magnitude in peak luminosity, total timescale (rise plus decline), and radiated energy (Figures~\ref{fig:bandP}, \ref{fig:rT}, and \ref{fig:Eradopt}).
\item Exhibit a distinct bimodality in their radiated energy distributions, separating into major energetic events ($\approx3\times10^{50}$\,erg) with characteristic timescales of $\approx100$ days, and a minor, less energetic population with timescales of $\approx50$ days (Figures~\ref{fig:Eradopt} and \ref{fig:Erad_mL50}).
\item Display a significantly more uniform light-curve morphology than SNe~IIn, closely resembling the luminous-slow IIn subgroup (Figure~\ref{fig:IInIbn}).
\item Reveal a diverse range of underlying SN~Ia explosion subtypes and CSM masses that are broadly consistent with the CD channel from CE evolution (Figure~\ref{fig:fIa}).
\end{itemize}

Our comprehensive exploration of SNe~Ia-CSM will continue in subsequent papers in this series. Future works will cover a population-level comparison to numerical light-curve simulations; semi-analytical modeling of individual light curves to extract SN and CSM parameters; a detailed spectroscopic analysis of SN continuum and Balmer line morphologies; and an evaluation of host galaxy environments and explosion sites. Ultimately, establishing correlations between these various photometric, spectroscopic, and environmental properties will provide critical insights into the progenitor systems that drive this intriguing class of SN~Ia-CSM mapping SN~Ia and CSM formation pathways.

\section*{Acknowledgments}

We thank the time-domain astronomy community for providing public transient discovery and classification reports.
We gratefully acknowledge helpful discussions on related topics with members of the Transient Theory group at Harvard University, and with A. Twum, E. Ramirez-Ruiz, and E. Hueichapan. 

The Hiramatsu Time-Domain Explosions (Hi-TiDEs) group at the University of Florida is supported by STScI grants HST-GO-17770.002, JWST-GO-12468.001, and JWST-GO-09964.001.
D.T. is supported by Harvard University through the Institute for Theory and Computation Fellowship.

TNS is supported by funding from the Weizmann Institute of Science, as well as grants from the Israeli Institute for Advanced Studies and the European Union via ERC grant No. 725161. 
We acknowledge WISeREP (\url{https://wiserep.org}).

This work has made use of data from the Zwicky Transient Facility (ZTF).  ZTF is supported by NSF grant No. AST-1440341 and a collaboration including Caltech, IPAC, the Weizmann Institute for Science, the Oskar Klein Center at Stockholm University, the University of Maryland, the University of Washington, Deutsches Elektronen-Synchrotron and Humboldt University, Los Alamos National Laboratories, the TANGO Consortium of Taiwan, the University of Wisconsin at Milwaukee, and Lawrence Berkeley National Laboratories. Operations are conducted by COO, IPAC, and UW. The ZTF forced-photometry service was funded under the Heising-Simons Foundation grant
No. 12540303 (PI: Graham).

This work has made use of data from the Asteroid Terrestrial-impact Last Alert System (ATLAS) project. ATLAS is primarily funded to search for near-Earth asteroids through NASA grant Nos. NN12AR55G, 80NSSC18K0284, and 80NSSC18K1575; byproducts of the NEO search include images and catalogs from the survey area. This work was partially funded by Kepler/K2 grant No. J1944/80NSSC19K0112 and HST grant No. GO-15889, and STFC grant Nos. ST/T000198/1 and ST/S006109/1. The ATLAS science products have been made possible through the contributions of the University of Hawaii Institute for Astronomy, the Queen’s University Belfast, the Space Telescope Science Institute, the South African Astronomical Observatory, and The Millennium Institute of Astrophysics (MAS), Chile.

We acknowledge ESA \textit{Gaia}, DPAC, and the Photometric Science Alerts Team (\url{http://gsaweb.ast.cam.ac.uk/alerts}).

This research has made use of the NASA Astrophysics Data System (ADS, which is funded by NASA under Cooperative Agreement 80NSSC21M00561) and NASA/IPAC Infrared Science Archive (IRSA, which is funded by NASA and operated by the California Institute of Technology).


\textit{Facilities:} 
ADS, ATLAS, Gaia, IRSA, TNS, ZTF

\textit{Software:}
\texttt{ALeRCE} \citep{ALeRCE2021AJ....161..242F},
\texttt{Astropy} \citep{AstropyCollaboration2013,AstropyCollaboration2018,AstropyCollaboration2022}, 
\texttt{astroquery} \citep{Ginsburg2019AJ....157...98G},
\texttt{corner} \citep{Foreman-Mackey2016JOSS....1...24F}, 
\texttt{emcee} \citep{Foreman-Mackey2013PASP..125..306F}, 
\texttt{Matplotlib} \citep{Hunter2007CSE.....9...90H}, 
\texttt{NumPy} \citep{Oliphant2006}, 
\texttt{scikit-learn} \citep{Pedregosa2011JMLR...12.2825P}, 
\texttt{SciPy} \citep{SciPy2020NatMe..17..261V}, 
\texttt{seaborn} \citep{seaborn2020zndo...3629446W}.

\appendix

\section{Additional Sample Description}
\label{sec:extra}

Detailed information for the SN~Ia-CSM sample is compiled in Table~\ref{tab:sample}. For completeness, the objects excluded from the primary analysis, alongside the specific reasons for their omission, are summarized in Table~\ref{tab:exc}.

\begin{deluxetable*}{cccccc}
\tabletypesize{\scriptsize}
\tablecaption{SN~Ia-CSM Sample (N=50) \label{tab:sample}}
\tablehead{
\colhead{SN} & \colhead{Redshift} & \colhead{TNS} & \colhead{WISeREP} & \colhead{Discovery} & \colhead{Classfication} \\
\colhead{} & \colhead{(z)} & \colhead{Type} & \colhead{Type} & \colhead{Report} & \colhead{Report}
}
2025aaxv & 0.0798 & Ia-CSM & Ia-CSM & \cite{2025TNSTR4191....1S} & \cite{2026TNSCR.240....1W} \\
2025xtq & 0.054 & Ia-CSM & Ia-CSM & \cite{2025TNSTR3746....1S} & \cite{2026TNSCR1413....1D} \\
2025ril & 0.0658 & Ia-CSM & Ia-CSM & \cite{2025TNSTR2739....1T} & \cite{2025TNSCR3306....1F} \\
2025qim & 0.117 & Ia-CSM & Ia-CSM & \cite{2025TNSTR2573....1X} & \cite{2025TNSCR2879....1M} \\
2025nfb & 0.059 & Ia-CSM & Ia-CSM & \cite{2025TNSTR2170....1T} & \cite{2025TNSCR2682....1S} \\
2025ndr & 0.182 & Ia-CSM & Ia-CSM & \cite{2025TNSTR2160....1S} & \cite{2025TNSCR3531....1P} \\
2025fry & 0.1344 & Ia-CSM & Ia-CSM & \cite{2025TNSTR1112....1M} & \cite{2025TNSCR1669....1C} \\
2024abpg & 0.154 & Ia-CSM & Ia-CSM & \cite{2024TNSTR4570....1S} & \cite{2025TNSCR.380....1W} \\
2024ztk & 0.096 & Ia-CSM & Ia-CSM & \cite{2024TNSTR4213....1S} & \cite{2024TNSCR4471....1T} \\
2024zjp & 0.15 & Ia-CSM & Ia-CSM & \cite{2024TNSTR4143....1T} & \cite{2024TNSCR4425....1H} \\
2024tyc & 0.0355 & Ia-CSM & Ia-CSM & \cite{2024TNSTR3181....1P} & \cite{2024TNSCR3896....1T} \\
2024knu & 0.15 & Ia-CSM & Ia-CSM & \cite{2024TNSTR1849....1Z} & \cite{2024TNSCR2424....1D} \\
2024gvz & 0.055 & Ia-CSM & Ia-CSM & \cite{2024TNSTR1130....1T} & \cite{2024TNSCR3256....1M} \\
2024bex & 0.1423 & Ia-CSM & Ia-CSM & \cite{2024TNSTR.286....1M} & \cite{2024TNSCR.999....1P} \\
2023qtr & 0.033 & Ia-CSM & Ia-CSM & \cite{2023TNSTR2083....1F} & \cite{2023TNSCR2509....1J} \\
2023ocx & 0.076 & Ia-CSM & Ia-CSM & \cite{2023TNSTR1800....1F} & \cite{2023TNSCR2467....1A} \\
2023khp & 0.0908 & Ia-CSM & -- & \cite{2023TNSTR1357....1T} & \cite{2023TNSCR1829....1S} \\
2023ggb & 0.08 & Ia-CSM & Ia & \cite{2023TNSTR.860....1F} & \cite{2023TNSCR1311....1W} \\
2023ayq & 0.185 & Ia-CSM & -- & \cite{2023TNSTR.227....1F} & \cite{2023TNSCR.591....1S} \\
2022acgi & 0.17 & Ia-CSM & Ia-CSM & \cite{2022TNSTR3516....1T} & \cite{2022TNSCR3533....1G} \\
2022xlm & 0.136 & Ia-CSM & Ia-CSM & \cite{2022TNSTR2967....1M} & \cite{2022TNSCR3182....1R} \\
2022ujn & 0.12 & Ia-CSM & Ia-CSM & \cite{2022TNSTR2674....1F} & \cite{2022TNSCR2854....1D} \\
2022pja & 0.213 & Ia-CSM & Ia-CSM & \cite{2022TNSTR2076....1C} & \cite{2023TNSCR.136....1G} \\
2022nen\tablenotemark{a} & 0.1165 & SLSN-II & SLSN-II & \cite{2022TNSTR1719....1P} & \cite{2022TNSCR2509....1G} \\
2022igm & 0.077 & Ia-CSM & Ia-CSM & \cite{2022TNSTR1048....1F} & \cite{2022TNSCR3135....1T} \\
2022erq\tablenotemark{b} & 0.066 & Ia-CSM & Ia-CSM & \cite{2022TNSTR.679....1T} & \cite{2022TNSCR.701....1L} \\
2021aekv\tablenotemark{a} & 0.187 & SLSN-II & SLSN-II & \cite{2021TNSTR3894....1T} & \cite{2021TNSCR4234....1P} \\
2021uiq\tablenotemark{c} & 0.126 & Ia-91T-like & Ia-91T-like & \cite{2021TNSTR2617....1J} & \cite{2021TNSCR2747....1T} \\
2021tqt\tablenotemark{a,d} & 0.06 & I & IIn & \cite{2021TNSTR2498....1T} & \cite{2021TNSCR2619....1M} \\
2021jmm\tablenotemark{a} & 0.19 & IIn & IIn & \cite{2021TNSTR1192....1Y} & \cite{2021TNSCR1647....1G} \\
2020aekp\tablenotemark{c,e} & 0.046 & Ia-CSM & Ia & \cite{2020TNSTR3948....1H} & \cite{2021TNSCR1634....1D} \\
2020aeur & 0.0506 & Ia-CSM & Ia-CSM & \cite{2021TNSTR..82....1C} & \cite{2021TNSCR..86....1G} \\
2020abfe\tablenotemark{f} & 0.093 & Ia-CSM & Ia-CSM & \cite{2020TNSTR3612....1F} & \cite{2021TNSCR1234....1D} \\
2020zos\tablenotemark{a} & 0.14 & IIn & IIn & \cite{2020TNSTR3437....1F} & \cite{2020TNSCR3730....1B} \\
2020xtg\tablenotemark{e} & 0.0612 & II & II & \cite{2020TNSTR3195....1H} & \cite{2021TNSCR.223....1D} \\
2020qxz\tablenotemark{e,f} & 0.0964 & Ia-CSM & Ia-CSM & \cite{2020TNSTR2427....1N} & \cite{2022TNSCR.277....1C} \\
2020onv\tablenotemark{e} & 0.095 & Ia-CSM & IIn & \cite{2020TNSTR2100....1T} & \cite{2020TNSCR2479....1Q} \\
2020kro\tablenotemark{a} & 0.127 & IIn & IIn & \cite{2020TNSTR1440....1T} & \cite{2020TNSCR2394....1B} \\
2020kre\tablenotemark{c} & 0.136 & Ia-CSM & Ia & \cite{2020TNSTR1440....1T} & \cite{2020TNSCR2258....1D} \\
2020mad\tablenotemark{a} & 0.123 & SLSN-II & SLSN-II & \cite{2020TNSTR1716....1S} & \cite{2020TNSCR3149....1G} \\
2019rvb\tablenotemark{e} & 0.1835 & Ia-CSM & Ia & \cite{2019TNSTR1990....1N} & \cite{2020TNSCR2653....1Y} \\
2018gkx\tablenotemark{e} & 0.137 & Ia-CSM & Ia-CSM & \cite{2018TNSTR1402....1F} & \cite{2018TNSCR1870....1F} \\
2018evt\tablenotemark{e,g} & 0.0254 & Ia & Ia & \cite{2018TNSTR1144....1N} & \cite{2018TNSCR1168....1S} \\
2018crl\tablenotemark{e,h} & 0.097 & Ia-pec & Ia-CSM & \cite{2018TNSTR.851....1F} & \cite{2018TNSCR1404....1F} \\
2018lxa\tablenotemark{a} & 0.202 & SLSN-II & SLSN-II & \cite{2021TNSTR1713....1N} & \cite{2024TNSCR3447....1P} \\
2017hzw & 0.0514 & Ia-CSM & Ia-CSM & \cite{2017TNSTR1246....1F} & \cite{2017TNSCR1330....1F} \\
2017ifu & 0.183 & Ia-CSM & Ia-CSM & \cite{2017TNSTR1263....1T} & \cite{2017TNSCR1436....1A} \\
2017eni\tablenotemark{a} & 0.0816 & IIn & IIn & \cite{2017TNSTR.632....1D} & \cite{2017TNSCR.833....1S} \\
2017eby & 0.081 & Ia-CSM & Ia-CSM & \cite{2017TNSTR.560....1T} & \cite{2017TNSCR1526....1P} \\
2016iks & 0.063 & Ia-CSM & Ia-CSM & \cite{2016TNSTR.948....1D} & \cite{2016TNSCR1051....1B} \\

\tablenotetext{a}{Ia-CSM classification from \citet[see their Table A3]{Hiramatsu2026ApJ..1005...82H}.} 
\tablenotetext{b}{See \cite{Zhai2026ApJ..1007..184Z} for a detailed study.} 
\tablenotetext{c}{In YSE DR1 \citep[][see their Table C1]{Aleo2023ApJS..266....9A}.} 
\tablenotetext{d}{See also \cite{2021TNSAN.206....1D}.} 
\tablenotetext{e}{In the ZTF BTS Ia-CSM sample \citep[][see their Table 1]{Sharma2023ApJ...948...52S}.} 
\tablenotetext{f}{See \cite{Terwel2025A&A...702A..21T} for a detailed study.} 
\tablenotetext{g}{See \cite{Yang2023MNRAS.519.1618Y,Wang2024NatAs...8..504W} for detailed studies.} 
\tablenotetext{h}{See also \cite{Yao2019ApJ...886..152Y}.} 
\end{deluxetable*}
\begin{deluxetable*}{cccccc}
\tabletypesize{\scriptsize}
\tablecaption{Excluded Sample (N=9) \label{tab:exc}}
\tablehead{
\colhead{SN} & \colhead{TNS} & \colhead{WISeREP} & \colhead{Discovery} & \colhead{Classfication} & \colhead{Reason} \\
\colhead{} & \colhead{Type} & \colhead{Type} & \colhead{Report} & \colhead{Report} & \colhead{} 
}
\multicolumn{6}{c}{Spectroscopic Cuts} \\
\hline
2025kxz & Ia-CSM & Ia-CSM & \cite{2025TNSTR1877....1S} & \cite{2025TNSCR1906....1C} & SN Ia  \\
2023vtj & Ia-CSM & Ia-CSM & \cite{2023TNSTR2714....1F} & \cite{2023TNSCR3011....1R} & Galaxy contamination\\
2022esa\tablenotemark{a} & Ia-CSM & Ia-CSM & \cite{2022TNSTR.685....1T} & \cite{2022TNSCR.733....1L} & Ic-CSM \\
\hline
\multicolumn{6}{c}{Photometric Cuts} \\
\hline
2023cpq & Ia-CSM & Ia-CSM & \cite{2023TNSTR.442....1P} & \cite{2023TNSCR1029....1G} & No peak coverage \\
2021admm & Ia-CSM & Ia-CSM & \cite{2021TNSTR3767....1M} & \cite{2021TNSCR4071....1N} & No peak coverage \\
2020uem\tablenotemark{b,c} & IIP & Ia & \cite{2020TNSTR2930....1T} & \cite{2020TNSCR3122....1D} & No peak coverage \\
2019ibk\tablenotemark{c} & II & II & \cite{2019TNSTR1070....1D} & \cite{2019TNSCR1402....1D} & No peak coverage \\
2019fcg\tablenotemark & IIn & Ia-CSM & \cite{2019TNSTR.767....1S} & \cite{2019TNSCR.782....1N} & No peak coverage  \\
2019agi\tablenotemark{c} & Ia-CSM & Ia-CSM & \cite{2019TNSTR.157....1N} & \cite{2019TNSCR.343....1P} & No peak coverage \\

\tablenotetext{a}{See \cite{Griffith2025ApJ...995...54G,Maeda2026PASJ...78L...1M} for detailed studies.} 
\tablenotetext{b}{See \cite{Uno2023ApJ...944..203U,Uno2023ApJ...944..204U} detailed studies.} 
\tablenotetext{c}{In the ZTF BTS Ia-CSM sample \citep[][see their Table 1]{Sharma2023ApJ...948...52S}.} 
\end{deluxetable*}

\section{Timescale and Radiated Energy in Luminosity Space}
\label{sec:Ltime}

\begin{figure*}
    \centering
    \includegraphics[width=0.49\textwidth]{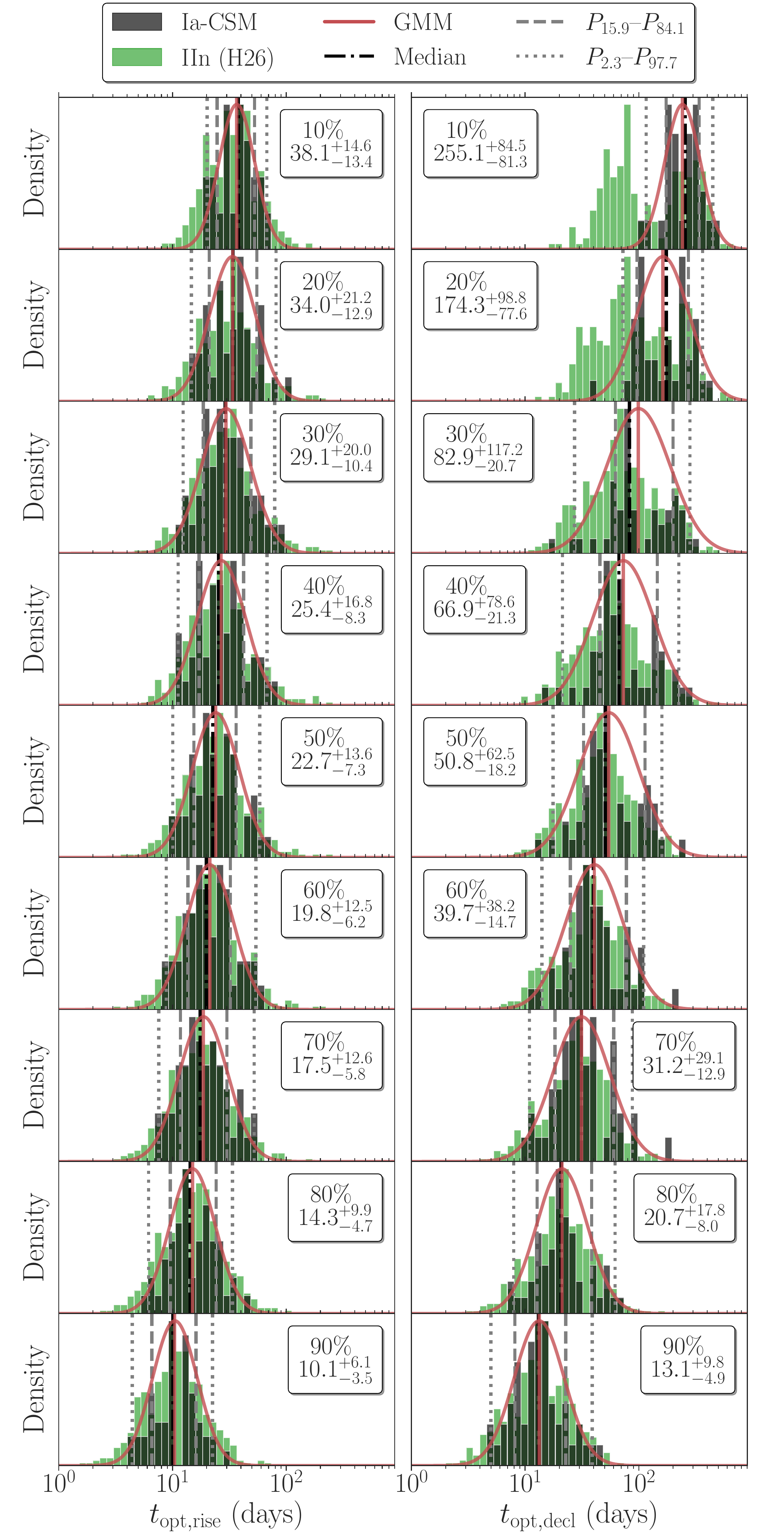}
    \includegraphics[width=0.49\textwidth]{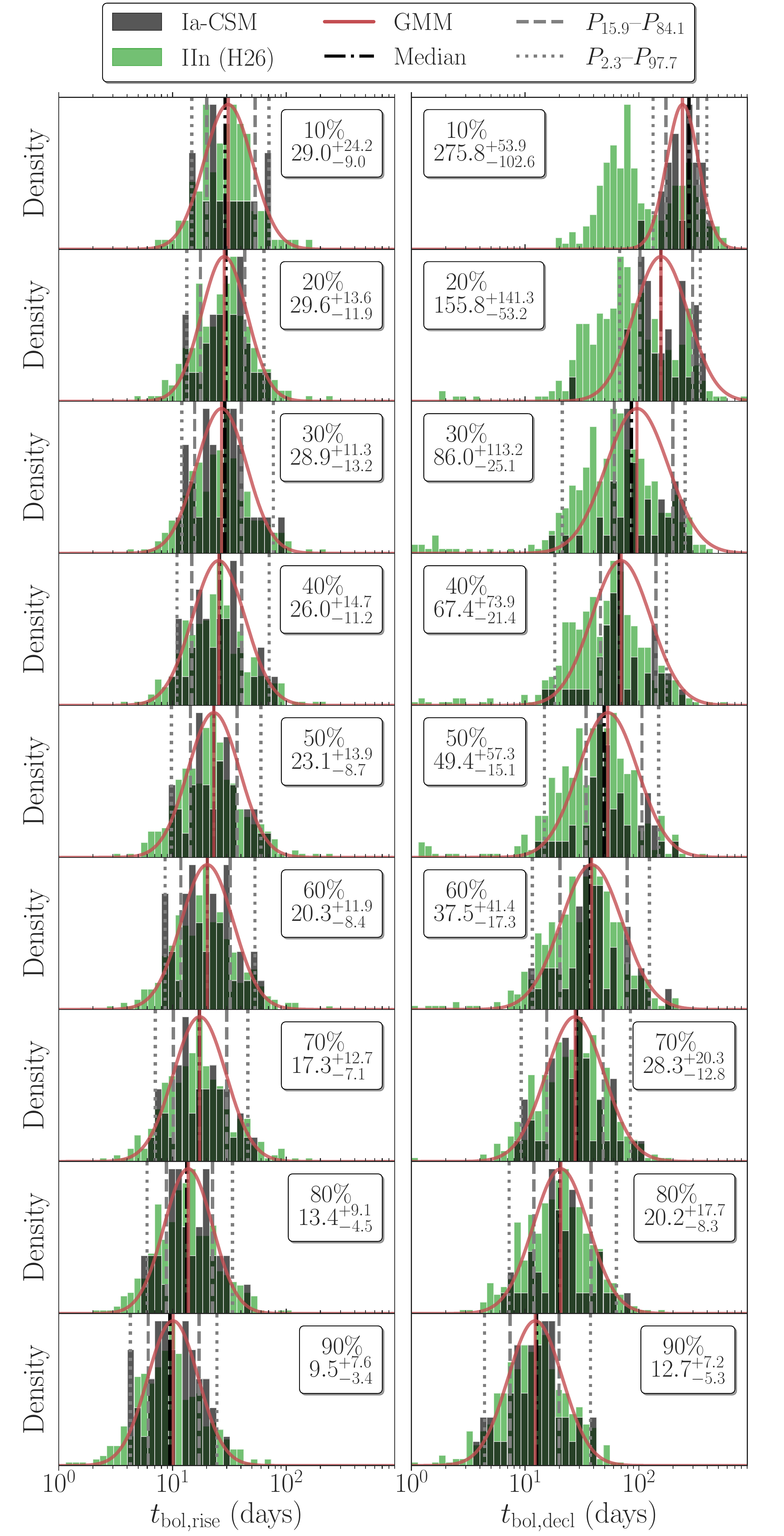}
\caption{Similar to Fig.~\ref{fig:rT}, but for optical (\textit{left}) and bolometric (\textit{right}) luminosity.
}
    \label{apfig:optbolT}
\end{figure*}

\begin{figure*}
    \centering
    \includegraphics[width=0.46\textwidth]{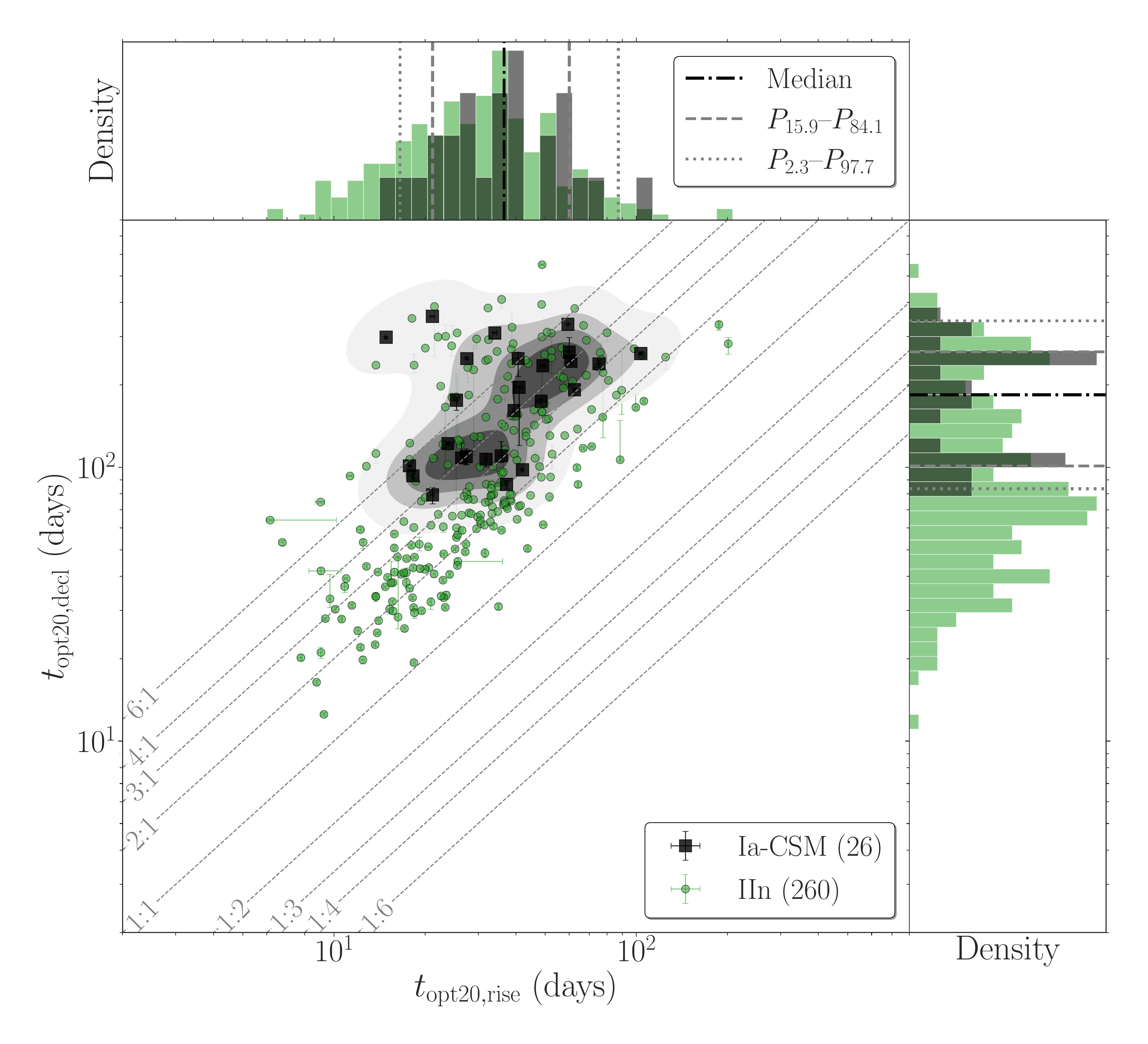}
    \includegraphics[width=0.46\textwidth]{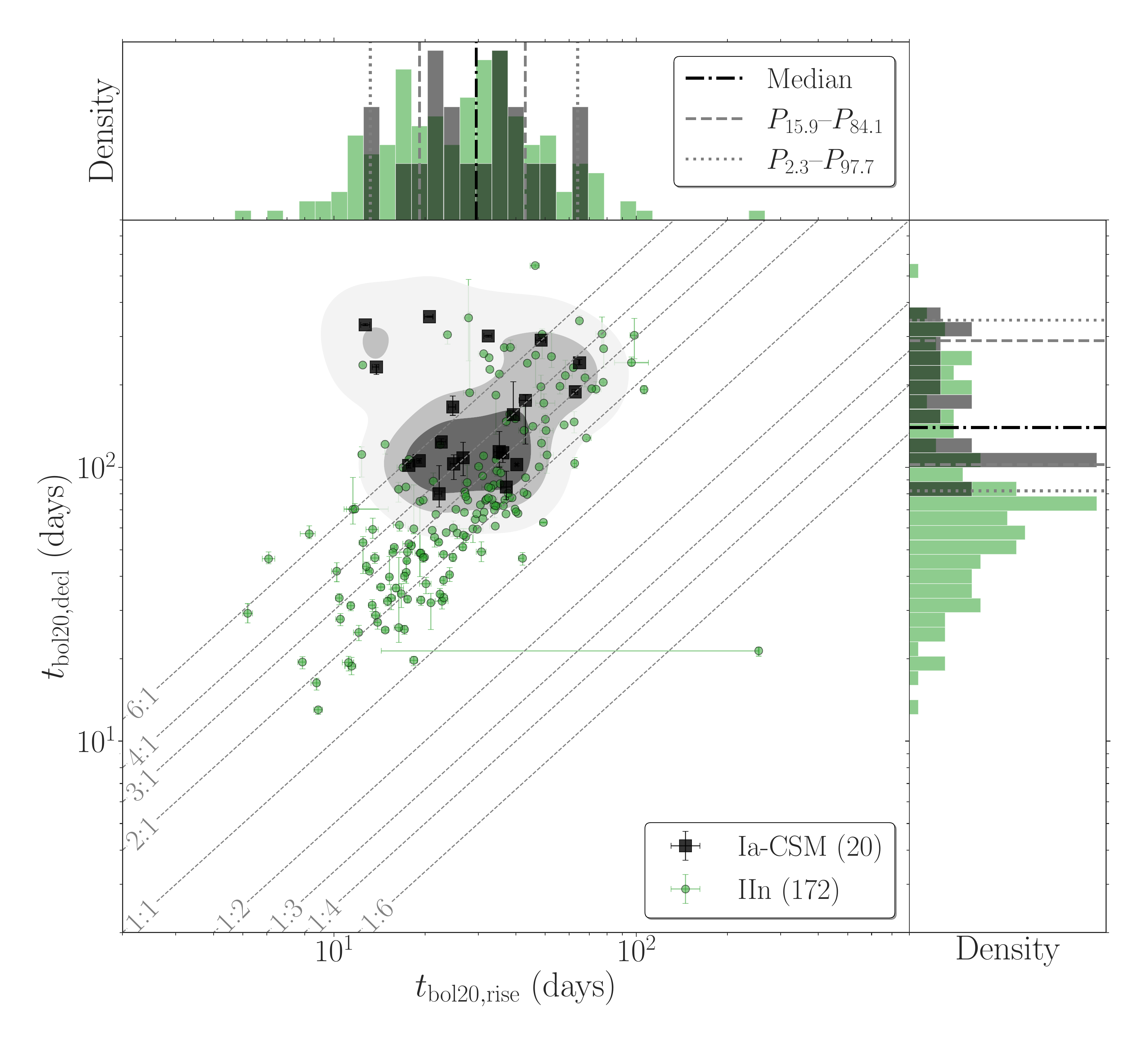}
    \includegraphics[width=0.46\textwidth]{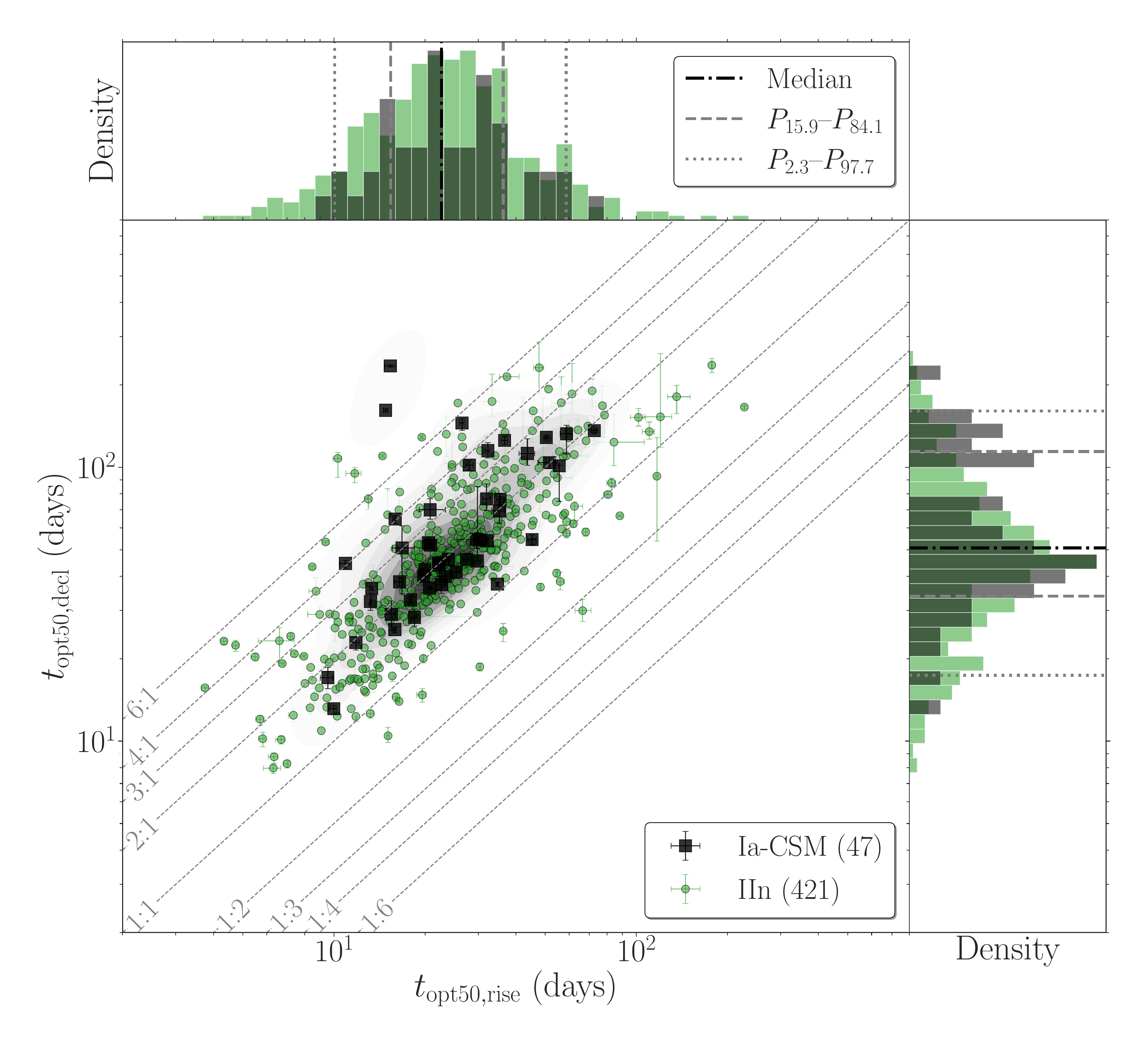}
    \includegraphics[width=0.46\textwidth]{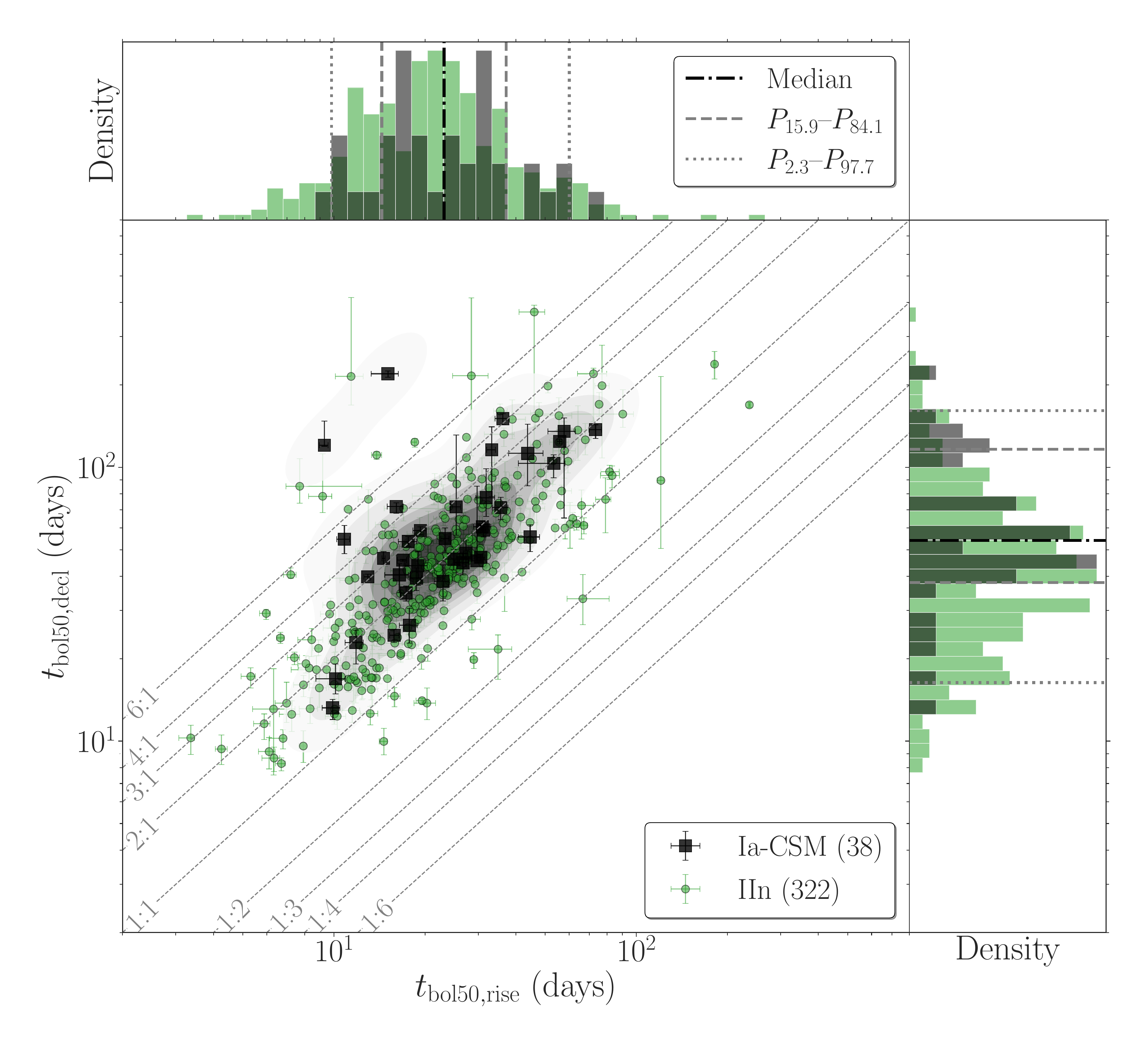}
    \includegraphics[width=0.46\textwidth]{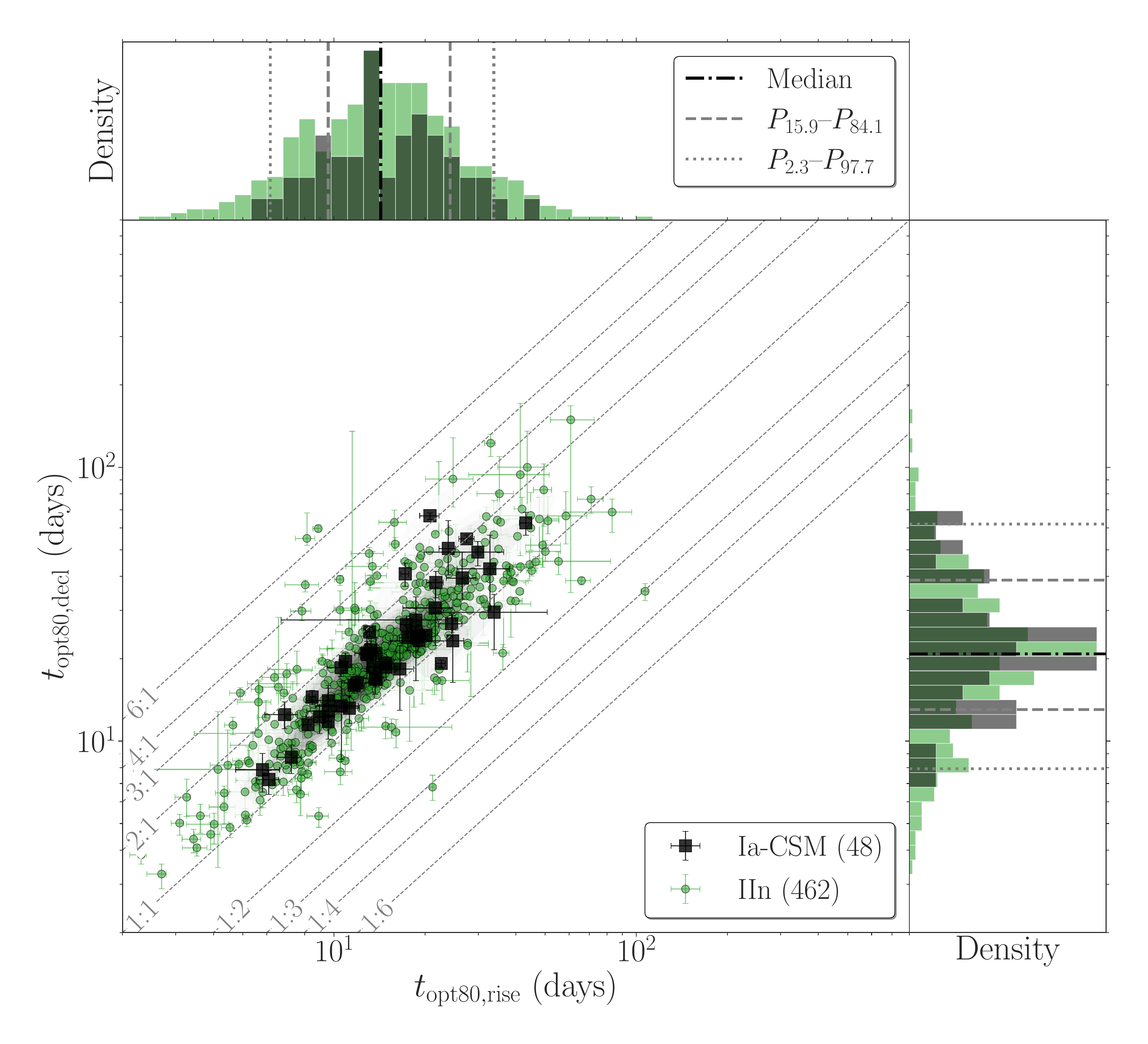} 
    \includegraphics[width=0.46\textwidth]{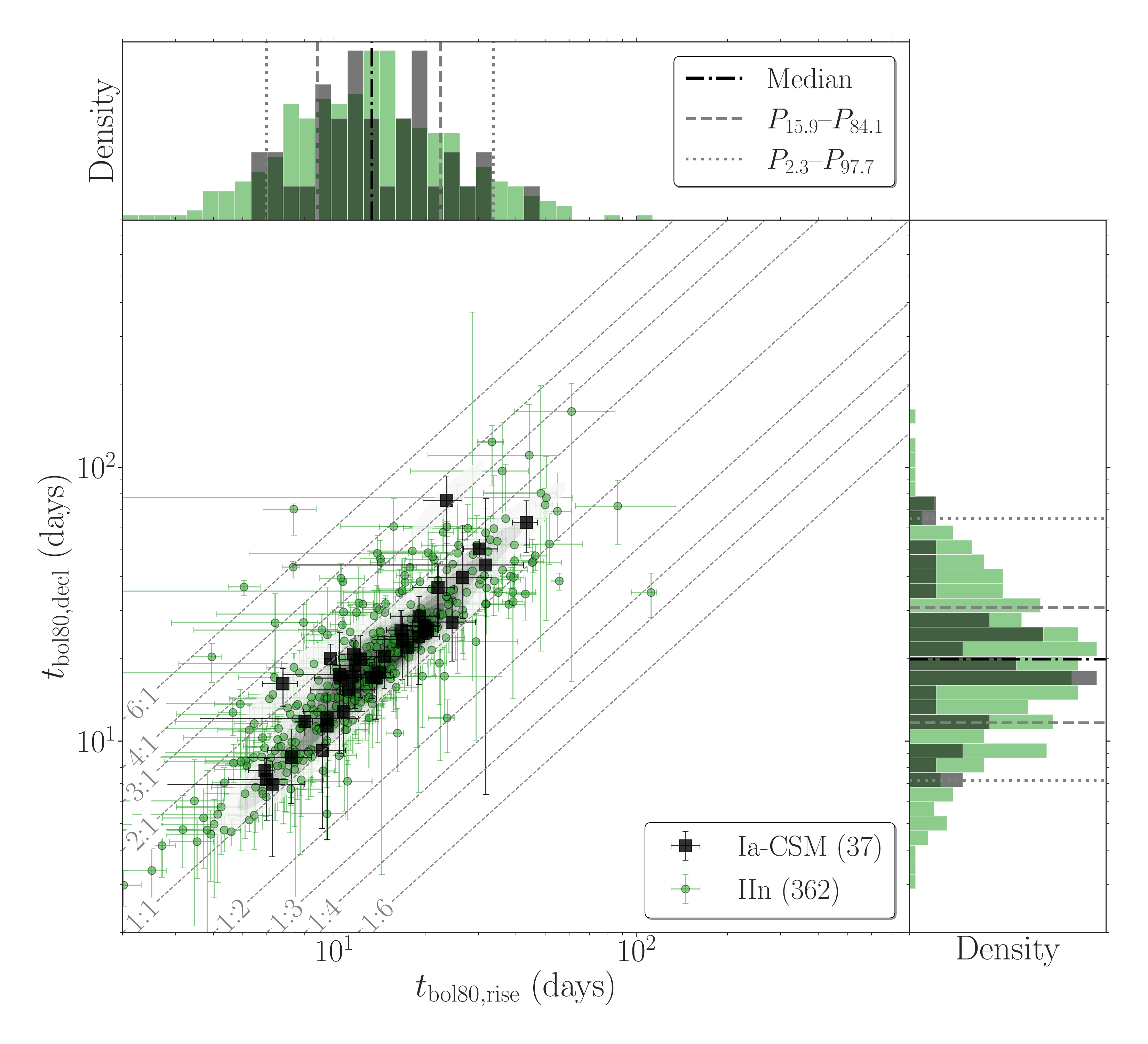} 
\caption{Similar to Figure~\ref{fig:rT205080}, but for optical (\textit{left}) and bolometric (\textit{right}) luminosity.
}
    \label{apfig:optbolT205080}
\end{figure*}

\begin{figure*}
    \centering
    \includegraphics[width=0.49\textwidth]{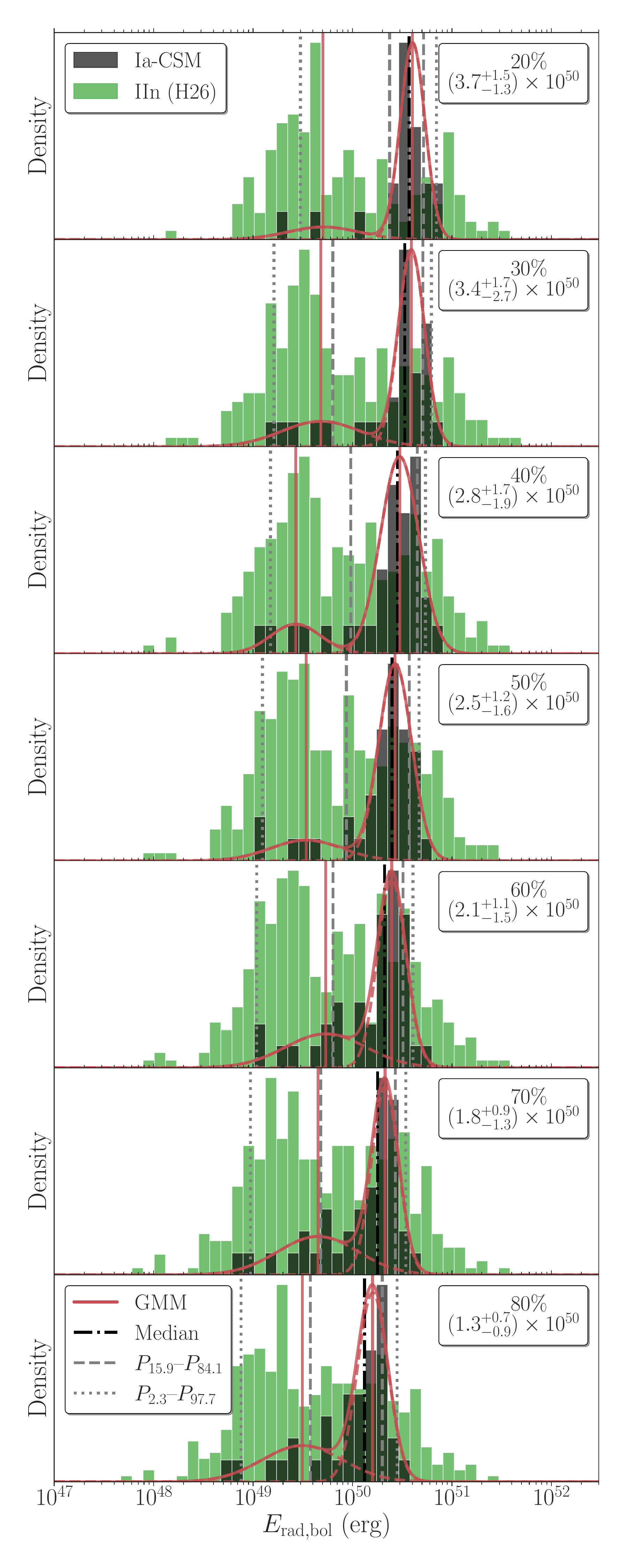}
    \includegraphics[width=0.49 \textwidth]{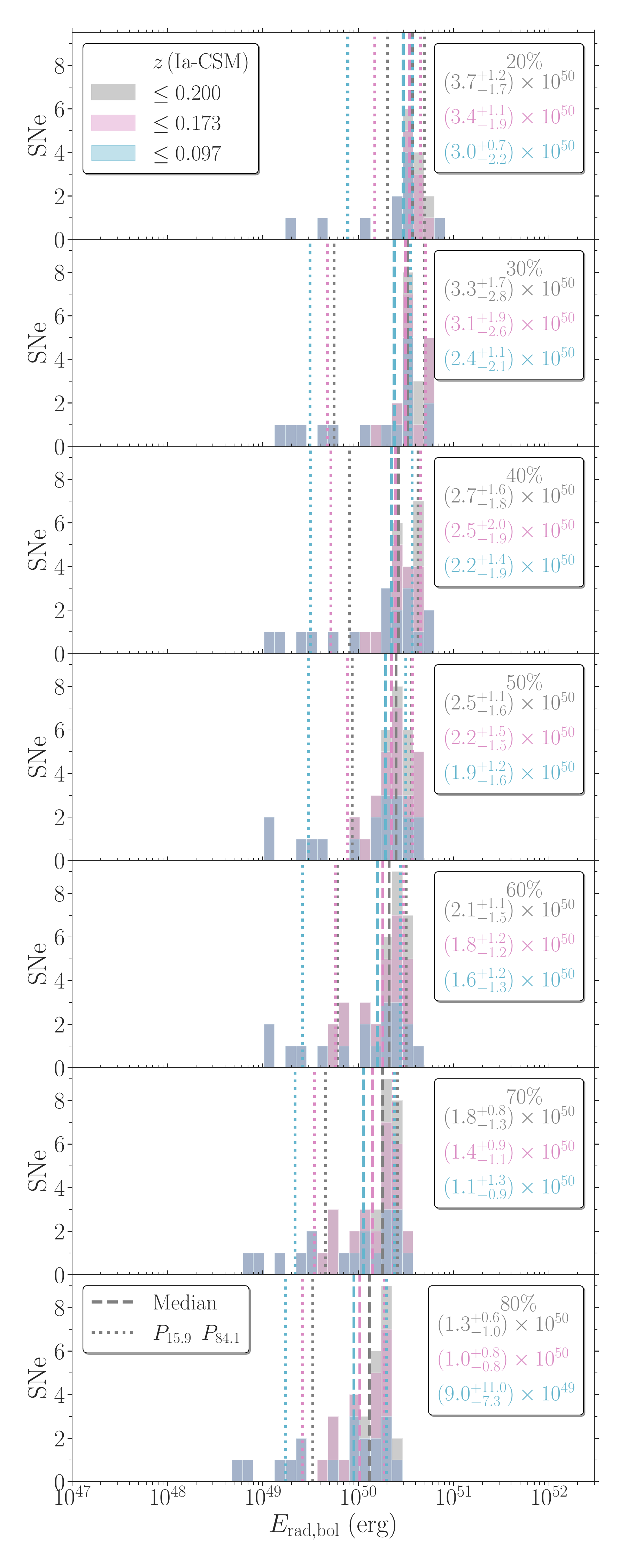}
\caption{Similar to Figure~\ref{fig:Eradopt}, but for bolometric luminosity.
}
    \label{apfig:Eradbol}
\end{figure*}

The sample timescale evolution and resulting correlations in the bolometric and pseudo-bolometric (optical: 3250--8900$\,\Angstrom$) light curves are displayed in Figures~\ref{apfig:optbolT} and \ref{apfig:optbolT205080}. Additionally, the distributions of the bolometric radiated energy are presented in Figure~\ref{apfig:Eradbol}.

\bibliographystyle{aasjournal}
\bibliography{main,tns}

\end{document}